\documentclass[aps,prb,twocolumn,superscriptaddress,showpacs]{revtex4}
\usepackage{feyn}
\usepackage{graphicx}
\usepackage{latexsym}
\usepackage{amssymb}
\usepackage{amsmath}
\usepackage{amsfonts}
\usepackage{bm}
\usepackage{multirow}
\usepackage{color}
\usepackage{comment}
\newcommand{\ii}{\mathrm{i}}
\newcommand{\dd}{\mathrm{d}}

\renewcommand{\Im}{\mathop{\mathrm{Im}}}

\newcommand{\U}{\mathrm{U}}

\newcommand{\vect}[1]{{\bm{#1}}}

\newcommand{\beq}{\begin{equation}}
\newcommand{\eeq}{\end{equation}}
\newcommand{\beqn}{\begin{eqnarray}}
\newcommand{\eeqn}{\end{eqnarray}}

\DeclareMathAlphabet{\mathbbold}{U}{bbold}{m}{n}

\def\U{{\rm U}}

\newcommand{\cxd}[1]{\textcolor{black}{#1}}

\newcommand{\ua}{\uparrow}
\newcommand{\da}{\downarrow}

\usepackage{ulem}

\begin{document}

\title{Interaction driven Metal-Insulator Transition with Charge Fractionalization}

\author{Yichen Xu}
\affiliation{Department of Physics, University of California,
Santa Barbara, CA 93106, USA}

\author{Xiao-Chuan Wu}
\affiliation{Department of Physics, University of California,
Santa Barbara, CA 93106, USA}

\author{Mengxing Ye}
\affiliation{Kavli Institute for Theoretical Physics, University
of California, Santa Barbara, CA 93106, USA}

\author{Zhu-Xi Luo}
\affiliation{Kavli Institute for Theoretical Physics, University
of California, Santa Barbara, CA 93106, USA}

\author{Chao-Ming Jian}
\affiliation{Department of Physics, Cornell University, Ithaca,
New York 14853, USA}

\author{Cenke Xu}
\affiliation{Department of Physics, University of California,
Santa Barbara, CA 93106, USA}

\begin{abstract}

It has been proposed that an extended version of the Hubbard model
which potentially hosts rich correlated physics may be well
simulated by the transition metal dichalcogenide (TMD) moir\'{e}
heterostructures. Motivated by recent reports of continuous
metal-insulator transition (MIT) at half filling, as well as
correlated insulators at various fractional fillings in TMD
moir\'{e} heterostructures, we propose a theory for the
potentially continuous MIT with fractionalized electric charges.
The charge fractionalization at the MIT will lead to various
experimental observable effects, such as a large critical
resistivity as well as large universal resistivity jump at the
continuous MIT. These predictions are different from previously
proposed theory for interaction-driven continuous MIT. Physics in
phases near the MIT will also be discussed.

\end{abstract}

\maketitle

\section{Introduction}

Many correlated phenomena have been observed in graphene-based
moir\'{e} systems, such as high temperature superconductivity
(compared with the bandwidth of the moir\'{e} bands), correlated
insulators~\cite{pablo1,pablo2,wang1,young1,young2,efetov,wang2,kim1,pablo3},
and the strange metal phase~\cite{pablostrange,youngstrange}, etc.
The most fundamental reason for the emergence of these correlated
physics is that the slow modulating moir\'{e} potential leads to
very narrow bandwidths~\cite{tbgmodel1,tbgmodel2}. Great
theoretical interests and efforts have been devoted to the
graphene based moir\'{e}
systems~\cite{xubalents,fu1,fu2,fu3,subir1,steve1,senthil1,vafek1,senthil2,you,zaletel,zaletel2,wu,lian,lee2019,skyrmion,xunematic,fernandesnematic}.
But the theoretical description and understanding of the graphene
based moir\'{e} systems may be complicated by the fact that in the
noninteracting limit the moir\'{e} mini bands can have various
types of either robust or fragile nontrivial
topologies~\cite{wangtopo,youngtopo,senthiltopo1,wangtopo2,repellintopo,efetovtopo,dean1,yacobytopo,liantopo,wuchern},
although the exact role of the band topology to the interacting
physics at fractional filling is not entirely clear. Hence similar
narrow band systems with trivial band topology and unambiguous
concise theoretical framework would be highly desirable. It was
proposed that much of the correlated physics of the transition
metal dichalcogenide (TMD) moir\'{e} heterostructure can be
captured by an extended Hubbard model with an effective spin-1/2
electron on a triangular moir\'{e} lattice~\cite{tmdhubbard} \beqn
H = \sum_{\vect{r}, \vect{r}', \alpha} - t_{\vect{r}, \vect{r}'}
c^\dagger_{\vect{r},\alpha} c_{\vect{r}',\alpha} + H.c. +
\sum_\vect{r} U n_{\vect{r},\ua}n_{\vect{r},\da} + \cdots
\label{hubbard}\eeqn The electron operator $c_{\vect{r},\alpha}$
is constructed by states within a topologically trivial moir\'{e}
mini band. Due to the strong spin-orbit coupling, the spin and
valley degrees of freedom are locked with each other in the TMD
moir\'{e} system. We will use $\alpha = \ua,\da$ or $1,2$ to
denote two spin or equivalently two valley flavors. When a
moir\'{e} band is partially filled, the correlated physics within
the partially filled moir\'{e} mini bands may be well described by
Eq.~\ref{hubbard}, which only contains half of the degrees of
freedom of a mini band in a graphene based moir\'{e} system. The
ellipsis in Eq.~\ref{hubbard} can include further neighbor
hopping, ``spin-orbit" coupling terms allowed by
symmetry~\cite{sarmatmd}, and further neighbor interaction. Note
the ``spin-orbit" coupling here refers to the hopping terms in
Eq.~\ref{hubbard} that depend on the spin index $\alpha$ and
should not be confused with the bare spin-orbit coupling within
the TMD system before the moir\'{e} superlattice is imposed. The
TMD moir\'{e} systems are hence considered as a simulator for the
extended Hubbard model on a triangular lattice~\cite{tmdhubbard2}.

Like the graphene-based moir\'{e} systems, the TMD moir\'{e}
heterostructure is a platform for many correlated physics. This
manuscript mainly concerns the metal-insulator transition (MIT)
driven by interaction. The MIT of the Hubbard model on a
triangular lattice has attracted much numerical efforts
recently~\cite{joelmike,szasz}. The symmetry of the TMD moir\'{e}
heterostructure is different from the simplest version of the
Hubbard model, hence even richer physics can happen in the system.
Continuous MIT has been reported at half-filling of the moir\'{e}
bands (electron filling $\nu = 1/2$, or one electron per moir\'{e}
unit cell on average) in the TMD moir\'{e}
system~\cite{tmdmit1,tmdmit2}. The experimental tuning parameter
of the MIT in the TMD heterostructure is the displacement field,
i.e. an out-of-plane electric field, which tunes the width of the
mini moir\'{e} bands, and hence the ratio between the kinetic and
interaction energies in the effective Hubbard model. Correlated
insulators have also been observed at various other fractional
electron fillings, though the nature of the MITs at these
fractional fillings have not been thoroughly inspected
experimentally~\cite{tmdinsulator1,tmdinsulator2,tmdinsulator3,tmdinsulator4}.
In this manuscript we will mainly focus on $\nu = 1/2$, but other
fractional fillings will also be briefly discussed.

The nature of an interaction driven MIT depends on the nature of
the insulator phase near the MIT. The Hubbard model on the
triangular lattice has one site per unit cell, which based on the
generalized Lieb-Shultz-Matthis theorem~\cite{LSM,hastings}
demands that the insulator phase at half-filling should not be a
trivial incompressible (gapped) state which preserves the
translation symmetry. If the insulator phase has a semiclassical
spin order that breaks the translation symmetry, the evolution
between the metal and insulator could involve two transitions: at
the first transition a semiclassical spin order develops, which
reduces the Fermi surface to several Fermi pockets; and at the
second transition the size of the Fermi pockets shrink to zero,
and the system enters an insulator phase. A more interesting
scenario of the MIT only involves one single
transition~\cite{lee2005,senthilmit1,senthilmit2}, but then the
insulator phase is not a semiclassical spin order, instead it is a
spin liquid state with a spinon Fermi surface. An intuitive
picture for this transition is that, at the MIT, the charge
degrees of freedom are gapped, but the spins still behave as if
there is a ``ghost" Fermi surface. The spinon Fermi surface can
lead to the Friedel oscillation just like the metal
phase~\cite{friedelspinliquid}. The structure of the Fermi surface
does not change drastically across the transition.

In a purely two dimensional system, conductivity (or resistivity)
is a dimensionless quantity, hence it can take universal value at
the order of $e^2/h$ (or $h/e^2$) in various scenarios. For
example, the Hall conductivity of the quantum Hall state is
precisely $\sigma_H = \nu e^2/h$; the conductivity (or
resistivity) at a $(2+1)d$ quantum critical point also takes a
universal value at the order of $e^2/h$ (or $h/e^2$)~\cite{UCt1}.
One central prediction given by the theory above for interaction
driven continuous MIT is that, there is a universal resistivity
jump at the order of $\sim h/e^2$ at the MIT compared with the
metal phase; and the critical resistivity at the MIT should also
be close to the order of $h/e^2$ (we will review these predictions
in the next section). In this manuscript we will argue that the
current experimental observations suggest that the nature of the
MIT in MoTe$_2$/WSe$_2$ moir\'{e} superlattice without
twisting~\cite{tmdmit1} is beyond the previous
theory~\cite{lee2005,senthilmit1,senthilmit2}, and we propose an
alternative candidate theory of MIT with further charge
fractionalizations. We will discuss how the alternative theory can
potentially address the experimental puzzles, and more predictions
based on our theory will be made. Our assumption is that the MIT
in this system is indeed driven by electron-electron interaction
(as was suggested by Ref.~\onlinecite{tmdmit1}); If the disorder
plays the dominant role in this system, the MIT may be described
by the picture discussed in Ref.~\onlinecite{kivelsonreview}.

The paper is organized as follows: In section~\ref{parton} we
introduce an alternative parton construction for systems described
by the extended Hubbard model with a spin-orbit coupling, which
naturally leads to charge fractionalization at the
interaction-driven MIT even at half-filling; we also give an
intuitive argument of physical effects caused by charge
fractionalization at the MIT. In section~\ref{densitywave}, we
will discuss the theory for MIT when the insulating phase
spontaneously breaks the translation symmetry. Section~\ref{TO}
studies the theory of MIT when the insulating phase has different
types of topological orders. In section~\ref{experiment} we
discuss various experimental predictions based on our theory, for
the MIT and also the phases nearby. We present the details of our
theory in the appendix, including the projective symmetry group,
field theories, and calculation of DC resistivity, etc.

\section{Two Parton constructions}
\label{parton}

The previous theory for the interaction-driven continuous MIT for
correlated electrons on frustrated lattices was based on a parton
construction. The parton construction splits the quantum number of
an electron into a bosonic parton which carries the electric
charge, and a fermionic parton which carries the spin. In the
current manuscript we compare two different parton constructions:
\beqn \mathrm{I}: c_{\vect{r},\alpha} = b_\vect{r}
f_{\vect{r},\alpha}, \ \ \ \ \mathrm{II}: c_{\vect{r},\alpha} =
b_{\vect{r},\alpha} f_{\vect{r},\alpha}. \label{parton0}\eeqn In
parton construction-I only one species of charged bosonic parton
$b$ is introduced for electrons with both spin/valley flavors;
while in parton construction-II a separate charged bosonic parton
$b_\alpha$ is introduced for each spin/valley flavor. As we will
see later, the two different parton constructions will lead to
very different observable effects. The construction-I is the
standard starting point of the theory of MIT that was used in
previous literature~\cite{lee2005,senthilmit1,senthilmit2};
construction-II is usually unfavorable for systems with a full
spin SU(2) invariance, because the parton construction itself
breaks the spin rotation symmetry. But the construction-II is a
legitimate parton construction for the system under study, whose
band structure in general does not have full rotation symmetry
between the two spin/valley flavors.

The time-reversal symmetry of the microscopic TMD system relates
the two spin/valley flavors. But it is not enough to guarantee a
full SU(2) rotation symmetry between the two flavors. In fact,
since the two flavors can be tied to the two valleys of the TMD
material, the trigonal warping of the TMD bands, which takes {\it
opposite} signs for the two different valleys, can lead to the
breaking of such an SU(2) rotation symmetry. To estimate the
trigonal warping effect in the Hubbard model, one can compare the
$k^2$ term and the $k_x^3 - 2 k_x k^2_y$ term in the electron
dispersion of one of the two layers in the heterostructure
expanded at one valley. Then the relative strength of the trigonal
warping compared to the SU(2)-invariant hopping in
Eq.~\ref{hubbard} is given by the ratio between the lattice
constant of the original TMD material and that of the mori\'{e}
superlattice. In addition, the natural microscopic origin of the
interactions in the Hamiltonian Eq.~\ref{hubbard} is the Coulomb
interaction between the electrons. The Coulomb interaction when
projected to the low-energy bands relevant to the moir\'{e}-scale
physics is expected to contain SU(2)-breaking interaction terms.
The momentum conservation only guarantees the valley U(1)
symmetry. Assuming the unscreened Coulomb interaction between
electrons before the projection to the low-energy bands, further
neighbor interaction will appear in the extended Hubbard model.
The relative strength of the SU(2)-breaking interaction terms
obtained from the projection compared to the SU(2)-invariant
interactions can again be estimated by the ratio between the
lattice constant of the original TMD material and the moir\'{e}
superlattice, as the Fourier transform of unscreened Coulomb
interaction in $2d$ space is $V_q \sim 1/q$.

The most important difference between these two parton theories
resides in the filling of the bosonic partons. Since each bosonic
parton carries the same electric charge as an electron, the total
number of bosonic partons should equal to the number of electrons.
Hence at electron filling $\nu$ (meaning $2\nu$ electrons per unit
cell), the filling factor of boson $b$ in construction-I is $\nu_b
= 2\nu$, i.e. $2\nu$ bosonic parton per unit cell; in
construction-II the filling factor of boson $b_\alpha$ has filling
factor $\nu^\alpha_b = \nu$ for each spin/valley flavor. Hence
even with one electron per site (half-filing or $\nu = 1/2$ of the
extended Hubbard model), the bosonic parton in construction-II is
already at half filling for each spin/valley flavor. The
half-filling will lead to nontrivial features inside the Mott
insulator phase, as well as at the MIT. Another more theoretical
difference is that, in construction-I there is one dynamical
emergent $\U(1)$ gauge field $a_\mu$ which couples to both $b$ and
$f_\alpha$; while in construction-II there are two dynamical
$\U(1)$ gauge fields $a_{\alpha,\mu}$, one for each spin/valley
flavor.

%The current experimental observation of the MIT in
%MoTe$_2$/WSe$_2$ moir\'{e} heterostructure at $\nu = 1/2$ also
%suggests that construction-II may be a more desirable formalism in
%the system.
In construction-I, the bosonic parton $b$ is at
integer filling, and the MIT is naturally interpreted as a
superfluid to Mott insulator (SF-MI) transition of boson $b$. At
the MIT, using the Ioffe-Larkin rule~\cite{larkin}, the DC
resistivity of system is $\rho = \rho_b + \rho_f$, where $\rho_b$
and $\rho_f$ are the resistivity contributed by the bosonic and
fermionic partons respectively. $\rho_f$ caused by disorder or
interaction such as the Umklapp process is a smooth function of
the tuning parameter, the drastic change of $\rho$ across the MIT
arises from $\rho_b$. In the metal phase, i.e. the ``superfluid
phase" of $b$, $\rho_b$ is zero, and the total resistivity is just
given by $\rho_f$. Also, in the superfluid phase of $b$, the
$\U(1)$ gauge field $a_\mu$ that couples to both $b$ and
$f_\alpha$ is rendered massive due to the Higgs mechanism caused
by the condensate of $b$. In the insulator phase, $\rho_b$ and
$\rho$ are both infinity, and the system enters a spin liquid
phase with a spinon Fermi surface of $f_\alpha$ that couples to
the dynamical $\U(1)$ gauge field $a_\mu$. The MIT which
corresponds to the condensation of $b$ belongs to the 3D XY
universality class. The dynamical gauge field $a_\mu$ is argued to
be irrelevant at the transition due to the overdamping of the
gauge field that arises from the spinon Fermi
surface~\cite{senthilmit1,senthilmit2}, and hence does not change
the universality class of the SF-MI transition of $b$.

In parton construction-I, at the MIT the bosonic parton
contribution to the resistivity $\rho_b$ is given by $\rho_b = R
h/e^2$, where $R$ is an order-1 universal constant. In the order
of limit $T \rightarrow 0$ before $\omega \rightarrow 0$, $R$ is
associated to the 3D XY universality class~\cite{UC1}, because the
gauge field $a_\mu$ is irrelevant as mentioned above. This
universal conductivity at the 3D XY transition has been studied
through various analytical and numerical
methods~\cite{UCt1,UCt2,UCt3,cond3,UCt4,UCt5,UCe1,UCe2,UCe3}. At
finite $T$ and zero frequency, the gauge field $a_\mu$ can
potentially enhance the value $R$ to $R' > R$, based on a
large-$\bf{N}$ calculation in Ref.~\onlinecite{resistivity2}
($\bf{N}$ is different from $N$ in our work). The evaluation in
Ref.~\onlinecite{resistivity2} gave $R' \sim 7.92$, while we
evaluate the same quantity to be $R' \sim 7.44$. Hence the
prediction of the construction-I is that, the DC resistivity of
the system right at the MIT has a universal jump compared with the
resistivity at the metallic phase close to the
MIT~\cite{senthilmit1,senthilmit2}, i.e. $\Delta \rho = \rho_b =
R' h /e^2$. With moderate disorder, at the MIT $\rho_b$ of the
bosonic parton is supposed to dominate the resistivity $\rho_f$ of
the fermionic parton $f_\alpha$, hence the total resistivity $\rho
= \rho_b + \rho_f$ should be close to $\rho_b$.

In the experiment on the MoTe$_2$/WSe$_2$ moir\'{e} superlattice,
it was reported that disorder in the system is playing a
perturbative role, and the continuous MIT is mainly driven by the
interaction~\cite{tmdmit1}. However, the reported resistivity
$\rho$ increases rapidly with the tuning parameter (the
displacement field) near the MIT. The bare value of $\rho$ near
and at the MIT is significantly greater than $h/e^2$ (and
significantly larger than the computed value of $\rho_b \sim R'
h/e^2$ mentioned above), and it is clearly beyond the
Mott-Ioffe-Regel limit, i.e. the system near and at the MIT is a
very ``bad metal"~\cite{badmetal,badmetal2}.
%Hence the resistivity jump
%$\Delta \rho_b$ at the MIT is likely significantly larger than
%$h/e^2$.
This suggests that the MIT is not a simple SF-MI transition of
$b$, or in other words $b$ should be ``much less conductive"
compared with what was predicted in construction-I considered in
previous literature. We will demonstrate that construction-II can
potentially address the large resistivity at the MIT. The most
basic picture is that, since $b_1$ and $b_2$ are both at
half-filling, the LSM theorem~\cite{LSM,hastings} dictates that
the Mott insulator phase of each flavor of boson cannot be a
trivial insulator, namely the Mott insulator must either be a
density wave that spontaneously breaks the translation symmetry,
or have topological order. In either case, the MIT is not a simple
3D XY transition, and the most prominent feature of the transition
is that, the bosonic parton number (or the electric charge) must
further fractionalize.

The MIT with charge fractionalization will be discussed in detail
in the next section using the dual vortex formalism, but the
consequence of this charge fractionalization can be understood
from a rather intuitive picture. Suppose $b$ fractionalizes into
$N$ parts at the MIT, meaning the charge carriers at the MIT have
charge $e_\ast = e/N$, then each charge carrier will approximately
contribute a resistivity at the order of $h/e_\ast^2 \sim N^2 h /
e^2$ at the MIT; and if there are in total $N_{b}$ species of the
fractionalized charge carriers, at the MIT the bosonic parton will
approximately contribute resistivity \beqn \rho_b \sim \frac{N^2
h}{N_{b} e^2}. \label{rhob0} \eeqn There is a factor of $N_b$ in
the denominator because intuitively the total conductivity of $b$
will be a sum of the conductivity of each species of
fractionalized charge carriers, i.e. $\sigma_b = \sum_{j =
1}^{N_b} \sigma_j$, in the unit of $e^2/h$ (a more rigorous rule
of combining transport from different partons will be discussed
later). Hence when $N^2/N_b > 1$, the construction-II with
inevitable charge fractionalization can serve as a natural
explanation for the large $\rho$ at the MIT, and it will also
predict a large jump of resistivity $\Delta \rho$ at the MIT.

\section{Mott insulator with translation symmetry breaking}

\label{densitywave}

\subsection{General Formalism}

In this section we will discuss the MIT following the parton
construction-II discussed in the previous section. The MIT is
still interpreted as the SF-MI transition of both spin/valley
flavors of the bosonic parton $b_\alpha$, although as we discussed
previously the insulator cannot be a trivial incompressible state
of $b_\alpha$. In the superfluid phase of $b_\alpha$, both $\U(1)$
gauge fields $a_{1,\mu}$ and $a_{2,\mu}$ that couple to the two
flavors of partons are gapped out by the Higgs mechanism, and the
system enters a metal phase of the electrons; $b_1$ and $b_2$ must
undergo the SF-MI transition simultaneously, since the
time-reversal or spatial reflection symmetries both interchange
the two flavors of partons due to the spin-valley locking.

\begin{figure}
\includegraphics[width=0.4\textwidth]{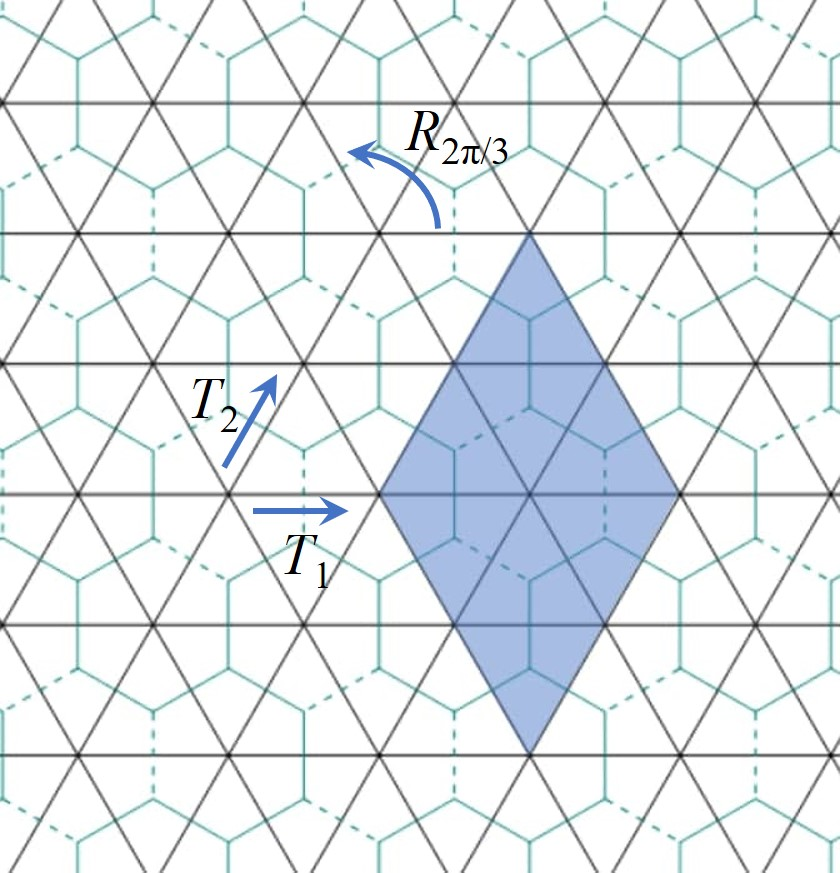}
\caption{The triangular moir\'{e} lattice, and its dual honeycomb
lattice. In the parton construction-II, the bosonic parton
$b_\alpha$ is at half-filling for each spin/valley flavors, which
becomes a $\pi-$flux of the dual gauge field $A_\mu$ through the
hexagon of the dual honeycomb lattice. Hence the vortex $\psi$
defined on the dual honeycomb lattice does not have a uniform
hopping amplitude, the dashed links on the dual honeycomb lattice
have negative hopping amplitudes. The symmetry of the lattice will
be realized as a projective symmetry group. There are eight dual
sites per unit cell (shaded area) in this gauge choice. At each
spin/valley flavor, there are translation symmetries $T_{1,2}$, a
rotation symmetry $R_{\frac{2\pi}{3}}$, and a product of
reflection $P_x (x \rightarrow -x)$ and time-reversal
$\mathcal{T}$. We also argue that $P_y$ is a symmetry of the
system as long as there is no valley mixing; and the six-fold
rotation $R_{\pi/3}$ becomes a good approximate symmetry of the
Hubbard model in the case of long moir\'{e} lattice constant.}
\label{lattice}
\end{figure}

The dual vortex theory~\cite{peskindual,halperindual,leedual} is
the most convenient formalism that describes a transition between
a superfluid and a nontrivial insulator of a boson at fractional
filling. If we start with a boson $b$, after the boson-vortex
duality, a vortex of the superfluid phase of $b$ becomes a point
particle that couples to a dynamical $\U(1)$ gauge field $A_\mu$,
which is the dual of the Goldstone mode of the superfluid (not to
be confused with the $\U(1)$ gauge field $a_\mu$ mentioned before
that couples to the bosonic parton $b$). In the dual picture, the
superfluid phase of $b$ (which corresponds to the metal phase of
the electron) is the insulator phase of the vortex field; while
the Mott insulator phase of $b$ corresponds to the condensate of
the vortices, which ``Higgses" the $\U(1)$ gauge field $A_\mu$,
and drives the boson $b$ into a gapped insulator phase. If at low
energy there is only one component of vortex field with gauge
charge $1$ under $A_\mu$ (which corresponds to integer filling of
boson $b$), the insulator phase of $b$ is a trivial insulator
without any further symmetry breaking or topological order; if
there are more than one component of the vortex fields at low
energy, or if the vortex field carries multiple gauge charges of
$A_\mu$, the insulator must be of nontrivial nature.

For example, when $b$ has a fractional filling $\nu_b = 1/q$ with
integer $q$, Ref.~\onlinecite{vortexbalents,burkovbalents} studied
the quantum phase transition between the bosonic SF and various
MIs with commensurate density waves which spontaneously break the
translation symmetry but have no topological order. The study is
naturally generalized to filling factor $\nu_b = p/q$ with coprime
integers $(p,q)$. We can use this formalism in our system.
Hereafter we focus on one spin/valley flavor $\alpha$, and the
index $\alpha$ will be hidden for conciseness. In this case the
theory for the SF-MI transition at one spin/valley flavor is:
\beqn \mathcal{L}^{(1)} &=& \sum_{j = 0}^{N-1} (|(\partial_\mu -
\ii A_\mu)\psi_j|^2 + r |\psi_j|^2) + u (\sum_{j = 0}^{N-1}
|\psi_j|^2)^2 \cr\cr && + \frac{\ii}{2\pi} A \wedge d (a + e
A_{\mathrm{ext}}) + \cdots \label{cpn}\eeqn Here $\psi_j$ with $j
\in\{0, \cdots N-1\}$ are $N$ flavors of vortex fields of the
boson $b$ at low energy, and $A_\mu$ is the dual gauge field of
boson $b$: $\frac{1}{2\pi} dA = J_b $, where $J_b$ is the current
of boson $b$. $a_\mu$ is the gauge field that couples to both $b$
and $f$, and $A_{\mathrm{ext}}$ is the external electromagnetic
field. The reason there are $N$ flavors of the vortex field is
that, the vortex which is defined on a dual honeycomb lattice will
view the partially filled boson density as a fractional background
flux of the dual gauge field $A_\mu$ through each hexagon, and the
band structure of the vortex will have multiple minima in the
momentum space. The degeneracy of the multiple minima is protected
by the symmetry of the triangular lattice. $\psi_j$ transforms as
a representation of the projective symmetry group (PSG) of the
lattice. Notice that since Eq.~\ref{cpn} describes one of the two
spin/valley flavors, the PSG that constrains Eq.~\ref{cpn} should
include translation, and $2\pi/3$ rotation of the lattice
($R_{\frac{2\pi}{3}}$). There is another more subtle symmetry $P_x
\mathcal{T}$ for each spin/valley flavor of the boson and vortex
fields. $P_x$ that takes $x \rightarrow -x$, and time-reversal
$\mathcal{T}$ both exchange the two spin/valley indices, but their
product will act on the same spin/valley species, and part of its
role is to take momentum $k_y$ to $-k_y$.

In the appendix we will argue that $P_y$ which takes $y$ to $-y$
within each valley is also a good symmetry of the system, as long
as valley mixing is negligible. One consequence of the $P_y$
symmetry is that the expectation value of gauge flux $da$ can be
set to zero for the theory Eq.~\ref{cpn}, or equivalently the
$P_y$ symmetry ensures that the ``chemical potential" term
$\psi_j^* \partial_\tau \psi_j$ does not appear in Eq.~\ref{cpn},
as $P_y$ transforms a vortex to anti-vortex: $\psi_a \rightarrow
U_{ab}\psi^\ast_b$. Also, with long moir\'{e} lattice constant,
the trigonal warping $k_x^3 - 3 k_x k_y^2$ in each valley of the
original BZ of the system becomes less important compared with the
leading order quadratic dispersion expanded at each valley, hence
the six-fold rotation $R_{\pi/3}$ becomes a good approximate
symmetry of the effective Hubbard model with long moir\'{e}
lattice constant.

The theory in Eq.~\ref{cpn} also has an emergent particle-hole
symmetry. The simplest choice of the particle-hole symmetry is
$\psi_a \rightarrow U_{ab} \psi^*_b$, $A\rightarrow -A$,
$a\rightarrow -a$ and $A_{\rm ext} \rightarrow - A_{\rm ext} $.
Although we used the same transformation matrix $U_{ab}$ as $P_y$,
this emergent particle-hole symmetry is different from $P_y$ as it
does not involve any spatial transformations. Note that any
(spatially uniform) $P_y$-symmetric terms involving only the
``matter fields" $\psi_j$ must also preserve this emergent
particle-hole symmetry. Another potentially relevant
particle-hole-symmetry-breaking perturbation that needs to be
examined is given by the finite density of the fluxes $dA$. $dA$
is tied to the physical U(1) charge density (compared to the
charge density set by the fixed electron filling $\nu=1/2$) and
hence should have a vanishing spatial average. At the SF-MI
transition point, the translation symmetry of the theory
Eq.~\ref{cpn} and the fact that $dA$ has a vanishing spatial
average guarantee that $dA$ has a vanishing expectation value
everywhere, which respects the particle-hole symmetry. Therefore,
the particle-hole symmetry is a valid emergent symmetry at the
SF-MI critical point described by Eq.~\ref{cpn}. The same argument
would also conclude the emergent particle-hole symmetry at the
ordinary SF-MI transition in the Bose-Hubbard model.

For parton construction-II, when the electron has filling $\nu =
1/2$, both $b_1$ and $b_2$ are at filling $\nu^\alpha_b = 1/2$.
For each flavor of $b_\alpha$, the formalism in
Ref.~\onlinecite{burkovbalents} would lead to a dual vortex theory
with $N = 4$ components of vortex fields, i.e. there are four
degenerate minima of the vortex band structure in the momentum
space for each spin/valley index. This calculation is analogous to
the frustrated Ising model on the honeycomb
lattice~\cite{sondhiising,xusachdevtriangle}. Using the gauge
choice of Fig.~\ref{lattice}, the four minima are located at the
$K$ and $K'$ points of the reduced Brillouin zone (BZ), with two
fold degeneracy at each point.

\subsection{From $N = 4$ to ``$N = \infty$"}

\begin{figure}
\includegraphics[width=0.5\textwidth]{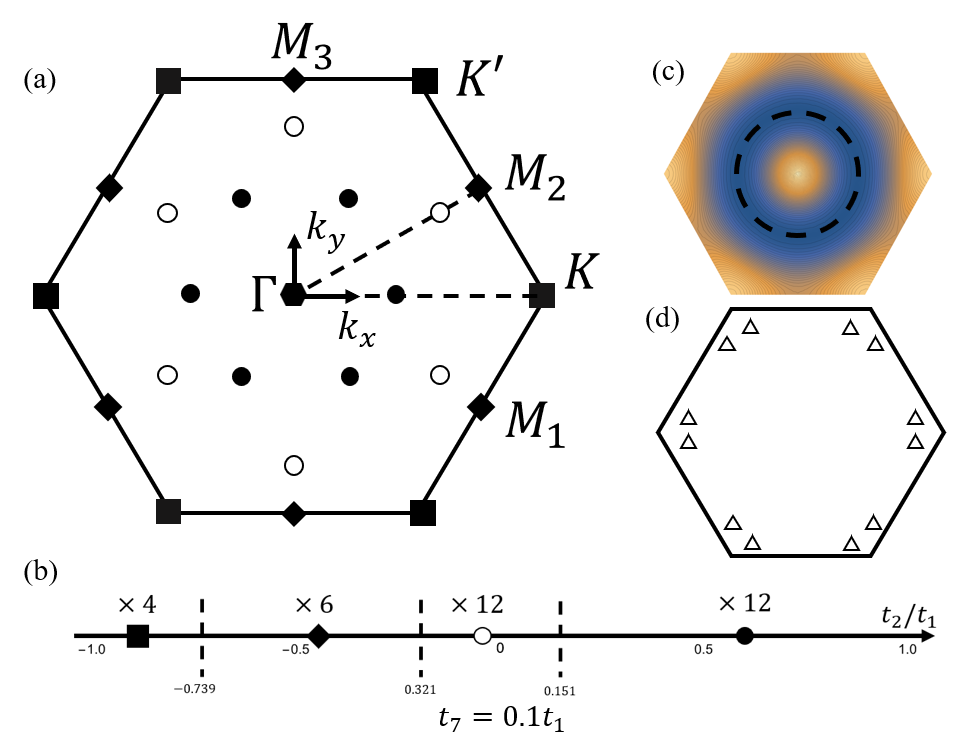}
\caption{(a) The minima of the vortex band structure. With nearest
neighbor vortex hopping on Fig.~\ref{lattice}, the minima locate
at the $K$ and $K'$ points of the Brillouin zone, each $K$ point
has two fold degeneracy; with further neighbor hoppings, the
minima can shift to the three $M$ points, still with two fold
degeneracy at each $M$ point. (b) The phase diagram of vortex
modes with seventh neighbor hopping $t_7=0.1t_1$, and by tuning
$t_2$ there are two regions in the phase diagram with $N = 12$
vortex modes at low energy. The 12 vortex modes are located either
on the lines between $\Gamma$ and $K/K'$ or $\Gamma$ and $M$. (c)
With only $t_1$ and $t_2$, there is a large region of the phase
diagram where there is a ring degeneracy of the vortex band
structure. (d) All the symmetries (including approximate
symmetries) of the system can protect up to 24 degenerate vortex
modes, which locate at 12 incommensurate momenta in the BZ.}
\label{BZ}
\end{figure}

Ref.~\onlinecite{burkovbalents} considered a specific band
structure of the vortex, which only involved the nearest neighbor
hopping of vortices on the dual honeycomb lattice. But there is no
fundamental reason that further neighbor hopping of vortices
should be excluded. Indeed, once we take into account of further
neighbor hopping, the dual vortex theory has a much richer
possibility. We have explored the phase diagram of the dual vortex
theory up to seventh neighbor hopping, and we obtained the phase
diagram in Fig.~\ref{BZ}$a$. Further neighbor hopping of the
vortex field can modify the band structure, and lead to $N = 6$ or
$N = 12$ components of vortex fields by choosing different hopping
amplitudes. The $N = 6$ minima are located at three inequivalent
$M$ points of the reduced BZ (Fig.~\ref{BZ}), each $M$ point again
has two-fold degeneracy. The two-fold degeneracy at each $M$ point
is protected by the translation symmetry of the triangular
moir\'{e} lattice only, which is required by the LSM theorem. The
shift of the vortex field minima from the $K$ points to $M$ points
is similar to what was discussed in the context of frustrated
quantum Ising models with further neighbor
couplings~\cite{kevintriangle,xubalents2011}. With symmetries
$T_{1,2}$, $R_{\frac{2\pi}{3}}$ and $P_x \mathcal{T}$ at each
spin/valley flavor, the degeneracy of the $N=6$ minima at the $M$
points are protected.

There are two regions in the phase diagram in Fig.~\ref{BZ}$b$
with $N = 12$ modes of vortex, two at each momentum. The six
incommensurate momenta at the minima of the vortex band structure
can be located either on the lines between $\Gamma$ and $K/K'$ or
$\Gamma$ and $M$. With the $R_{\pi/3}$ symmetry that becomes a
good approximate symmetry with long moir\'{e} lattice constant,
the degeneracy of the $N = 12$ vortex modes is protected. In
principle, all the symmetries together including $R_{\pi/3}$ can
protect up to $N = 24$ degenerate minima, as shown in
Fig.~\ref{BZ}$d$.

For a theory with $N$ components of vortex fields, the electric
charge carried by the boson $b$ will fractionalize. Under the
boson-vortex duality $\frac{1}{2\pi} dA = J_b $, the boson number
of $b$ becomes the flux number of the dual gauge field $A_\mu$.
The gauge flux of $A_\mu$ is trapped at the vortex core of each
field $\psi_j$ (we denote the vortex of $\psi_j$ as $\varphi_j$).
With $N$ components of the vortex fields, the vortex of each
$\psi_j$ field will carry $1/N$ flux quantum of the gauge field
$A_\mu$, hence the charge $e_\ast$ of each fractionalized charge
carrier should be $e/N$ at the MIT. And there are in total $N_{b}
= 2N$ species of the charge carriers (the factor of 2 comes from
the two spin/valley flavors).

With just $t_1$ and $t_2$ (first and second neighbor vortex
hopping), there is a large region of the parameter space where the
minima of the vortex band structure form a ring. This one
dimensional ring degeneracy is not protected by the symmetry of
the system, but its effect may still be observable for a finite
energy range. A ring degeneracy is analogous to $N = \infty$ in
Eq.~\ref{cpn}. Condensed matter systems with a ring degeneracy
have attracted considerable
interests~\cite{ringwu,ringzhai,ringchen,ringlake}. By integrating
out the vortices with ring degeneracy, a ``mass term" for the
transverse component of $A_\mu$ is generated in the infrared
limit~\cite{ringlake} (in the limit of momentum goes to zero
before frequency), meaning the fluctuation of $A_\mu$ is highly
suppressed, which is consistent with the intuition of $N =
\infty$.

The ellipsis in Eq.~\ref{cpn} includes other terms allowed by the
PSG of the triangular lattice, but break the enlarged flavor
symmetry of the CP$^{N-1}$ model field theory. More details about
PSG, extra terms in the Lagrangian, coupling to fermionic parton
$f_\alpha$~\cite{debanjan2021}, and the possible valence bond
solid orders with $N = 6$ will be discussed in appendix A and B.
The exact fate of the critical theory in the infrared is
complicated by these extra perturbations. It was shown previously
that nonlocal interactions can drive a transition to a new fixed
point~\cite{groveredge,edgexu1,edgexu2}, and here nonlocal
interactions arise from coupling to the fermionic
partons~\cite{debanjan2021}. Hence the transition may eventually
flow to a CFT different from the CP$^{N-1}$ theory in
Eq.~\ref{cpn}, or be driven to a first order transition
eventually. But as long as the first order nature is not strong,
the charge fractionalization and large resistivity to be discussed
in the next subsection is expected to hold at least for a
considerable energy/temperature window.

So far we have not paid much attention to the dynamical gauge
fields $a_\mu$ in parton construction-I or $a_{\alpha,\mu}$ in
construction-II shared by the bosonic and fermionic partons, as
the gauge coupling between $b$ ($b_\alpha$) and the gauge field is
irrelevant at the MIT with a background spinon Fermi surface. Here
we briefly discuss the fate of the spinon Fermi surface in the
insulator phase. When the bosonic parton $b$ is gapped, the theory
of spinon Fermi surface coupled with the dynamical $\U(1)$ gauge
field is a problem that has attracted a great deal of theoretical
efforts~\cite{polchinskinfl,nayaknfl1,nayaknfl2,nfl2,nfl3,nfl4,nfl5}.
These studies mostly rely on a ``patch" theory approximation of
the problem, which zooms in one or two patches of the Fermi
surface. Then an interacting fixed point with a nonzero gauge
coupling is found in the IR limit based on various analytical
perturbative expansion methods.

Previous studies have also shown that the non-Fermi liquid
obtained through coupling a Fermi surface to a dynamical bosonic
field can be instable against BCS pairing of
fermions~\cite{nflpairing1,nflpairing2,nflpairing3,nflpairing4,nflpairing5,nflpairing6,zounfl}.
If there is only one flavor of $\U(1)$ gauge field, the low energy
interacting fixed point is expected to be robust against this
pairing instability, because the $\U(1)$ gauge field leads to
repulsive interaction between the spinons. However, when there are
two flavors of $\U(1)$ gauge fields~\cite{zounfl,mandal}, like the
case in our parton construction-II, the two $\U(1)$ gauge fields
can lead to interflavor spinon pairing instability. This
interflavor pairing can still happen at the MIT. But depending on
the microscopic parameters this instability can happen at rather
low energy scale.

\subsection{Resistivity at the MIT}

For low frequency and temperature, the resistivity of a system is
usually written as $\rho(x)$ with $x = \omega /T$. The DC
conductivity at zero temperature corresponds to $x = 0$, i.e. the
limit $\omega \rightarrow 0$ before $T \rightarrow 0$. As we have
mentioned, the interaction driven MIT has a jump of resistivity at
the MIT compared with the metal phase near MIT, and this jump is
given by the resistivity $\rho_b$ of the bosonic parton
$b_\alpha$. For a bosonic system with an emergent particle-hole
symmetry in the infrared, $\rho_b(x)$ with $x = 0$ or $x = \infty$
have attracted most studies. In general both $\rho_b(0)$ and
$\rho_b(\infty)$ should be universal numbers at the order of $\sim
h/e^2$. The reason $\rho_b(0)$ could be finite even without
considering disorder and Umklapp process is that, with an emergent
particle-hole symmetry in the infrared discussed in the previous
subsection, there is zero overlap between the electric current and
the conserved momentum density (extra subtleties about this from
hydrodynamics will be discussed in section VI). The universal
$\rho_b(0)$ was evaluated in Ref.~\onlinecite{resistivity2} for
the interaction-driven MIT without charge fractionalization. The
calculation therein was based on Boltzmann equation in a
theoretical large$-\bf{N}$ limit and eventually $\bf{N}$ was taken
to 1 (we remind the readers that the $\bf{N}$ introduced in
Ref.~\onlinecite{resistivity2} was for technical reasons, it is
not to be confused with $N$ used in this work).

We have generalized the computation in
Ref.~\onlinecite{resistivity2} to our case with $N-$components of
vortex fields and charge fractionalization. To proceed with the
computation we need to turn on ``easy plane" anisotropy to
Eq.~\ref{cpn} and perform duality to the basis of fractional
charge carriers $\varphi_j$ (Eq.~\ref{cpndual}). The $\varphi_j$
will be coupled to multiple gauge fields which are the dual of the
$\psi_j$ fields. Eventually the total resistivity $\rho_b(0)$ is
obtained through a generalized Ioffe-Larkin rule, which combines
the resistivity of each parton $\varphi_j$ into $\rho_b$: \beqn
\rho_b = \frac{\hbar}{e^2} \left( \sum_{j = 0}^{N-1} \rho_{b,j}
\right). \label{rhob1} \eeqn $\rho_{b,j}$ is the resistivity of
each charge carrier $\varphi_j$ when its charge is taken to be
$1$. The detail of the computation is presented in the appendix,
and we summarize the results here. For $N$ flavors of vortices in
Eq.~\ref{cpn}, the resistivity $\rho_b(0)$ at the MIT roughly
increases linearly with $N$, as was expected through the intuitive
argument we gave before: \beqn \rho_b(0) = \Delta \rho = \left(
R^{(0)} + R^{(1)} (N - 1) \right) \frac{h}{e^2}, \eeqn where
$R^{(0)} \sim 3.62$, $R^{(1)} \sim 1.68$. We would like to compare
our prediction with the previous theory of MIT without charge
fractionalization. In the previous theory, the DC resistivity jump
is evaluated to be $\Delta \rho \sim 7.92 h/e^2$
~\cite{resistivity2} (we reproduced this calculation and our
result at $N = N_b = 1$ is $7.44 h/e^2$). Eq.~\ref{rhob0} suggests
that when $N \geq 4$, the resistivity jump in our case is indeed
larger than that predicted by the previous theory of MIT.

We would also like to discuss the AC resistivity $\rho_b(\infty)$.
One way to evaluate $\rho_b(\infty)$ is to again start with
Eq.~\ref{cpndual}, and follow the same strategy as the calculation
of the DC resistivity. According to the generalized Ioffe-Larkin
rule, the AC resistivity contributed by {\it each valley} is given
by \beqn \rho_b = N \frac{1}{\sigma_\varphi} \frac{\hbar}{e^2}, \
\ \ \sigma_\varphi = \lim_{\omega \rightarrow 0} \frac{1}{\ii
\omega}\langle J^\varphi_{\omega} J^\varphi_{-\omega}
\rangle_{\vec{p}=0},  \eeqn where $J^\varphi = \ii \varphi^\ast_j
\nabla \varphi_j + h.c.$ is the current of the charge carrier
$\varphi_j$. With the theoretical large-$\bf{N}$ limit mentioned
above, the effects of all the dynamical gauge fields are
suppressed, and $\varphi_j$ will contribute conductivity
$\sigma_\varphi(\infty) = \frac{1}{16}$ (contrary to DC transport,
$\sigma_\varphi(\infty)$ does not need collisions; the effects of
dynamical gauge fields can be included through the $1/{\bf N}$
expansion). Eventually one would obtain resistivity from each
valley \beqn \rho_b = \frac{8N}{\pi} \frac{h}{e^2},
\label{rhobac1} \eeqn the final resistivity of the system is half
of Eq.~\ref{rhobac1} due to the two spin/valley flavors. With $N =
1$, the transition should belong to the ordinary 3D XY
universality class, and the value given by Eq.~\ref{rhobac1} is
not far from what was obtained through more sophisticated methods
(see for instance Ref.~\onlinecite{cond3,UCt3,UCt4}, $\rho_b \sim
2.8 h/e^2$). This should not be surprising as the 3D XY
universality class can be obtained perturbatively from the free
boson theory. In our current case with charge fractionalization,
with $N \geq 4$, the total AC resistivity which is half of the
value in Eq.~\ref{rhobac1} is larger than the universal
resistivity at the 3D XY transition.

Another way to evaluate the resistivity of Eq.~\ref{cpn} is by
integrating out $\psi_j$ from Eq.~\ref{cpn}, and an effective
Lagrangian for $A_\mu$ is generated \beqn \mathcal{L} &=&
\sum_{p_\mu} \frac{N p}{16}\left( \delta_{\mu\nu} - \frac{p_\mu
p_\nu}{p^2} \right) A_\mu(p)A_\nu(-p). \label{Aeff} \eeqn This
effective action is supposed to be accurate in the limit of $N
\rightarrow \infty$. The electric current carried by $b$ is $J^b =
\frac{e}{2\pi} dA$, hence the current-current correlation can be
extracted from the photon Green's function based on the effective
action Eq.~\ref{Aeff}: \beqn \rho_{b, N \rightarrow \infty} =
\frac{\pi N}{8} \frac{h}{e^2}. \label{rhobac2} \eeqn Again the
final resistivity of the system is half of Eq.~\ref{rhobac2} due
to the two spin/valley flavors. The evaluation Eq.~\ref{rhobac2}
is still proportional to $N$ just like Eq.~\ref{rhobac1}. These
two different evaluations discussed above give different values
for $N = N_b = 1$, and compared with the known value of the
universal resistivity at the 3D XY transition, the evaluation in
Eq.~\ref{rhobac1} is much more favorable, though the evaluation
Eq.~\ref{rhobac2} based on Eq.~\ref{Aeff} is supposed to be
accurate with large $N$.

When there is a ring of degeneracy in the vortex band structure,
as we mentioned before the gauge field $A_\mu$ will acquire a
``mass term" after integrating out $\psi_j$~\cite{ringlake}. In
this case the resistivity of the system at the MIT will be
infinity, as the dynamics of $A_\mu$ is fully suppressed by the
mass term in the infrared. One can also integrate out the action
of $A_\mu$ with the mass term, and verify that the response theory
of $A_{\mathrm{ext}}$ is no different from that of an insulator in
the infrared limit. This is consistent with both
Eq.~\ref{rhobac1},\ref{rhobac2} by naively taking $N$ to infinity.
In Ref.~\onlinecite{ringlake} when the boson field has a ring
degeneracy, the phase is identified as a bose metal; this is
because in Ref.~\onlinecite{ringlake} it is the boson with ring
degeneracy that carries charges. But in Eq.~\ref{cpn} the electric
charge is carried by the flux of $A_\mu$.

\section{Mott insulator with topological order}

\label{TO}

As we explained in the previous subsection, due to the fractional
filling of boson $b_\alpha$, the vortex dynamics is frustrated by
the background fractional flux through the hexagons. To drive the
system into an insulator phase, the vortex can either condense at
multiple minima in the BZ as was discussed in the previous
section, or form a bound state that carries multiple gauge charge
of $A_\mu$ and become ``blind" to the background flux. In parton
construction-II, with electron filling $\nu = 1/2$, each flavor of
boson is at filling $\nu_b = 1/2$. The double-vortex, i.e. bound
state of two vortices, or more generally the bound state of $N$
vortices with even integer $N$, no longer see the background flux.
Hence the $N$-vortex can condense at zero momentum, and its
condensate will drive the system into a $Z_N$ topological order.

After the boson-vortex duality, the theory for the $N$-vortex
condensation at one of the two spin/valley flavors is \beqn
\mathcal{L}^{(2)} &=& |(\partial_\mu - \ii N A_\mu)\psi|^2 + r
|\psi|^2 + g |\psi|^4 \cr\cr &+& \frac{\ii}{2\pi} A \wedge d (a +
e A_{\mathrm{ext}}) + \cdots \label{z2}\eeqn The condensate of
$\psi$ will break the $\U(1)$ gauge field to a $Z_N$ gauge field,
whose deconfined phase has a nontrivial $Z_N$ topological order.
In the $Z_N$ topological order as well as at the MIT, the charge
carrier is an anyon of the $Z_N$ topological order, and it carries
charge $e^\ast = e/N$. We still label the fractional charge
carrier as $\varphi$. $\varphi$ carries charge $e/N$, and is
coupled to a $Z_N$ gauge field originated from the $Z_N$
topological order discussed in the previous paragraph.

In our case, in order to preserve the time-reversal symmetry, both
spin/valley flavors should form a $Z_N$ topological order
simultaneously. Hence there is one species of $\varphi_\alpha$
field for each spin/valley flavor. The MIT can equally be
described as the condensation of the $\varphi_\alpha$ field, and
since the $Z_N$ gauge field does not lead to singular correction
in the infrared, the condensation of $\varphi_\alpha$ is a 3D
XY$^\ast$ transition, and the transition for $N = 2$ was discussed
in
Ref.~\onlinecite{vicari,isakov,hastings,deconfine1,deconfine2,chengfrac}.
The $b_\alpha$ field is now a composite operator of
$\varphi_\alpha$. In the condensate of $\varphi_\alpha$, the
electron operator $c_\alpha$ is related to the fermionic parton
operator $f_{\alpha}$ through $c_\alpha \sim \langle b_\alpha
\rangle f_\alpha \sim \langle \varphi^N_\alpha \rangle f_\alpha$.
The coupling between the two flavors of $\varphi_\alpha$, i.e. the
coupling $|\varphi_1|^2|\varphi_2|^2$ is irrelevant at the
decoupled 3D XY$^\ast$ transition according to the known critical
exponents of the 3D XY$^\ast$ transition. There are also couplings
such as $|\varphi_\alpha|^2 f^\dagger_\alpha f_\alpha$ allowed by
all the symmetries, but after formally integrating out the
fermions, the generated couplings for $\varphi_\alpha$ is also
irrelevant at the two decoupled 3D XY$^\ast$ universality class.
The reason is that after formally integrating out the fermions,
terms such as
$\frac{|\omega|}{q}|\varphi_\alpha|^2_{\omega,\vec{q}}|\varphi_\beta|^2_{-\omega,-\vec{q}}$
can be generated, but this term is irrelevant knowing that the
standard critical exponent $\nu > 2/3$ for the 3D XY$^{\ast}$
transition.

Following the large$-\bf{N}$ calculation discussed before, the DC
resistivity jump $\rho_b(0)$ would be $N^2/2$ times that of the
previous theory~\cite{resistivity2}, namely \beqn \rho_b(0) \sim
R^{(2)} N^2 \frac{h}{e^2}, \eeqn where $R^{(2)} = R' /2 \sim 3.7$
based on our evaluation. The AC resistivity jump at the MIT is
enhanced by the same factor compared with the previous theory. We
also note that the fractional universal conductivity at the
transition between the superfluid and a $Z_2$ topological order
was observed numerically in Ref.~\onlinecite{chengfrac}.

Another set of natural topological orders a boson at fractional
filling can form are bosonic fractional quantum Hall (bFQH) states
which are close analogues to the bosonic Laughlin's wave function.
We would like to discuss this possibility as a general
exploration, although this state breaks the $P_y$ symmetry (but it
still preserves the product $P_x \mathcal{T}$ symmetry). If we
interpret the half-filled boson at each site as a quantum spin-1/2
system, this set of states are analogous to a chiral spin
liquid~\cite{csl1,csl2}. The Chern-Simons theory for this set of
states at each valley reads \beqn \mathcal{L}_{\mathrm{cs}} =
-\frac{\ii k}{4\pi} A\wedge dA + \frac{\ii}{2\pi}A \wedge d(a +
A_{\mathrm{ext}}), \label{eq:SUklevel1} \eeqn with an even integer
$k$ and a dynamical Spin$_c$ U(1) gauge field $A$. The topological
order characterized by this theory is the SU$(k)_1$ topological
order. Here, the integer $k$ needs to be even so that this theory
is compatible with the LSM constraint imposed by the boson filling
$1/2$ on the lattice~\cite{LSMtopo}. This is because the boson
filling 1/2 requires the topological phase to contain an Abelian
anyon that carries a fractional charge 1/2 (modulo integer). There
should be one such anyon per unit cell to account for the boson
filling 1/2 on the lattice. The fact that such an anyon carries a
fractional charge 1/2 implies that this anyon should generate
under fusion an Abelian group $\mathbb{Z}_{p}$ with $p$ an even
number. Such a fusion rule is incompatible with any odd value of
$k$. Therefore, $k$ needs to be even in the theory given by Eq.
\ref{eq:SUklevel1}. The time-reversal of the TMD moir\'{e} system
demands that the bosonic parton $b_\alpha$ with opposite
spin/valley index $\alpha$ forms a pair of time-reversal conjugate
bFQH states. Or in other words if we take both spin/valley flavors
together, this state is a fractional topological insulator, like
the state discussed in Ref.~\onlinecite{FQSH1}.

The MIT is now a direct transition between the bFQH state and the
superfluid of $b_\alpha$. When the even integer $k$ is $k = 2n^2$
with odd integer $n$, there is a natural theory for this direct
continuous transition, and its simplest version with $n = 1$ was
proposed in Ref.~\onlinecite{mcgreevy}. The transition is a 3D QED
with two flavors of Dirac fermions coupled to the dynamical
$\U(1)$ Spin$_c$ gauge field $A_\mu$ (the dual of the Goldstone
mode of the boson superfluid) with a Chern-Simons term at
level-$n^2$, and the fermions have gauge charge-$n$: \beqn
\mathcal{L}^{(3)} &=& \sum_{j = 1}^2 \bar{\chi}_j \gamma \cdot
(\partial - \ii n A) \chi_j + M \bar{\chi}_j \chi_j - \frac{\ii
n^2}{4\pi} A \wedge d A \cr\cr &+& \frac{\ii}{2\pi} A \wedge d (a
+ e A_{\mathrm{ext}}) + \cdots \label{qedcs} \eeqn In this theory,
the fact that $A$ is a Spin$_c$ U(1) gauge field and that $n$ is
odd guarantee that this theory describes the phases of a boson.
\cxd{A Spin$_c$ connection $A_\mu$ means a U(1) gauge field with a
``charge-statistics relation": there is no fermionic object that
is neutral under $A_\mu$. When $A_\mu$ is a Spin$_c$ U(1) gauge
field, and $n$ is an odd integer in Eq.~\ref{qedcs},
Eq.~\ref{qedcs} describes an interacting state of bosons that
carries electric charge $e$. The charge$-e$ object of
Eq.~\ref{qedcs} that is also neutral under $A_\mu$, is a composite
of $2\pi$ flux of $A_\mu$ and $n$ fermions $\chi$. This composite
is a boson as long as $n$ being an odd integer, and this composite
should be identified as $b_\alpha$ in Eq.~\ref{parton0}.} The
ellipsis in this Lagrangian includes other terms such as the
Maxwell term of the gauge field $A_\mu$. Please note that this
equation is for one of the two spin/valley flavors of the physical
system. The mass $M$ of the Dirac fermions is the tuning parameter
of the transition. With one sign of the mass term, after
integrating out the Dirac fermions, the Spin$_c$ U(1) gauge field
$A$ will acquire a Chern-Simons term at level $-2n^2$, which
describes the SU$(k)_1$ topological order with $k=2n^2$. With the
opposite sign of $M$, there is no Chern-Simons term of the gauge
field $A$ after integrating out the Dirac fermions, and the
Maxwell term of the gauge field $A$ is the dual description of the
superfluid phase. Hence by tuning $M$ the system undergoes a
transition between the $k=2n^2$ bFQH state and the superfluid
state of $b$ (the metal phase of the original electron system).

The translation symmetry of the system actually guarantees that
the two flavors of Dirac fermions are degenerate in
Eq.~\ref{qedcs}. If these two Dirac fermions are not degenerate,
an intermediate topological order is generated by changing the
sign of the mass of one of the Dirac fermions in Eq.~\ref{qedcs}.
Then after integrating out the fermions, the gauge field $A$
acquires a total CS term with an odd level $-n^2$, which violates
the LSM constraint imposed by the boson filling 1/2. Therefore,
the masses of the two flavors of the Dirac fermions in
Eq.~\ref{qedcs} should be the same. In fact, for the simplest case
with $n = 1$ ($k = 2$), an explicit parton construction of this
transition can be given following the strategy in
Ref.~\onlinecite{mcgreevy}, and the two Dirac fermions in
Eq.~\ref{qedcs} are two Dirac cones of a $\pi-$flux state of
$\chi$ on the triangular lattice. The degeneracy of these two
Dirac fermions is protected by the translation symmetry of the
triangular lattice. From the parton formalism one can also see
that the boson $b$ is constructed as a product of the two fermions
$\chi_i$.

At the transition $M = 0$, though it is difficult to compute the
resistivity of Eq.~\ref{qedcs} exactly, the resistivity $\rho(x)$
should scale as $1/k$ with large $k \sim n^2$, as after
integrating out $\chi_j$ the entire effective action of $A$ scales
linearly as $k$. Then after integrating out $A$, the response
theory to $A_{\mathrm{ext}}$ is proportional to $1/k$.

\section{Summary of Predictions}

\label{experiment}

So far we have discussed three different kinds of possible Mott
insulators at half filling of the extended Hubbard model, based on
the parton construction-II: (1) Mott insulators with translation
symmetry breaking; (2) a $Z_N$ topological order at each
spin/valley flavor with even integer $N \geq 2$; and (3) a pair of
conjugate bFQH states at two spin/valley flavors. For all
scenarios, we have evaluated the bosonic parton contribution to
the resistivity $\rho_b$ at the MIT, which is also the universal
jump of resistivity $\Delta \rho$. The predicted resistivity jump
for the three scenarios are summarized in the table below.

\begin{center}
\begin{tabular}{ |c|c|c|}
 \hline  Nature of Insulator  & $\Delta \rho$, or $\rho_b$
 \\
 \hline
 (1) Density wave  & $ \rho_b(0) \sim ( R^{(0)} + R^{(1)} (N-1) ) \frac{h}{e^2} $ \\
 \hline
 (2) $Z_N$ TO each flavor & $\rho_b(0) = R^{(2)} N^2 \frac{ h }{e^2}$
 \\
 \hline
 (3) Conjugate bFQH & $\rho_b(x) \sim \frac{1}{k} \frac{h}{e^2}$
 \\
 \hline
\end{tabular}
\label{prediction}
\end{center}

Another observable effect predicted by the previous theory of
interaction-driven MIT is the scaling of quasi-particle weight
$\sqrt{Z}$ near the MIT~\cite{senthilmit1,senthilmit2}, where
$\sqrt{Z} \sim r^{\beta_1} \sim |r|^{0.33}$. Our theory also gives
a different prediction of the quasi-particle weight compared with
the previous theory, and this is most conveniently evaluated for
scenario (2). In the metal phase but close to the MIT, the
quasi-particle weight scales as \beqn \sqrt{Z} \sim \langle
\varphi^N_\alpha \rangle \sim |r|^{\beta_N}, \label{weight}\eeqn
where $\beta_N = \nu\Delta_N$. $\nu\sim 0.67$ is the standard
correlation length exponent at the 3D XY$^\ast$ transition (it is
the same as the 3D XY transition) and $\Delta_N$ is the scaling
dimension of $\varphi^N$ at the 3D XY transition. These exponents
can be extracted from numerical simulation on the 3D XY and
XY$^\ast$ transitions. For example, when $N = 2$, $\beta_2$ should
be close to $0.8$~\cite{vicari,isakov,hastingsz2}, hence $\sqrt{Z}
\sim |r|^{0.8}$. The scaling of quasi-particle weight can be
checked in future experiments through the measurement of local
density of states of electrons.

For scenario (1), i.e. where the insulator has translation
symmetry breaking, the scaling of quasiparticle weight can be
estimated with large-$N$ in Eq.~\ref{cpn}. The boson creation
operator $b^\dagger$ is a monopole operator of $A_\mu$ which
creates a $2\pi$ gauge flux. With large-$N$ in Eq.~\ref{cpn} the
monopole operator has scaling dimension proportional to
$N$~\cite{monopole2,monopole3}, hence the critical exponent
$\beta$ in the quasiparticle weight $\sqrt{Z} \sim |r|^{\beta}$ is
expected to be proportional to $N$. The similar evaluation applies
to Eq.~\ref{qedcs}, and the creation operator $b^\dagger$ has a
scaling dimension proportional to $k$, which is also proportional
to $\sqrt{Z}$.

As we explained, our theory provides a natural explanation of the
anomalously large resistivity at the MIT. Another qualitative
experimental feature reported in Ref.~\onlinecite{tmdmit1} is
that, the resistivity drops rapidly as a function of temperature
at the MIT where the charge gap vanishes.
%A similar rapid drop of
%longitudinal resistivity as a function of temperature at the MIT
%in magnetic field has also been observed in another TMD moir\'{e}
%sample recently~\cite{tmdmit3}.
Our theory also provides a natural explanation for the temperature
dependence of the critical resistivity. At zero temperature the
bosonic chargeon parton $b$ fractionalizes into multiple partons
with smaller charges, and these partons will couple to extra gauge
fields. These extra gauge fields will all confine at finite
temperature. Hence at finite temperature, there is a crossover
from transport with fractionalized charge to unfractionalized
charge, which will cause a significant drop of resistivity with
increasing temperature.

In the following paragraphs we discuss physics in phases near the
MIT, based on our theory. These analysis can distinguish the three
possible scenarios discussed to this point. Let us first discuss
the insulator phase at fixed electron filling $\nu = 1/2$. The
scenario (3) describes a topological order that is essentially a
topological fractional quantum spin Hall insulator, hence this
insulator phase, if does exists, must have nonchiral gapless modes
localized at the boundary of the system. This nonchiral edge
gapless modes should lead to similar experimental phenomena as the
experiments on quantum spin Hall insulator~\cite{qshex}; but
rather than edge conductance $2e^2/h$, the edge conductance of the
fractional quantum spin Hall insulator should be $2e^2/(k h)$,
which is twice of the edge conductance of the bFQH state with CS
level-$k$. Also, the edge conductance should be suppressed by
external magnetic field, also analogous to what was observed in
Ref.~\onlinecite{qshex}.

The insulating phase of scenario (1) and scenario (2) also lead to
distinctive predictions. In scenario (1), the electric charges are
only deconfined at the MIT, but still confined in the insulating
phase, which has no topological order. Hence the charge
deconfinement of scenario (1) is analogous to the original
deconfined quantum critical point discussed in
Ref.~\onlinecite{deconfine1,deconfine2}. The confinement of
fractional charges in scenario (1) happens even at zero
temperature in the insulating phase. However, in scenario (2), the
insulator phase has a $Z_N$ topological order that supports
deconfined fractional charge at zero temperature even in the
insulator phase. While at finite temperature, the $Z_N$ gauge
field will lead to confinement of fractional charges with
confinement length $\xi \sim \exp(c \Delta_m/T)$, where $\Delta_m$
is the gap of the fractionalized $Z_N$ gauge fluxes, which is an
anyon with nontrivial statistics with the fractional charges. If
we look at the insulator phase close to the MIT, the gap of the
fractional charge, i.e. the $e-$anyon of the $Z_N$ topological
order is suppsosed to be smaller than $\Delta_m$, as the MIT
corresponds to the condensation of the $e-$anyon, hence at very
low temperature the thermally activated $e-$anyon has a much
smaller distance $l_e$ with each other compared with $\xi$. Then
at low but finite temperature the transport is governed by charge
carriers with gap $\Delta_e$ and charge $e_\ast = e/N$. The gap
$\Delta_e$ can be extracted from fitting the low temperature
transport data versus temperature. However, if one measures the
tunnelling gap through tunnelling spectroscopy, since the external
device can only inject a single electron which fractionalizes into
multiple $e-$anyons, the tunneling gap should be approximately $N
\Delta_e$. This contrast between tunneling gap and the thermally
activated transport gap happens in scenario (2) but not scenario
(1).

We also consider the metallic phase next to the insulator after
charge doping, and we will see the scenario (2) also leads to very
nontrivial predictions due to the deconfined nature of the $Z_N$
topological order. In scenario (2), after some charge doping, we
expect a metallic state with charge fractionalization at low
temperature. The bosonic charge carriers are coupled to the $Z_N$
gauge field as well as the U(1) gauge field $a_\mu$ that are
shared with the fermionic partons $f_\alpha$. When the temperature
is increased, the $Z_N$ gauge field will confine, and due to the
time-reversal symmetry, the confine-deconfine crossover should
happen for both spin/valley flavors simultaneously. In the
following, we shall only focus on one spin/valley. According to
the Ioffe-Larkin composition rule, the total resistivity is
composed of contributions from both bosonic and fermionic partons
$\rho=\sigma^{-1}=\sigma_{b}^{-1}+\sigma_{f}^{-1}$. Let us assume
the resistivity of both the bosonic and fermionic sectors are
dominated by the scattering with the gauge field $a_\mu$ (this of
course assumes that the momentum of the gauge field $a_\mu$ can
relax through other mechanism such as disorder). This scattering
mechanism was first evaluated in Ref.~\onlinecite{leenagaosa}. The
gauge-field propagator can be written as
$D(\omega,\boldsymbol{q})^{-1}=i\gamma \omega/q+\chi_{d}q^{2}$,
where the $\omega/q$ term is due to the Landau damping from the
fermi-surface, and the ``diamagnetic" $\chi_{d}$ is roughly a
constant within the temperature window of interest. The scattering
rate can then be estimated using the imaginary part of the
boson/fermion self-energy:
\begin{flalign*}
&\textrm{Im}\Sigma_{b,f}(\omega,\boldsymbol{k})=\int_{0}^{\infty}d\omega^{\prime}\int
\frac{d^{2}\boldsymbol{k}^{\prime}}{(2\pi)^{2}}(1+n_{b}(\omega^{\prime}))(1\pm
n_{b,f}(\omega_{\boldsymbol{k}^{\prime}}))\\
&\frac{(k_{\alpha}+k_{\alpha}^{\prime})(k_{\beta}+k_{\beta}^{\prime})}{2m_{b,f}}
\frac{\delta_{\alpha\beta}-q_{\alpha}q_{\beta}}{\boldsymbol{q}^{2}}
\delta(\omega-\omega_{\boldsymbol{k}^{\prime}}-\omega^{\prime})\textrm{Im}D(\omega^{\prime},\boldsymbol{q}),
\end{flalign*}
where $\boldsymbol{q}=\boldsymbol{k}^{\prime}-\boldsymbol{k}$,
$n_{b,f}(\omega)$ denotes the Bose-Einstein (Fermi-Dirac)
distribution function, and $m_{b,f}$ is the boson/fermion mass. We
must stress that the expression of $\Sigma_{b,f}$ is valid for
partons with gauge charge-1. When the $Z_N$ gauge field is
deconfined, each boson carries the gauge charge-$1/N$ of the gauge
field $a_\mu$, and therefore there is an additional factor $1/N^2$
in the self-energy. The integral was evaluated in
Ref.~\onlinecite{leenagaosa}, and the time-scale responsible for
transport has an extra factor proportional to $q^2$ in the
integral. After taking these into account, we obtain the
``transport" scattering rate for boson/fermion \beqn
\frac{1}{\tau_{f}}\sim T^{4/3},\qquad\frac{1}{\tau_{b}}\approx
\frac{k_{B}T}{m_{b}\chi_{d}}. \label{scateringrate} \eeqn
Comparing $1/\tau_{b}$ and $1/\tau_{f}$, we can see that the
resistivity is dominated by the boson-gauge scattering at low
temperature, and the bosonic partons are in a disordered phase
rather than a quasi long range order at finite temperature due to
their coupling to the dynamical gauge field $a_\mu$. We take the
Drude formula for the dilute Bose gas that we use to model the
bosonic partons at finite temperature: \beqn \rho \sim
\frac{m_{b}}{n_{*}e_{*}^{2}}\frac{1}{\tau_{b}} \sim
\frac{g_{*}^{2}}{n_{*}e_{*}^{2}}\frac{k_{B}T}{\chi_{d}}, \eeqn
where $e_{*}=e/N$ and $g_{*}=1/N$ denote the electric and gauge
charges of bosons, and $n_{*}e_{*}$ is the doped physical electric
charge density. Here, we have assumed that the resistivity $\rho$
is dominated by the boson contribution because ($i.$) the
scattering rate of the boson is bigger compared to the fermions at
low temperature as shown in Eq.~\ref{scateringrate}, and ($ii.$)
the bosons have much lower density at low charge doping compared
to the fermions which already has finite fermi surface at zero
charge doping. In the following discussion, we will work under
these assumptions at least up to the temperature scale $T_c$
around which the $Z_N$ gauge becomes fully confined.

The $Z_N$ gauge field is fully confined when $\xi$ is at the same
order as the lattice constant; $i.e.$ $T > T_c \sim \Delta_m$.
Here we assume that the gauge field $a_\mu$ that is coupled to the
fermionic parton is less prone to confinement due to its coupling
to the large density of gapless fermoins. Above $T_c$, the charge
carriers in the system carry charge-$e$. The equation above still
hold with the substitutions $e_{*}\rightarrow e =
Ne_{*},g_{*}\rightarrow g = Ng_{*},n_{*}\rightarrow n = n_{*}/N$.
We expect there is a crossover from the deconfined value of
resistivity $\rho(T \sim 0)$ to the confined value $\rho(T \geq
T_c)$: \beqn \frac{(d\rho/dT)_{T \geq T_c}}{(d\rho/dT)_{T \sim
0}}\sim N, \eeqn This is an observable effect of scenario (2) that
can be experimentally verified. Note that the crossover caused by
confinement at the metallic phase is different from the critical
point of the MIT; as transport at the critical point originates
from rather different physics; for example both particles and
holes will contribute to the charge transport at the critical
point~\cite{HoloQ}.

Contrary to the Ioffe-Larkin rule, the total thermal conductivity
of the system is a sum of the contribution from the bosonic
parton, fermionic parton, and also the gauge boson. With low
charge doping away from $\nu = 1/2$, we expect the fermionic
partons dominates the thermal transport according to
Ref.~\onlinecite{lee2007}: $\kappa_{f}\sim T^{1/3}$. As we
discussed above, in scenario (2) the low-temperature charge
transport is dominated by the boson contribution
$\sigma_{b}\sim1/T$, while the thermal transport is dominated by
the fermion contribution $\kappa_{f}\sim T^{1/3}$. Due to the
crossover of charge transport at finite temperature caused by the
confinement of the $Z_N$ gauge field in scenario (2), there is
also an observable prediction one can make for the Lorentz number
$L=\kappa/(T\sigma) \approx\kappa_{f}/(T\sigma_{b}) $: \beqn
\frac{(L/T^{1/3})_{T \geq T_c}}{(L/T^{1/3})_{T \sim 0}} \sim N.
\eeqn

\section{Summary, Discussion, and Other fractional fillings}

In this work we proposed a theory for a potentially continuous
metal-insulator transition for the extended Hubbard model on the
triangular lattice at half-filling (one electron per unit cell).
The extended Hubbard model is simulated by the TMD moir\'{e}
systems. We introduce a different parton construction from the
previous literature, which leads to a series of observable
predictions. We demonstrated that our theory is more favorable
given the current experiments on the heterobilayer TMD moir\'{e}
systems. Although our theory was motivated by the recent
experiments on MoTe$_2$/WSe$_2$ moir\'{e}
superlattice~\cite{tmdmit1}, we envision our theory can have broad
application given the recent rapid progresses in synthesizing pure
two dimensional systems.

The moir\'{e} potential in the MoTe$_2$/WSe$_2$ moir\'{e}
superlattice with no twisting is formed due to the mismatch of the
lattice constants of the two layers. There is another experiment
on MIT in twisted WSe$_2$~\cite{tmdmit2}. The situation in twisted
WSe$_2$ seems rather different from MoTe$_2$/WSe$_2$ moir\'{e}
superlattice. Inside the ``insulator phase", the resistivity
$\rho(T)$ at some displacement fields first increases with
decreasing temperature, and eventually the plot seems to saturate
at a finite value, which is much lower than the resistivity
observed in the MoTe$_2$/WSe$_2$ moir\'{e} superlattice near the
MIT. Hence the MIT of twisted WSe$_2$ could be of a different
nature, between the metallic phase and the insulator phase, there
could be an intermediate phase with an order at nonzero momentum
and reduced size of electron Fermi pockets.

Correlated insulators at other fractional fillings $\nu = p/q$
have been reported in various TMD moir\'{e}
systems~\cite{tmdinsulator1,tmdinsulator2,tmdinsulator3,tmdinsulator4}.
Although the nature of the MIT at these fillings has not been
looked into carefully, here we briefly discuss the theory for the
possible continuous MIT at general fractional filling $\nu = p/q$.
As long as $q > 2$, even for parton construction-I, the bosonic
parton $b$ will have fractional filling, and hence the insulator
phase of $b$ cannot be a trivial incompressible state without
translation symmetry breaking or topological order. Here we would
like to acknowledge that charge fractionalization for interacting
electron system at fractional electron number per unit cell was
discussed in previous literature~\cite{gang2014}, using similar
formalism as the parton construction-I. At electron filling $\nu =
1/q$, the boson filling $\nu_b = 2/q$; if we only consider nearest
neighbor hopping of the vortex, the insulator has commensurate
density wave that spontaneously breaks the translation symmetry,
and the MIT is described by Eq.~\ref{cpn} with $N = q$ for odd
integer $q$; $N = q/2$ for $q = 4k+2$; and $N = q$ for $q = 4k$.
The electron charge will further fractionalize at the continuous
MIT. In parton construction-I, there are in total $N$ species of
the charge carriers each carrying electric charge $e^\ast = e/N$.
Hence the estimate of $\rho_b$ is $\rho_b \sim N h/e^2$.

For parton construction-II, with electron filling $\nu = 1/q$, the
boson filling for each spin/valley flavor is $\nu_b = 1/q$. Again,
if only nearest neighbor hopping of the vortices is considered,
the MIT is described by Eq.~\ref{cpn} with $N = q$ for odd integer
$q$; $N = 2q$ for even integer $q$. The field theory describing
the MIT is two copies of Eq.~\ref{cpn}: $\psi_j$, $A_\mu$ and
$a_\mu$ should all carry a spin index $\alpha$. There are in total
$N_b = 2N$ species of the charge carriers each carrying electric
charge $e^\ast = e/N$. Hence the estimate of $\rho_b$ is $\rho_b
\sim N h/(2e^2)$. If we consider further neighbor hopping like
section~\ref{densitywave}, the charge carriers may carry even
smaller fractional charge, and hence larger $\rho_b$.

Here, we would like to discuss some subtlety regarding the
conductivity $\sigma_b$ of the bosonic parton. In a generic theory
with momentum conservation, one expects a finite overlap between
the electric current and the conversed momentum. Such a finite
overlap would lead to a Drude peak in the (optical) conductivity
(see Ref.~\onlinecite{HoloQ} for a review) $\sigma(\omega) =
\sigma_Q + \mathcal{D}\left( \frac{i}{\omega} + \delta(\omega)
\right)$ where $\mathcal{D}>0$ is the Drude weight and $\omega$ is
the frequency. In a theory with an exact particle-hole symmetry,
this overlap between the electric current and momentum is strictly
zero and, consequently, the Drude weight $\mathcal{D}$ vanishes.
In the MIT considered in this paper and previous literature such
as Ref.~\onlinecite{lee2005,senthilmit1,resistivity2}, the
theories that govern the bosonic partons all have an emergent
particle-hole symmetry. This emergent particle-hole symmetry is
expected to produce a Drude weight that vanishes at zero
temperature, namely $\mathcal{D} \rightarrow 0 $ as $T\rightarrow
0$. If there is a finite momentum relaxation time $\tau_p$ induced
by for example disorder, the Drude peak should take the form
$\frac{\mathcal{D}}{\tau_p^{-1} - i\omega }$ and should be viewed
as an extra correction, when we take $\omega\rightarrow 0$, to the
bosonic parton DC conductivity $\sigma_b$ calculated for the MIT.
%Note that the momentum relaxation time $\tau_p$ can be treated as
%independent from the temperature.
Since $\mathcal{D}$ vanishes as $T\rightarrow 0$ due to the
emergent particle-hole symmetry, the DC limit, i.e. $\omega
\rightarrow 0$, of the Drude peak becomes a small correction to
the bosonic parton DC conductivity $\sigma_b$ at low temperature.

There is another subtlety associated with the bosonic parton
conductivity due to extra hydrodynamical corrections and the
purely two dimensional nature of the system. It was known (see,
for example, Ref.~\onlinecite{Kovtun2012} for a review) that, when
momentum is strictly conversed, even in the presence of
particle-hole symmetry, hydrodynamical fluctuations lead to a
logarithmic correction to the optical conductivity that scale as
$\log (\tau_{\rm th} \omega) $. Here, $\tau_{\rm th}$ is the time
scale of local thermalization~\cite{luca} and can be estimated as
$\sim T^{-1}$. This hydrodynamical correction to the conductivity
diverges in the DC limit. This divergence is due to the long-lived
hydrodynamical mode associated with the conserved momentum. As we
mentioned before, in real systems disorder and Umklapp process
always induce a finite momentum relaxation time $\tau_p$. The
diverging hydrodynamical correction is only valid when $\tau_p \gg
\tau_{\rm th} \sim T^{-1}$, meaning momentum is strictly conserved
over the thermalization time scale, where the hydrodynamical
description becomes applicable. When the temperature $T$ is low
compared to $\tau_p^{-1}$, hydrodynamical corrections are cut-off
by $\tau_p^{-1}$ and are again expected to be small corrections to
the bosonic parton conductivity calculated in the rest parts of
this paper. In fact the divergent hydrodynamical correction may be
already cut-off at a higher temperature scale that is favorable to
us, as the crossover scale is suppressed by a large factor
depending on the dimensionless entropy density of the
system~\cite{luca}.

We would like to stress that the optical conductivity
$\sigma(\infty)$ which is much easier to evaluate theoretically
(see section.III for an example) is free of these subtleties, and
we encourage future experiments to measure the optical
conductivity at the MIT as well.

In recent years very impressive progresses have been made on
numerically simulating interacting fermionic systems (for examples
see
Ref.~\onlinecite{numerical1,numerical2,numerical3,numerical4}). It
is conceivable that an extended Hubbard model with spin-orbit
coupling can be constructed on the triangular lattice, and by
changing the parameter (for example the strength of the spin-orbit
coupling), two types of interaction-driven MIT may be realized,
one described by the original theory~\cite{lee2005,senthilmit2},
the other described by our current theory. Predictions made in
these two theories, such as different universality classes and
transport properties at the MIT, different scalings of
quasiparticle weight, and the existence of the spinon Fermi
surface in the insulator phase, can potentially be directly tested
through various numerical methods on the extended Hubbard model.
We will leave this to future exploration.

The authors thank L. Balents, Luca Delacretaz, Sung-Sik Lee, C.
Nayak, T. Senthil, and Kevin Slagle for very helpful discussions.
C.X. is supported by NSF Grant No. DMR-1920434, and the Simons
Investigator program; Z.L. is supported by the Simons
Collaborations on Ultra-Quantum Matter, grant 651440 (LB); M.Y.
was supported in part by the Gordon and Betty Moore Foundation
through Grant GBMF8690 to UCSB, and by the NSF Grant No.
PHY-1748958.

\bibliographystyle{apsrev}

\bibliography{mit,moireref}

\begin{thebibliography}{131}
\expandafter\ifx\csname natexlab\endcsname\relax\def\natexlab#1{#1}\fi
\expandafter\ifx\csname bibnamefont\endcsname\relax
  \def\bibnamefont#1{#1}\fi
\expandafter\ifx\csname bibfnamefont\endcsname\relax
  \def\bibfnamefont#1{#1}\fi
\expandafter\ifx\csname citenamefont\endcsname\relax
  \def\citenamefont#1{#1}\fi
\expandafter\ifx\csname url\endcsname\relax
  \def\url#1{\texttt{#1}}\fi
\expandafter\ifx\csname urlprefix\endcsname\relax\def\urlprefix{URL }\fi
\providecommand{\bibinfo}[2]{#2}
\providecommand{\eprint}[2][]{\url{#2}}

\bibitem[{\citenamefont{Cao et~al.}(2018{\natexlab{a}})\citenamefont{Cao,
  Fatemi, Demir, Fang, Tomarken, Luo, Sanchez-Yamagishi, Watanabe, Taniguchi,
  Kaxiras et~al.}}]{pablo1}
\bibinfo{author}{\bibfnamefont{Y.}~\bibnamefont{Cao}},
  \bibinfo{author}{\bibfnamefont{V.}~\bibnamefont{Fatemi}},
  \bibinfo{author}{\bibfnamefont{A.}~\bibnamefont{Demir}},
  \bibinfo{author}{\bibfnamefont{S.}~\bibnamefont{Fang}},
  \bibinfo{author}{\bibfnamefont{S.~L.} \bibnamefont{Tomarken}},
  \bibinfo{author}{\bibfnamefont{J.~Y.} \bibnamefont{Luo}},
  \bibinfo{author}{\bibfnamefont{J.~D.} \bibnamefont{Sanchez-Yamagishi}},
  \bibinfo{author}{\bibfnamefont{K.}~\bibnamefont{Watanabe}},
  \bibinfo{author}{\bibfnamefont{T.}~\bibnamefont{Taniguchi}},
  \bibinfo{author}{\bibfnamefont{E.}~\bibnamefont{Kaxiras}},
  \bibnamefont{et~al.}, \bibinfo{journal}{Nature}
  \textbf{\bibinfo{volume}{556}}, \bibinfo{pages}{80}
  (\bibinfo{year}{2018}{\natexlab{a}}), ISSN \bibinfo{issn}{1476-4687},
  \urlprefix\url{http://dx.doi.org/10.1038/nature26154}.

\bibitem[{\citenamefont{Cao et~al.}(2018{\natexlab{b}})\citenamefont{Cao,
  Fatemi, Fang, Watanabe, Taniguchi, Kaxiras, and Jarillo-Herrero}}]{pablo2}
\bibinfo{author}{\bibfnamefont{Y.}~\bibnamefont{Cao}},
  \bibinfo{author}{\bibfnamefont{V.}~\bibnamefont{Fatemi}},
  \bibinfo{author}{\bibfnamefont{S.}~\bibnamefont{Fang}},
  \bibinfo{author}{\bibfnamefont{K.}~\bibnamefont{Watanabe}},
  \bibinfo{author}{\bibfnamefont{T.}~\bibnamefont{Taniguchi}},
  \bibinfo{author}{\bibfnamefont{E.}~\bibnamefont{Kaxiras}}, \bibnamefont{and}
  \bibinfo{author}{\bibfnamefont{P.}~\bibnamefont{Jarillo-Herrero}},
  \bibinfo{journal}{Nature} \textbf{\bibinfo{volume}{556}}, \bibinfo{pages}{43}
  (\bibinfo{year}{2018}{\natexlab{b}}), ISSN \bibinfo{issn}{1476-4687},
  \urlprefix\url{http://dx.doi.org/10.1038/nature26160}.

\bibitem[{\citenamefont{Chen et~al.}(2019{\natexlab{a}})\citenamefont{Chen,
  Jiang, Wu, Lyu, Li, Chittari, Watanabe, Taniguchi, Shi, Jung et~al.}}]{wang1}
\bibinfo{author}{\bibfnamefont{G.}~\bibnamefont{Chen}},
  \bibinfo{author}{\bibfnamefont{L.}~\bibnamefont{Jiang}},
  \bibinfo{author}{\bibfnamefont{S.}~\bibnamefont{Wu}},
  \bibinfo{author}{\bibfnamefont{B.}~\bibnamefont{Lyu}},
  \bibinfo{author}{\bibfnamefont{H.}~\bibnamefont{Li}},
  \bibinfo{author}{\bibfnamefont{B.~L.} \bibnamefont{Chittari}},
  \bibinfo{author}{\bibfnamefont{K.}~\bibnamefont{Watanabe}},
  \bibinfo{author}{\bibfnamefont{T.}~\bibnamefont{Taniguchi}},
  \bibinfo{author}{\bibfnamefont{Z.}~\bibnamefont{Shi}},
  \bibinfo{author}{\bibfnamefont{J.}~\bibnamefont{Jung}}, \bibnamefont{et~al.},
  \bibinfo{journal}{Nature Physics} \textbf{\bibinfo{volume}{15}},
  \bibinfo{pages}{237} (\bibinfo{year}{2019}{\natexlab{a}}), ISSN
  \bibinfo{issn}{1745-2481},
  \urlprefix\url{http://dx.doi.org/10.1038/s41567-018-0387-2}.

\bibitem[{\citenamefont{Yankowitz et~al.}(2019)\citenamefont{Yankowitz, Chen,
  Polshyn, Zhang, Watanabe, Taniguchi, Graf, Young, and Dean}}]{young1}
\bibinfo{author}{\bibfnamefont{M.}~\bibnamefont{Yankowitz}},
  \bibinfo{author}{\bibfnamefont{S.}~\bibnamefont{Chen}},
  \bibinfo{author}{\bibfnamefont{H.}~\bibnamefont{Polshyn}},
  \bibinfo{author}{\bibfnamefont{Y.}~\bibnamefont{Zhang}},
  \bibinfo{author}{\bibfnamefont{K.}~\bibnamefont{Watanabe}},
  \bibinfo{author}{\bibfnamefont{T.}~\bibnamefont{Taniguchi}},
  \bibinfo{author}{\bibfnamefont{D.}~\bibnamefont{Graf}},
  \bibinfo{author}{\bibfnamefont{A.~F.} \bibnamefont{Young}}, \bibnamefont{and}
  \bibinfo{author}{\bibfnamefont{C.~R.} \bibnamefont{Dean}},
  \bibinfo{journal}{Science} \textbf{\bibinfo{volume}{363}},
  \bibinfo{pages}{1059} (\bibinfo{year}{2019}), ISSN \bibinfo{issn}{1095-9203},
  \urlprefix\url{http://dx.doi.org/10.1126/science.aav1910}.

\bibitem[{\citenamefont{Saito et~al.}(2020)\citenamefont{Saito, Ge, Watanabe,
  Taniguchi, and Young}}]{young2}
\bibinfo{author}{\bibfnamefont{Y.}~\bibnamefont{Saito}},
  \bibinfo{author}{\bibfnamefont{J.}~\bibnamefont{Ge}},
  \bibinfo{author}{\bibfnamefont{K.}~\bibnamefont{Watanabe}},
  \bibinfo{author}{\bibfnamefont{T.}~\bibnamefont{Taniguchi}},
  \bibnamefont{and} \bibinfo{author}{\bibfnamefont{A.~F.} \bibnamefont{Young}},
  \bibinfo{journal}{Nature Physics} \textbf{\bibinfo{volume}{16}},
  \bibinfo{pages}{926} (\bibinfo{year}{2020}), ISSN \bibinfo{issn}{1745-2481},
  \urlprefix\url{http://dx.doi.org/10.1038/s41567-020-0928-3}.

\bibitem[{\citenamefont{Stepanov
  et~al.}(2020{\natexlab{a}})\citenamefont{Stepanov, Das, Lu, Fahimniya,
  Watanabe, Taniguchi, Koppens, Lischner, Levitov, and Efetov}}]{efetov}
\bibinfo{author}{\bibfnamefont{P.}~\bibnamefont{Stepanov}},
  \bibinfo{author}{\bibfnamefont{I.}~\bibnamefont{Das}},
  \bibinfo{author}{\bibfnamefont{X.}~\bibnamefont{Lu}},
  \bibinfo{author}{\bibfnamefont{A.}~\bibnamefont{Fahimniya}},
  \bibinfo{author}{\bibfnamefont{K.}~\bibnamefont{Watanabe}},
  \bibinfo{author}{\bibfnamefont{T.}~\bibnamefont{Taniguchi}},
  \bibinfo{author}{\bibfnamefont{F.~H.~L.} \bibnamefont{Koppens}},
  \bibinfo{author}{\bibfnamefont{J.}~\bibnamefont{Lischner}},
  \bibinfo{author}{\bibfnamefont{L.}~\bibnamefont{Levitov}}, \bibnamefont{and}
  \bibinfo{author}{\bibfnamefont{D.~K.} \bibnamefont{Efetov}},
  \bibinfo{journal}{Nature} \textbf{\bibinfo{volume}{583}},
  \bibinfo{pages}{375} (\bibinfo{year}{2020}{\natexlab{a}}), ISSN
  \bibinfo{issn}{1476-4687},
  \urlprefix\url{http://dx.doi.org/10.1038/s41586-020-2459-6}.

\bibitem[{\citenamefont{Chen et~al.}(2019{\natexlab{b}})\citenamefont{Chen,
  Sharpe, Gallagher, Rosen, Fox, Jiang, Lyu, Li, Watanabe, Taniguchi
  et~al.}}]{wang2}
\bibinfo{author}{\bibfnamefont{G.}~\bibnamefont{Chen}},
  \bibinfo{author}{\bibfnamefont{A.~L.} \bibnamefont{Sharpe}},
  \bibinfo{author}{\bibfnamefont{P.}~\bibnamefont{Gallagher}},
  \bibinfo{author}{\bibfnamefont{I.~T.} \bibnamefont{Rosen}},
  \bibinfo{author}{\bibfnamefont{E.~J.} \bibnamefont{Fox}},
  \bibinfo{author}{\bibfnamefont{L.}~\bibnamefont{Jiang}},
  \bibinfo{author}{\bibfnamefont{B.}~\bibnamefont{Lyu}},
  \bibinfo{author}{\bibfnamefont{H.}~\bibnamefont{Li}},
  \bibinfo{author}{\bibfnamefont{K.}~\bibnamefont{Watanabe}},
  \bibinfo{author}{\bibfnamefont{T.}~\bibnamefont{Taniguchi}},
  \bibnamefont{et~al.}, \bibinfo{journal}{Nature}
  \textbf{\bibinfo{volume}{572}}, \bibinfo{pages}{215}
  (\bibinfo{year}{2019}{\natexlab{b}}), ISSN \bibinfo{issn}{1476-4687},
  \urlprefix\url{http://dx.doi.org/10.1038/s41586-019-1393-y}.

\bibitem[{\citenamefont{Liu et~al.}(2020)\citenamefont{Liu, Hao, Khalaf, Lee,
  Ronen, Yoo, Haei~Najafabadi, Watanabe, Taniguchi, Vishwanath et~al.}}]{kim1}
\bibinfo{author}{\bibfnamefont{X.}~\bibnamefont{Liu}},
  \bibinfo{author}{\bibfnamefont{Z.}~\bibnamefont{Hao}},
  \bibinfo{author}{\bibfnamefont{E.}~\bibnamefont{Khalaf}},
  \bibinfo{author}{\bibfnamefont{J.~Y.} \bibnamefont{Lee}},
  \bibinfo{author}{\bibfnamefont{Y.}~\bibnamefont{Ronen}},
  \bibinfo{author}{\bibfnamefont{H.}~\bibnamefont{Yoo}},
  \bibinfo{author}{\bibfnamefont{D.}~\bibnamefont{Haei~Najafabadi}},
  \bibinfo{author}{\bibfnamefont{K.}~\bibnamefont{Watanabe}},
  \bibinfo{author}{\bibfnamefont{T.}~\bibnamefont{Taniguchi}},
  \bibinfo{author}{\bibfnamefont{A.}~\bibnamefont{Vishwanath}},
  \bibnamefont{et~al.}, \bibinfo{journal}{Nature}
  \textbf{\bibinfo{volume}{583}}, \bibinfo{pages}{221} (\bibinfo{year}{2020}),
  ISSN \bibinfo{issn}{1476-4687},
  \urlprefix\url{http://dx.doi.org/10.1038/s41586-020-2458-7}.

\bibitem[{\citenamefont{Cao et~al.}(2020{\natexlab{a}})\citenamefont{Cao,
  Rodan-Legrain, Rubies-Bigorda, Park, Watanabe, Taniguchi, and
  Jarillo-Herrero}}]{pablo3}
\bibinfo{author}{\bibfnamefont{Y.}~\bibnamefont{Cao}},
  \bibinfo{author}{\bibfnamefont{D.}~\bibnamefont{Rodan-Legrain}},
  \bibinfo{author}{\bibfnamefont{O.}~\bibnamefont{Rubies-Bigorda}},
  \bibinfo{author}{\bibfnamefont{J.~M.} \bibnamefont{Park}},
  \bibinfo{author}{\bibfnamefont{K.}~\bibnamefont{Watanabe}},
  \bibinfo{author}{\bibfnamefont{T.}~\bibnamefont{Taniguchi}},
  \bibnamefont{and}
  \bibinfo{author}{\bibfnamefont{P.}~\bibnamefont{Jarillo-Herrero}},
  \bibinfo{journal}{Nature} \textbf{\bibinfo{volume}{583}},
  \bibinfo{pages}{215} (\bibinfo{year}{2020}{\natexlab{a}}), ISSN
  \bibinfo{issn}{1476-4687},
  \urlprefix\url{http://dx.doi.org/10.1038/s41586-020-2260-6}.

\bibitem[{\citenamefont{Cao et~al.}(2020{\natexlab{b}})\citenamefont{Cao,
  Chowdhury, Rodan-Legrain, Rubies-Bigorda, Watanabe, Taniguchi, Senthil, and
  Jarillo-Herrero}}]{pablostrange}
\bibinfo{author}{\bibfnamefont{Y.}~\bibnamefont{Cao}},
  \bibinfo{author}{\bibfnamefont{D.}~\bibnamefont{Chowdhury}},
  \bibinfo{author}{\bibfnamefont{D.}~\bibnamefont{Rodan-Legrain}},
  \bibinfo{author}{\bibfnamefont{O.}~\bibnamefont{Rubies-Bigorda}},
  \bibinfo{author}{\bibfnamefont{K.}~\bibnamefont{Watanabe}},
  \bibinfo{author}{\bibfnamefont{T.}~\bibnamefont{Taniguchi}},
  \bibinfo{author}{\bibfnamefont{T.}~\bibnamefont{Senthil}}, \bibnamefont{and}
  \bibinfo{author}{\bibfnamefont{P.}~\bibnamefont{Jarillo-Herrero}},
  \bibinfo{journal}{Phys. Rev. Lett.} \textbf{\bibinfo{volume}{124}},
  \bibinfo{pages}{076801} (\bibinfo{year}{2020}{\natexlab{b}}),
  \urlprefix\url{https://link.aps.org/doi/10.1103/PhysRevLett.124.076801}.

\bibitem[{\citenamefont{Polshyn et~al.}(2019)\citenamefont{Polshyn, Yankowitz,
  Chen, Zhang, Watanabe, Taniguchi, Dean, and Young}}]{youngstrange}
\bibinfo{author}{\bibfnamefont{H.}~\bibnamefont{Polshyn}},
  \bibinfo{author}{\bibfnamefont{M.}~\bibnamefont{Yankowitz}},
  \bibinfo{author}{\bibfnamefont{S.}~\bibnamefont{Chen}},
  \bibinfo{author}{\bibfnamefont{Y.}~\bibnamefont{Zhang}},
  \bibinfo{author}{\bibfnamefont{K.}~\bibnamefont{Watanabe}},
  \bibinfo{author}{\bibfnamefont{T.}~\bibnamefont{Taniguchi}},
  \bibinfo{author}{\bibfnamefont{C.~R.} \bibnamefont{Dean}}, \bibnamefont{and}
  \bibinfo{author}{\bibfnamefont{A.~F.} \bibnamefont{Young}},
  \bibinfo{journal}{Nature Physics} \textbf{\bibinfo{volume}{15}},
  \bibinfo{pages}{1011} (\bibinfo{year}{2019}), ISSN \bibinfo{issn}{1745-2481},
  \urlprefix\url{http://dx.doi.org/10.1038/s41567-019-0596-3}.

\bibitem[{\citenamefont{Bistritzer and MacDonald}(2011)}]{tbgmodel1}
\bibinfo{author}{\bibfnamefont{R.}~\bibnamefont{Bistritzer}} \bibnamefont{and}
  \bibinfo{author}{\bibfnamefont{A.~H.} \bibnamefont{MacDonald}},
  \bibinfo{journal}{Proceedings of the National Academy of Sciences}
  \textbf{\bibinfo{volume}{108}}, \bibinfo{pages}{12233}
  (\bibinfo{year}{2011}), ISSN \bibinfo{issn}{0027-8424},
  \eprint{https://www.pnas.org/content/108/30/12233.full.pdf},
  \urlprefix\url{https://www.pnas.org/content/108/30/12233}.

\bibitem[{\citenamefont{Lopes~dos Santos et~al.}(2012)\citenamefont{Lopes~dos
  Santos, Peres, and Castro~Neto}}]{tbgmodel2}
\bibinfo{author}{\bibfnamefont{J.~M.~B.} \bibnamefont{Lopes~dos Santos}},
  \bibinfo{author}{\bibfnamefont{N.~M.~R.} \bibnamefont{Peres}},
  \bibnamefont{and} \bibinfo{author}{\bibfnamefont{A.~H.}
  \bibnamefont{Castro~Neto}}, \bibinfo{journal}{Phys. Rev. B}
  \textbf{\bibinfo{volume}{86}}, \bibinfo{pages}{155449}
  (\bibinfo{year}{2012}),
  \urlprefix\url{https://link.aps.org/doi/10.1103/PhysRevB.86.155449}.

\bibitem[{\citenamefont{Xu and Balents}(2018)}]{xubalents}
\bibinfo{author}{\bibfnamefont{C.}~\bibnamefont{Xu}} \bibnamefont{and}
  \bibinfo{author}{\bibfnamefont{L.}~\bibnamefont{Balents}},
  \bibinfo{journal}{Phys. Rev. Lett.} \textbf{\bibinfo{volume}{121}},
  \bibinfo{pages}{087001} (\bibinfo{year}{2018}),
  \urlprefix\url{https://link.aps.org/doi/10.1103/PhysRevLett.121.087001}.

\bibitem[{\citenamefont{Yuan and Fu}(2018)}]{fu1}
\bibinfo{author}{\bibfnamefont{N.~F.~Q.} \bibnamefont{Yuan}} \bibnamefont{and}
  \bibinfo{author}{\bibfnamefont{L.}~\bibnamefont{Fu}}, \bibinfo{journal}{Phys.
  Rev. B} \textbf{\bibinfo{volume}{98}}, \bibinfo{pages}{045103}
  (\bibinfo{year}{2018}),
  \urlprefix\url{https://link.aps.org/doi/10.1103/PhysRevB.98.045103}.

\bibitem[{\citenamefont{Isobe et~al.}(2018)\citenamefont{Isobe, Yuan, and
  Fu}}]{fu2}
\bibinfo{author}{\bibfnamefont{H.}~\bibnamefont{Isobe}},
  \bibinfo{author}{\bibfnamefont{N.~F.~Q.} \bibnamefont{Yuan}},
  \bibnamefont{and} \bibinfo{author}{\bibfnamefont{L.}~\bibnamefont{Fu}},
  \bibinfo{journal}{Phys. Rev. X} \textbf{\bibinfo{volume}{8}},
  \bibinfo{pages}{041041} (\bibinfo{year}{2018}),
  \urlprefix\url{https://link.aps.org/doi/10.1103/PhysRevX.8.041041}.

\bibitem[{\citenamefont{Koshino et~al.}(2018)\citenamefont{Koshino, Yuan,
  Koretsune, Ochi, Kuroki, and Fu}}]{fu3}
\bibinfo{author}{\bibfnamefont{M.}~\bibnamefont{Koshino}},
  \bibinfo{author}{\bibfnamefont{N.~F.~Q.} \bibnamefont{Yuan}},
  \bibinfo{author}{\bibfnamefont{T.}~\bibnamefont{Koretsune}},
  \bibinfo{author}{\bibfnamefont{M.}~\bibnamefont{Ochi}},
  \bibinfo{author}{\bibfnamefont{K.}~\bibnamefont{Kuroki}}, \bibnamefont{and}
  \bibinfo{author}{\bibfnamefont{L.}~\bibnamefont{Fu}}, \bibinfo{journal}{Phys.
  Rev. X} \textbf{\bibinfo{volume}{8}}, \bibinfo{pages}{031087}
  (\bibinfo{year}{2018}),
  \urlprefix\url{https://link.aps.org/doi/10.1103/PhysRevX.8.031087}.

\bibitem[{\citenamefont{Thomson et~al.}(2018)\citenamefont{Thomson, Chatterjee,
  Sachdev, and Scheurer}}]{subir1}
\bibinfo{author}{\bibfnamefont{A.}~\bibnamefont{Thomson}},
  \bibinfo{author}{\bibfnamefont{S.}~\bibnamefont{Chatterjee}},
  \bibinfo{author}{\bibfnamefont{S.}~\bibnamefont{Sachdev}}, \bibnamefont{and}
  \bibinfo{author}{\bibfnamefont{M.~S.} \bibnamefont{Scheurer}},
  \bibinfo{journal}{Phys. Rev. B} \textbf{\bibinfo{volume}{98}},
  \bibinfo{pages}{075109} (\bibinfo{year}{2018}),
  \urlprefix\url{https://link.aps.org/doi/10.1103/PhysRevB.98.075109}.

\bibitem[{\citenamefont{Dodaro et~al.}(2018)\citenamefont{Dodaro, Kivelson,
  Schattner, Sun, and Wang}}]{steve1}
\bibinfo{author}{\bibfnamefont{J.~F.} \bibnamefont{Dodaro}},
  \bibinfo{author}{\bibfnamefont{S.~A.} \bibnamefont{Kivelson}},
  \bibinfo{author}{\bibfnamefont{Y.}~\bibnamefont{Schattner}},
  \bibinfo{author}{\bibfnamefont{X.~Q.} \bibnamefont{Sun}}, \bibnamefont{and}
  \bibinfo{author}{\bibfnamefont{C.}~\bibnamefont{Wang}},
  \bibinfo{journal}{Phys. Rev. B} \textbf{\bibinfo{volume}{98}},
  \bibinfo{pages}{075154} (\bibinfo{year}{2018}),
  \urlprefix\url{https://link.aps.org/doi/10.1103/PhysRevB.98.075154}.

\bibitem[{\citenamefont{Po et~al.}(2018)\citenamefont{Po, Zou, Vishwanath, and
  Senthil}}]{senthil1}
\bibinfo{author}{\bibfnamefont{H.~C.} \bibnamefont{Po}},
  \bibinfo{author}{\bibfnamefont{L.}~\bibnamefont{Zou}},
  \bibinfo{author}{\bibfnamefont{A.}~\bibnamefont{Vishwanath}},
  \bibnamefont{and} \bibinfo{author}{\bibfnamefont{T.}~\bibnamefont{Senthil}},
  \bibinfo{journal}{Phys. Rev. X} \textbf{\bibinfo{volume}{8}},
  \bibinfo{pages}{031089} (\bibinfo{year}{2018}),
  \urlprefix\url{https://link.aps.org/doi/10.1103/PhysRevX.8.031089}.

\bibitem[{\citenamefont{Kang and Vafek}(2018)}]{vafek1}
\bibinfo{author}{\bibfnamefont{J.}~\bibnamefont{Kang}} \bibnamefont{and}
  \bibinfo{author}{\bibfnamefont{O.}~\bibnamefont{Vafek}},
  \bibinfo{journal}{Phys. Rev. X} \textbf{\bibinfo{volume}{8}},
  \bibinfo{pages}{031088} (\bibinfo{year}{2018}),
  \urlprefix\url{https://link.aps.org/doi/10.1103/PhysRevX.8.031088}.

\bibitem[{\citenamefont{Zou et~al.}(2018)\citenamefont{Zou, Po, Vishwanath, and
  Senthil}}]{senthil2}
\bibinfo{author}{\bibfnamefont{L.}~\bibnamefont{Zou}},
  \bibinfo{author}{\bibfnamefont{H.~C.} \bibnamefont{Po}},
  \bibinfo{author}{\bibfnamefont{A.}~\bibnamefont{Vishwanath}},
  \bibnamefont{and} \bibinfo{author}{\bibfnamefont{T.}~\bibnamefont{Senthil}},
  \bibinfo{journal}{Phys. Rev. B} \textbf{\bibinfo{volume}{98}},
  \bibinfo{pages}{085435} (\bibinfo{year}{2018}),
  \urlprefix\url{https://link.aps.org/doi/10.1103/PhysRevB.98.085435}.

\bibitem[{\citenamefont{You and Vishwanath}(2018)}]{you}
\bibinfo{author}{\bibfnamefont{Y.-Z.} \bibnamefont{You}} \bibnamefont{and}
  \bibinfo{author}{\bibfnamefont{A.}~\bibnamefont{Vishwanath}},
  \bibinfo{journal}{1805.06867}  (\bibinfo{year}{2018}), \eprint{1805.06867}.

\bibitem[{\citenamefont{Bultinck
  et~al.}(2020{\natexlab{a}})\citenamefont{Bultinck, Khalaf, Liu, Chatterjee,
  Vishwanath, and Zaletel}}]{zaletel}
\bibinfo{author}{\bibfnamefont{N.}~\bibnamefont{Bultinck}},
  \bibinfo{author}{\bibfnamefont{E.}~\bibnamefont{Khalaf}},
  \bibinfo{author}{\bibfnamefont{S.}~\bibnamefont{Liu}},
  \bibinfo{author}{\bibfnamefont{S.}~\bibnamefont{Chatterjee}},
  \bibinfo{author}{\bibfnamefont{A.}~\bibnamefont{Vishwanath}},
  \bibnamefont{and} \bibinfo{author}{\bibfnamefont{M.~P.}
  \bibnamefont{Zaletel}}, \bibinfo{journal}{Phys. Rev. X}
  \textbf{\bibinfo{volume}{10}}, \bibinfo{pages}{031034}
  (\bibinfo{year}{2020}{\natexlab{a}}),
  \urlprefix\url{https://link.aps.org/doi/10.1103/PhysRevX.10.031034}.

\bibitem[{\citenamefont{Bultinck
  et~al.}(2020{\natexlab{b}})\citenamefont{Bultinck, Chatterjee, and
  Zaletel}}]{zaletel2}
\bibinfo{author}{\bibfnamefont{N.}~\bibnamefont{Bultinck}},
  \bibinfo{author}{\bibfnamefont{S.}~\bibnamefont{Chatterjee}},
  \bibnamefont{and} \bibinfo{author}{\bibfnamefont{M.~P.}
  \bibnamefont{Zaletel}}, \bibinfo{journal}{Phys. Rev. Lett.}
  \textbf{\bibinfo{volume}{124}}, \bibinfo{pages}{166601}
  (\bibinfo{year}{2020}{\natexlab{b}}),
  \urlprefix\url{https://link.aps.org/doi/10.1103/PhysRevLett.124.166601}.

\bibitem[{\citenamefont{Wu et~al.}(2018{\natexlab{a}})\citenamefont{Wu,
  MacDonald, and Martin}}]{wu}
\bibinfo{author}{\bibfnamefont{F.}~\bibnamefont{Wu}},
  \bibinfo{author}{\bibfnamefont{A.~H.} \bibnamefont{MacDonald}},
  \bibnamefont{and} \bibinfo{author}{\bibfnamefont{I.}~\bibnamefont{Martin}},
  \bibinfo{journal}{Phys. Rev. Lett.} \textbf{\bibinfo{volume}{121}},
  \bibinfo{pages}{257001} (\bibinfo{year}{2018}{\natexlab{a}}),
  \urlprefix\url{https://link.aps.org/doi/10.1103/PhysRevLett.121.257001}.

\bibitem[{\citenamefont{Lian et~al.}(2019)\citenamefont{Lian, Wang, and
  Bernevig}}]{lian}
\bibinfo{author}{\bibfnamefont{B.}~\bibnamefont{Lian}},
  \bibinfo{author}{\bibfnamefont{Z.}~\bibnamefont{Wang}}, \bibnamefont{and}
  \bibinfo{author}{\bibfnamefont{B.~A.} \bibnamefont{Bernevig}},
  \bibinfo{journal}{Phys. Rev. Lett.} \textbf{\bibinfo{volume}{122}},
  \bibinfo{pages}{257002} (\bibinfo{year}{2019}),
  \urlprefix\url{https://link.aps.org/doi/10.1103/PhysRevLett.122.257002}.

\bibitem[{\citenamefont{Lee et~al.}(2019)\citenamefont{Lee, Khalaf, Liu, Liu,
  Hao, Kim, and Vishwanath}}]{lee2019}
\bibinfo{author}{\bibfnamefont{J.~Y.} \bibnamefont{Lee}},
  \bibinfo{author}{\bibfnamefont{E.}~\bibnamefont{Khalaf}},
  \bibinfo{author}{\bibfnamefont{S.}~\bibnamefont{Liu}},
  \bibinfo{author}{\bibfnamefont{X.}~\bibnamefont{Liu}},
  \bibinfo{author}{\bibfnamefont{Z.}~\bibnamefont{Hao}},
  \bibinfo{author}{\bibfnamefont{P.}~\bibnamefont{Kim}}, \bibnamefont{and}
  \bibinfo{author}{\bibfnamefont{A.}~\bibnamefont{Vishwanath}},
  \bibinfo{journal}{Nature Communications} \textbf{\bibinfo{volume}{10}},
  \bibinfo{pages}{5333} (\bibinfo{year}{2019}), ISSN \bibinfo{issn}{2041-1723},
  \urlprefix\url{http://dx.doi.org/10.1038/s41467-019-12981-1}.

\bibitem[{\citenamefont{Khalaf et~al.}(2021)\citenamefont{Khalaf, Chatterjee,
  Bultinck, Zaletel, and Vishwanath}}]{skyrmion}
\bibinfo{author}{\bibfnamefont{E.}~\bibnamefont{Khalaf}},
  \bibinfo{author}{\bibfnamefont{S.}~\bibnamefont{Chatterjee}},
  \bibinfo{author}{\bibfnamefont{N.}~\bibnamefont{Bultinck}},
  \bibinfo{author}{\bibfnamefont{M.~P.} \bibnamefont{Zaletel}},
  \bibnamefont{and}
  \bibinfo{author}{\bibfnamefont{A.}~\bibnamefont{Vishwanath}},
  \bibinfo{journal}{Science Advances} \textbf{\bibinfo{volume}{7}},
  \bibinfo{pages}{eabf5299} (\bibinfo{year}{2021}), ISSN
  \bibinfo{issn}{2375-2548},
  \urlprefix\url{http://dx.doi.org/10.1126/sciadv.abf5299}.

\bibitem[{\citenamefont{Xu et~al.}(2020{\natexlab{a}})\citenamefont{Xu, Wu,
  Jian, and Xu}}]{xunematic}
\bibinfo{author}{\bibfnamefont{Y.}~\bibnamefont{Xu}},
  \bibinfo{author}{\bibfnamefont{X.-C.} \bibnamefont{Wu}},
  \bibinfo{author}{\bibfnamefont{C.-M.} \bibnamefont{Jian}}, \bibnamefont{and}
  \bibinfo{author}{\bibfnamefont{C.}~\bibnamefont{Xu}}, \bibinfo{journal}{Phys.
  Rev. B} \textbf{\bibinfo{volume}{101}}, \bibinfo{pages}{205426}
  (\bibinfo{year}{2020}{\natexlab{a}}),
  \urlprefix\url{https://link.aps.org/doi/10.1103/PhysRevB.101.205426}.

\bibitem[{\citenamefont{Fernandes and Venderbos}(2020)}]{fernandesnematic}
\bibinfo{author}{\bibfnamefont{R.~M.} \bibnamefont{Fernandes}}
  \bibnamefont{and} \bibinfo{author}{\bibfnamefont{J.~W.~F.}
  \bibnamefont{Venderbos}}, \bibinfo{journal}{Science Advances}
  \textbf{\bibinfo{volume}{6}}, \bibinfo{pages}{eaba8834}
  (\bibinfo{year}{2020}), ISSN \bibinfo{issn}{2375-2548},
  \urlprefix\url{http://dx.doi.org/10.1126/sciadv.aba8834}.

\bibitem[{\citenamefont{Chittari et~al.}(2019)\citenamefont{Chittari, Chen,
  Zhang, Wang, and Jung}}]{wangtopo}
\bibinfo{author}{\bibfnamefont{B.~L.} \bibnamefont{Chittari}},
  \bibinfo{author}{\bibfnamefont{G.}~\bibnamefont{Chen}},
  \bibinfo{author}{\bibfnamefont{Y.}~\bibnamefont{Zhang}},
  \bibinfo{author}{\bibfnamefont{F.}~\bibnamefont{Wang}}, \bibnamefont{and}
  \bibinfo{author}{\bibfnamefont{J.}~\bibnamefont{Jung}},
  \bibinfo{journal}{Phys. Rev. Lett.} \textbf{\bibinfo{volume}{122}},
  \bibinfo{pages}{016401} (\bibinfo{year}{2019}),
  \urlprefix\url{https://link.aps.org/doi/10.1103/PhysRevLett.122.016401}.

\bibitem[{\citenamefont{Serlin et~al.}(2019)\citenamefont{Serlin, Tschirhart,
  Polshyn, Zhang, Zhu, Watanabe, Taniguchi, Balents, and Young}}]{youngtopo}
\bibinfo{author}{\bibfnamefont{M.}~\bibnamefont{Serlin}},
  \bibinfo{author}{\bibfnamefont{C.~L.} \bibnamefont{Tschirhart}},
  \bibinfo{author}{\bibfnamefont{H.}~\bibnamefont{Polshyn}},
  \bibinfo{author}{\bibfnamefont{Y.}~\bibnamefont{Zhang}},
  \bibinfo{author}{\bibfnamefont{J.}~\bibnamefont{Zhu}},
  \bibinfo{author}{\bibfnamefont{K.}~\bibnamefont{Watanabe}},
  \bibinfo{author}{\bibfnamefont{T.}~\bibnamefont{Taniguchi}},
  \bibinfo{author}{\bibfnamefont{L.}~\bibnamefont{Balents}}, \bibnamefont{and}
  \bibinfo{author}{\bibfnamefont{A.~F.} \bibnamefont{Young}},
  \bibinfo{journal}{Science} \textbf{\bibinfo{volume}{367}},
  \bibinfo{pages}{900} (\bibinfo{year}{2019}), ISSN \bibinfo{issn}{1095-9203},
  \urlprefix\url{http://dx.doi.org/10.1126/science.aay5533}.

\bibitem[{\citenamefont{Zhang et~al.}(2019)\citenamefont{Zhang, Mao, Cao,
  Jarillo-Herrero, and Senthil}}]{senthiltopo1}
\bibinfo{author}{\bibfnamefont{Y.-H.} \bibnamefont{Zhang}},
  \bibinfo{author}{\bibfnamefont{D.}~\bibnamefont{Mao}},
  \bibinfo{author}{\bibfnamefont{Y.}~\bibnamefont{Cao}},
  \bibinfo{author}{\bibfnamefont{P.}~\bibnamefont{Jarillo-Herrero}},
  \bibnamefont{and} \bibinfo{author}{\bibfnamefont{T.}~\bibnamefont{Senthil}},
  \bibinfo{journal}{Phys. Rev. B} \textbf{\bibinfo{volume}{99}},
  \bibinfo{pages}{075127} (\bibinfo{year}{2019}),
  \urlprefix\url{https://link.aps.org/doi/10.1103/PhysRevB.99.075127}.

\bibitem[{\citenamefont{Chen et~al.}(2020{\natexlab{a}})\citenamefont{Chen,
  Sharpe, Fox, Zhang, Wang, Jiang, Lyu, Li, Watanabe, Taniguchi
  et~al.}}]{wangtopo2}
\bibinfo{author}{\bibfnamefont{G.}~\bibnamefont{Chen}},
  \bibinfo{author}{\bibfnamefont{A.~L.} \bibnamefont{Sharpe}},
  \bibinfo{author}{\bibfnamefont{E.~J.} \bibnamefont{Fox}},
  \bibinfo{author}{\bibfnamefont{Y.-H.} \bibnamefont{Zhang}},
  \bibinfo{author}{\bibfnamefont{S.}~\bibnamefont{Wang}},
  \bibinfo{author}{\bibfnamefont{L.}~\bibnamefont{Jiang}},
  \bibinfo{author}{\bibfnamefont{B.}~\bibnamefont{Lyu}},
  \bibinfo{author}{\bibfnamefont{H.}~\bibnamefont{Li}},
  \bibinfo{author}{\bibfnamefont{K.}~\bibnamefont{Watanabe}},
  \bibinfo{author}{\bibfnamefont{T.}~\bibnamefont{Taniguchi}},
  \bibnamefont{et~al.}, \bibinfo{journal}{Nature}
  \textbf{\bibinfo{volume}{579}}, \bibinfo{pages}{56}
  (\bibinfo{year}{2020}{\natexlab{a}}), ISSN \bibinfo{issn}{1476-4687},
  \urlprefix\url{http://dx.doi.org/10.1038/s41586-020-2049-7}.

\bibitem[{\citenamefont{Repellin and Senthil}(2019)}]{repellintopo}
\bibinfo{author}{\bibfnamefont{C.}~\bibnamefont{Repellin}} \bibnamefont{and}
  \bibinfo{author}{\bibfnamefont{T.}~\bibnamefont{Senthil}},
  \bibinfo{journal}{1912.11469}  (\bibinfo{year}{2019}), \eprint{1912.11469}.

\bibitem[{\citenamefont{Stepanov
  et~al.}(2020{\natexlab{b}})\citenamefont{Stepanov, Xie, Taniguchi, Watanabe,
  Lu, MacDonald, Bernevig, and Efetov}}]{efetovtopo}
\bibinfo{author}{\bibfnamefont{P.}~\bibnamefont{Stepanov}},
  \bibinfo{author}{\bibfnamefont{M.}~\bibnamefont{Xie}},
  \bibinfo{author}{\bibfnamefont{T.}~\bibnamefont{Taniguchi}},
  \bibinfo{author}{\bibfnamefont{K.}~\bibnamefont{Watanabe}},
  \bibinfo{author}{\bibfnamefont{X.}~\bibnamefont{Lu}},
  \bibinfo{author}{\bibfnamefont{A.~H.} \bibnamefont{MacDonald}},
  \bibinfo{author}{\bibfnamefont{B.~A.} \bibnamefont{Bernevig}},
  \bibnamefont{and} \bibinfo{author}{\bibfnamefont{D.~K.}
  \bibnamefont{Efetov}}, \bibinfo{journal}{2012.15126}
  (\bibinfo{year}{2020}{\natexlab{b}}), \eprint{2012.15126}.

\bibitem[{\citenamefont{Chen et~al.}(2020{\natexlab{b}})\citenamefont{Chen, He,
  Zhang, Hsieh, Fei, Watanabe, Taniguchi, Cobden, Xu, Dean et~al.}}]{dean1}
\bibinfo{author}{\bibfnamefont{S.}~\bibnamefont{Chen}},
  \bibinfo{author}{\bibfnamefont{M.}~\bibnamefont{He}},
  \bibinfo{author}{\bibfnamefont{Y.-H.} \bibnamefont{Zhang}},
  \bibinfo{author}{\bibfnamefont{V.}~\bibnamefont{Hsieh}},
  \bibinfo{author}{\bibfnamefont{Z.}~\bibnamefont{Fei}},
  \bibinfo{author}{\bibfnamefont{K.}~\bibnamefont{Watanabe}},
  \bibinfo{author}{\bibfnamefont{T.}~\bibnamefont{Taniguchi}},
  \bibinfo{author}{\bibfnamefont{D.~H.} \bibnamefont{Cobden}},
  \bibinfo{author}{\bibfnamefont{X.}~\bibnamefont{Xu}},
  \bibinfo{author}{\bibfnamefont{C.~R.} \bibnamefont{Dean}},
  \bibnamefont{et~al.}, \bibinfo{journal}{Nature Physics}
  \textbf{\bibinfo{volume}{17}}, \bibinfo{pages}{374}
  (\bibinfo{year}{2020}{\natexlab{b}}), ISSN \bibinfo{issn}{1745-2481},
  \urlprefix\url{http://dx.doi.org/10.1038/s41567-020-01062-6}.

\bibitem[{\citenamefont{Pierce et~al.}(2021)\citenamefont{Pierce, Xie, Park,
  Khalaf, Lee, Cao, Parker, Forrester, Chen, Watanabe et~al.}}]{yacobytopo}
\bibinfo{author}{\bibfnamefont{A.~T.} \bibnamefont{Pierce}},
  \bibinfo{author}{\bibfnamefont{Y.}~\bibnamefont{Xie}},
  \bibinfo{author}{\bibfnamefont{J.~M.} \bibnamefont{Park}},
  \bibinfo{author}{\bibfnamefont{E.}~\bibnamefont{Khalaf}},
  \bibinfo{author}{\bibfnamefont{S.~H.} \bibnamefont{Lee}},
  \bibinfo{author}{\bibfnamefont{Y.}~\bibnamefont{Cao}},
  \bibinfo{author}{\bibfnamefont{D.~E.} \bibnamefont{Parker}},
  \bibinfo{author}{\bibfnamefont{P.~R.} \bibnamefont{Forrester}},
  \bibinfo{author}{\bibfnamefont{S.}~\bibnamefont{Chen}},
  \bibinfo{author}{\bibfnamefont{K.}~\bibnamefont{Watanabe}},
  \bibnamefont{et~al.}, \bibinfo{journal}{2101.04123}  (\bibinfo{year}{2021}),
  \eprint{2101.04123}.

\bibitem[{\citenamefont{Wu et~al.}(2019{\natexlab{a}})\citenamefont{Wu, Xu,
  Jian, and Xu}}]{liantopo}
\bibinfo{author}{\bibfnamefont{X.-C.} \bibnamefont{Wu}},
  \bibinfo{author}{\bibfnamefont{Y.}~\bibnamefont{Xu}},
  \bibinfo{author}{\bibfnamefont{C.-M.} \bibnamefont{Jian}}, \bibnamefont{and}
  \bibinfo{author}{\bibfnamefont{C.}~\bibnamefont{Xu}}, \bibinfo{journal}{Phys.
  Rev. B} \textbf{\bibinfo{volume}{100}}, \bibinfo{pages}{155138}
  (\bibinfo{year}{2019}{\natexlab{a}}),
  \urlprefix\url{https://link.aps.org/doi/10.1103/PhysRevB.100.155138}.

\bibitem[{\citenamefont{Wu et~al.}(2019{\natexlab{b}})\citenamefont{Wu, Xu,
  Jian, and Xu}}]{wuchern}
\bibinfo{author}{\bibfnamefont{X.-C.} \bibnamefont{Wu}},
  \bibinfo{author}{\bibfnamefont{Y.}~\bibnamefont{Xu}},
  \bibinfo{author}{\bibfnamefont{C.-M.} \bibnamefont{Jian}}, \bibnamefont{and}
  \bibinfo{author}{\bibfnamefont{C.}~\bibnamefont{Xu}}, \bibinfo{journal}{Phys.
  Rev. B} \textbf{\bibinfo{volume}{100}}, \bibinfo{pages}{155138}
  (\bibinfo{year}{2019}{\natexlab{b}}),
  \urlprefix\url{https://link.aps.org/doi/10.1103/PhysRevB.100.155138}.

\bibitem[{\citenamefont{Wu et~al.}(2018{\natexlab{b}})\citenamefont{Wu, Lovorn,
  Tutuc, and MacDonald}}]{tmdhubbard}
\bibinfo{author}{\bibfnamefont{F.}~\bibnamefont{Wu}},
  \bibinfo{author}{\bibfnamefont{T.}~\bibnamefont{Lovorn}},
  \bibinfo{author}{\bibfnamefont{E.}~\bibnamefont{Tutuc}}, \bibnamefont{and}
  \bibinfo{author}{\bibfnamefont{A.~H.} \bibnamefont{MacDonald}},
  \bibinfo{journal}{Phys. Rev. Lett.} \textbf{\bibinfo{volume}{121}},
  \bibinfo{pages}{026402} (\bibinfo{year}{2018}{\natexlab{b}}),
  \urlprefix\url{https://link.aps.org/doi/10.1103/PhysRevLett.121.026402}.

\bibitem[{\citenamefont{Pan et~al.}(2020)\citenamefont{Pan, Wu, and
  Das~Sarma}}]{sarmatmd}
\bibinfo{author}{\bibfnamefont{H.}~\bibnamefont{Pan}},
  \bibinfo{author}{\bibfnamefont{F.}~\bibnamefont{Wu}}, \bibnamefont{and}
  \bibinfo{author}{\bibfnamefont{S.}~\bibnamefont{Das~Sarma}},
  \bibinfo{journal}{Phys. Rev. Research} \textbf{\bibinfo{volume}{2}},
  \bibinfo{pages}{033087} (\bibinfo{year}{2020}),
  \urlprefix\url{https://link.aps.org/doi/10.1103/PhysRevResearch.2.033087}.

\bibitem[{\citenamefont{Tang et~al.}(2019)\citenamefont{Tang, Li, Li, Xu, Liu,
  Barmak, Watanabe, Taniguchi, MacDonald, Shan et~al.}}]{tmdhubbard2}
\bibinfo{author}{\bibfnamefont{Y.}~\bibnamefont{Tang}},
  \bibinfo{author}{\bibfnamefont{L.}~\bibnamefont{Li}},
  \bibinfo{author}{\bibfnamefont{T.}~\bibnamefont{Li}},
  \bibinfo{author}{\bibfnamefont{Y.}~\bibnamefont{Xu}},
  \bibinfo{author}{\bibfnamefont{S.}~\bibnamefont{Liu}},
  \bibinfo{author}{\bibfnamefont{K.}~\bibnamefont{Barmak}},
  \bibinfo{author}{\bibfnamefont{K.}~\bibnamefont{Watanabe}},
  \bibinfo{author}{\bibfnamefont{T.}~\bibnamefont{Taniguchi}},
  \bibinfo{author}{\bibfnamefont{A.~H.} \bibnamefont{MacDonald}},
  \bibinfo{author}{\bibfnamefont{J.}~\bibnamefont{Shan}}, \bibnamefont{et~al.},
  \bibinfo{journal}{1910.08673}  (\bibinfo{year}{2019}), \eprint{1910.08673}.

\bibitem[{\citenamefont{Szasz et~al.}(2020{\natexlab{a}})\citenamefont{Szasz,
  Motruk, Zaletel, and Moore}}]{joelmike}
\bibinfo{author}{\bibfnamefont{A.}~\bibnamefont{Szasz}},
  \bibinfo{author}{\bibfnamefont{J.}~\bibnamefont{Motruk}},
  \bibinfo{author}{\bibfnamefont{M.~P.} \bibnamefont{Zaletel}},
  \bibnamefont{and} \bibinfo{author}{\bibfnamefont{J.~E.} \bibnamefont{Moore}},
  \bibinfo{journal}{Phys. Rev. X} \textbf{\bibinfo{volume}{10}},
  \bibinfo{pages}{021042} (\bibinfo{year}{2020}{\natexlab{a}}),
  \urlprefix\url{https://link.aps.org/doi/10.1103/PhysRevX.10.021042}.

\bibitem[{\citenamefont{Szasz and Motruk}(2021)}]{szasz}
\bibinfo{author}{\bibfnamefont{A.}~\bibnamefont{Szasz}} \bibnamefont{and}
  \bibinfo{author}{\bibfnamefont{J.}~\bibnamefont{Motruk}},
  \bibinfo{journal}{Physical Review B} \textbf{\bibinfo{volume}{103}}
  (\bibinfo{year}{2021}), ISSN \bibinfo{issn}{2469-9969},
  \urlprefix\url{http://dx.doi.org/10.1103/PhysRevB.103.235132}.

\bibitem[{\citenamefont{Li et~al.}(2021)\citenamefont{Li, Jiang, Li, Zhang,
  Kang, Zhu, Watanabe, Taniguchi, Chowdhury, Fu et~al.}}]{tmdmit1}
\bibinfo{author}{\bibfnamefont{T.}~\bibnamefont{Li}},
  \bibinfo{author}{\bibfnamefont{S.}~\bibnamefont{Jiang}},
  \bibinfo{author}{\bibfnamefont{L.}~\bibnamefont{Li}},
  \bibinfo{author}{\bibfnamefont{Y.}~\bibnamefont{Zhang}},
  \bibinfo{author}{\bibfnamefont{K.}~\bibnamefont{Kang}},
  \bibinfo{author}{\bibfnamefont{J.}~\bibnamefont{Zhu}},
  \bibinfo{author}{\bibfnamefont{K.}~\bibnamefont{Watanabe}},
  \bibinfo{author}{\bibfnamefont{T.}~\bibnamefont{Taniguchi}},
  \bibinfo{author}{\bibfnamefont{D.}~\bibnamefont{Chowdhury}},
  \bibinfo{author}{\bibfnamefont{L.}~\bibnamefont{Fu}}, \bibnamefont{et~al.},
  \bibinfo{journal}{Nature} \textbf{\bibinfo{volume}{597}},
  \bibinfo{pages}{350C354} (\bibinfo{year}{2021}), ISSN
  \bibinfo{issn}{1476-4687},
  \urlprefix\url{http://dx.doi.org/10.1038/s41586-021-03853-0}.

\bibitem[{\citenamefont{Ghiotto et~al.}(2021)\citenamefont{Ghiotto, Shih,
  Pereira, Rhodes, Kim, Zang, Millis, Watanabe, Taniguchi, Hone
  et~al.}}]{tmdmit2}
\bibinfo{author}{\bibfnamefont{A.}~\bibnamefont{Ghiotto}},
  \bibinfo{author}{\bibfnamefont{E.-M.} \bibnamefont{Shih}},
  \bibinfo{author}{\bibfnamefont{G.~S. S.~G.} \bibnamefont{Pereira}},
  \bibinfo{author}{\bibfnamefont{D.~A.} \bibnamefont{Rhodes}},
  \bibinfo{author}{\bibfnamefont{B.}~\bibnamefont{Kim}},
  \bibinfo{author}{\bibfnamefont{J.}~\bibnamefont{Zang}},
  \bibinfo{author}{\bibfnamefont{A.~J.} \bibnamefont{Millis}},
  \bibinfo{author}{\bibfnamefont{K.}~\bibnamefont{Watanabe}},
  \bibinfo{author}{\bibfnamefont{T.}~\bibnamefont{Taniguchi}},
  \bibinfo{author}{\bibfnamefont{J.~C.} \bibnamefont{Hone}},
  \bibnamefont{et~al.}, \bibinfo{journal}{Nature}
  \textbf{\bibinfo{volume}{597}}, \bibinfo{pages}{345C349}
  (\bibinfo{year}{2021}), ISSN \bibinfo{issn}{1476-4687},
  \urlprefix\url{http://dx.doi.org/10.1038/s41586-021-03815-6}.

\bibitem[{\citenamefont{Regan et~al.}(2020)\citenamefont{Regan, Wang, Jin,
  Bakti~Utama, Gao, Wei, Zhao, Zhao, Zhang, Yumigeta et~al.}}]{tmdinsulator1}
\bibinfo{author}{\bibfnamefont{E.~C.} \bibnamefont{Regan}},
  \bibinfo{author}{\bibfnamefont{D.}~\bibnamefont{Wang}},
  \bibinfo{author}{\bibfnamefont{C.}~\bibnamefont{Jin}},
  \bibinfo{author}{\bibfnamefont{M.~I.} \bibnamefont{Bakti~Utama}},
  \bibinfo{author}{\bibfnamefont{B.}~\bibnamefont{Gao}},
  \bibinfo{author}{\bibfnamefont{X.}~\bibnamefont{Wei}},
  \bibinfo{author}{\bibfnamefont{S.}~\bibnamefont{Zhao}},
  \bibinfo{author}{\bibfnamefont{W.}~\bibnamefont{Zhao}},
  \bibinfo{author}{\bibfnamefont{Z.}~\bibnamefont{Zhang}},
  \bibinfo{author}{\bibfnamefont{K.}~\bibnamefont{Yumigeta}},
  \bibnamefont{et~al.}, \bibinfo{journal}{Nature}
  \textbf{\bibinfo{volume}{579}}, \bibinfo{pages}{359} (\bibinfo{year}{2020}),
  ISSN \bibinfo{issn}{1476-4687},
  \urlprefix\url{http://dx.doi.org/10.1038/s41586-020-2092-4}.

\bibitem[{\citenamefont{Jin et~al.}(2021)\citenamefont{Jin, Tao, Li, Xu, Tang,
  Zhu, Liu, Watanabe, Taniguchi, Hone et~al.}}]{tmdinsulator2}
\bibinfo{author}{\bibfnamefont{C.}~\bibnamefont{Jin}},
  \bibinfo{author}{\bibfnamefont{Z.}~\bibnamefont{Tao}},
  \bibinfo{author}{\bibfnamefont{T.}~\bibnamefont{Li}},
  \bibinfo{author}{\bibfnamefont{Y.}~\bibnamefont{Xu}},
  \bibinfo{author}{\bibfnamefont{Y.}~\bibnamefont{Tang}},
  \bibinfo{author}{\bibfnamefont{J.}~\bibnamefont{Zhu}},
  \bibinfo{author}{\bibfnamefont{S.}~\bibnamefont{Liu}},
  \bibinfo{author}{\bibfnamefont{K.}~\bibnamefont{Watanabe}},
  \bibinfo{author}{\bibfnamefont{T.}~\bibnamefont{Taniguchi}},
  \bibinfo{author}{\bibfnamefont{J.~C.} \bibnamefont{Hone}},
  \bibnamefont{et~al.}, \bibinfo{journal}{Nature Materials}
  \textbf{\bibinfo{volume}{20}}, \bibinfo{pages}{940} (\bibinfo{year}{2021}),
  ISSN \bibinfo{issn}{1476-4660},
  \urlprefix\url{http://dx.doi.org/10.1038/s41563-021-00959-8}.

\bibitem[{\citenamefont{Xu et~al.}(2020{\natexlab{b}})\citenamefont{Xu, Liu,
  Rhodes, Watanabe, Taniguchi, Hone, Elser, Mak, and Shan}}]{tmdinsulator3}
\bibinfo{author}{\bibfnamefont{Y.}~\bibnamefont{Xu}},
  \bibinfo{author}{\bibfnamefont{S.}~\bibnamefont{Liu}},
  \bibinfo{author}{\bibfnamefont{D.~A.} \bibnamefont{Rhodes}},
  \bibinfo{author}{\bibfnamefont{K.}~\bibnamefont{Watanabe}},
  \bibinfo{author}{\bibfnamefont{T.}~\bibnamefont{Taniguchi}},
  \bibinfo{author}{\bibfnamefont{J.}~\bibnamefont{Hone}},
  \bibinfo{author}{\bibfnamefont{V.}~\bibnamefont{Elser}},
  \bibinfo{author}{\bibfnamefont{K.~F.} \bibnamefont{Mak}}, \bibnamefont{and}
  \bibinfo{author}{\bibfnamefont{J.}~\bibnamefont{Shan}},
  \bibinfo{journal}{Nature} \textbf{\bibinfo{volume}{587}},
  \bibinfo{pages}{214} (\bibinfo{year}{2020}{\natexlab{b}}), ISSN
  \bibinfo{issn}{1476-4687},
  \urlprefix\url{http://dx.doi.org/10.1038/s41586-020-2868-6}.

\bibitem[{\citenamefont{Huang et~al.}(2021)\citenamefont{Huang, Wang, Miao,
  Wang, Li, Lian, Taniguchi, Watanabe, Okamoto, Xiao et~al.}}]{tmdinsulator4}
\bibinfo{author}{\bibfnamefont{X.}~\bibnamefont{Huang}},
  \bibinfo{author}{\bibfnamefont{T.}~\bibnamefont{Wang}},
  \bibinfo{author}{\bibfnamefont{S.}~\bibnamefont{Miao}},
  \bibinfo{author}{\bibfnamefont{C.}~\bibnamefont{Wang}},
  \bibinfo{author}{\bibfnamefont{Z.}~\bibnamefont{Li}},
  \bibinfo{author}{\bibfnamefont{Z.}~\bibnamefont{Lian}},
  \bibinfo{author}{\bibfnamefont{T.}~\bibnamefont{Taniguchi}},
  \bibinfo{author}{\bibfnamefont{K.}~\bibnamefont{Watanabe}},
  \bibinfo{author}{\bibfnamefont{S.}~\bibnamefont{Okamoto}},
  \bibinfo{author}{\bibfnamefont{D.}~\bibnamefont{Xiao}}, \bibnamefont{et~al.},
  \bibinfo{journal}{Nature Physics} \textbf{\bibinfo{volume}{17}},
  \bibinfo{pages}{715} (\bibinfo{year}{2021}), ISSN \bibinfo{issn}{1745-2481},
  \urlprefix\url{http://dx.doi.org/10.1038/s41567-021-01171-w}.

\bibitem[{\citenamefont{Lieb et~al.}(1961)\citenamefont{Lieb, Schultz, and
  Mattis}}]{LSM}
\bibinfo{author}{\bibfnamefont{E.}~\bibnamefont{Lieb}},
  \bibinfo{author}{\bibfnamefont{T.}~\bibnamefont{Schultz}}, \bibnamefont{and}
  \bibinfo{author}{\bibfnamefont{D.}~\bibnamefont{Mattis}},
  \bibinfo{journal}{Annals of Physics} \textbf{\bibinfo{volume}{16}},
  \bibinfo{pages}{407} (\bibinfo{year}{1961}), ISSN \bibinfo{issn}{0003-4916},
  \urlprefix\url{https://www.sciencedirect.com/science/article/pii/00034916619%
01154}.

\bibitem[{\citenamefont{Hastings}(2004)}]{hastings}
\bibinfo{author}{\bibfnamefont{M.~B.} \bibnamefont{Hastings}},
  \bibinfo{journal}{Phys. Rev. B} \textbf{\bibinfo{volume}{69}},
  \bibinfo{pages}{104431} (\bibinfo{year}{2004}),
  \urlprefix\url{https://link.aps.org/doi/10.1103/PhysRevB.69.104431}.

\bibitem[{\citenamefont{Lee and Lee}(2005)}]{lee2005}
\bibinfo{author}{\bibfnamefont{S.-S.} \bibnamefont{Lee}} \bibnamefont{and}
  \bibinfo{author}{\bibfnamefont{P.~A.} \bibnamefont{Lee}},
  \bibinfo{journal}{Phys. Rev. Lett.} \textbf{\bibinfo{volume}{95}},
  \bibinfo{pages}{036403} (\bibinfo{year}{2005}),
  \urlprefix\url{https://link.aps.org/doi/10.1103/PhysRevLett.95.036403}.

\bibitem[{\citenamefont{Senthil}(2008{\natexlab{a}})}]{senthilmit1}
\bibinfo{author}{\bibfnamefont{T.}~\bibnamefont{Senthil}},
  \bibinfo{journal}{Phys. Rev. B} \textbf{\bibinfo{volume}{78}},
  \bibinfo{pages}{035103} (\bibinfo{year}{2008}{\natexlab{a}}),
  \urlprefix\url{https://link.aps.org/doi/10.1103/PhysRevB.78.035103}.

\bibitem[{\citenamefont{Senthil}(2008{\natexlab{b}})}]{senthilmit2}
\bibinfo{author}{\bibfnamefont{T.}~\bibnamefont{Senthil}},
  \bibinfo{journal}{Phys. Rev. B} \textbf{\bibinfo{volume}{78}},
  \bibinfo{pages}{045109} (\bibinfo{year}{2008}{\natexlab{b}}),
  \urlprefix\url{https://link.aps.org/doi/10.1103/PhysRevB.78.045109}.

\bibitem[{\citenamefont{Mross and Senthil}(2011)}]{friedelspinliquid}
\bibinfo{author}{\bibfnamefont{D.~F.} \bibnamefont{Mross}} \bibnamefont{and}
  \bibinfo{author}{\bibfnamefont{T.}~\bibnamefont{Senthil}},
  \bibinfo{journal}{Phys. Rev. B} \textbf{\bibinfo{volume}{84}},
  \bibinfo{pages}{041102} (\bibinfo{year}{2011}),
  \urlprefix\url{https://link.aps.org/doi/10.1103/PhysRevB.84.041102}.

\bibitem[{\citenamefont{Cha et~al.}(1991)\citenamefont{Cha, Fisher, Girvin,
  Wallin, and Young}}]{UCt1}
\bibinfo{author}{\bibfnamefont{M.-C.} \bibnamefont{Cha}},
  \bibinfo{author}{\bibfnamefont{M.~P.~A.} \bibnamefont{Fisher}},
  \bibinfo{author}{\bibfnamefont{S.~M.} \bibnamefont{Girvin}},
  \bibinfo{author}{\bibfnamefont{M.}~\bibnamefont{Wallin}}, \bibnamefont{and}
  \bibinfo{author}{\bibfnamefont{A.~P.} \bibnamefont{Young}},
  \bibinfo{journal}{Phys. Rev. B} \textbf{\bibinfo{volume}{44}},
  \bibinfo{pages}{6883} (\bibinfo{year}{1991}),
  \urlprefix\url{https://link.aps.org/doi/10.1103/PhysRevB.44.6883}.

\bibitem[{\citenamefont{Spivak et~al.}(2010)\citenamefont{Spivak, Kravchenko,
  Kivelson, and Gao}}]{kivelsonreview}
\bibinfo{author}{\bibfnamefont{B.}~\bibnamefont{Spivak}},
  \bibinfo{author}{\bibfnamefont{S.~V.} \bibnamefont{Kravchenko}},
  \bibinfo{author}{\bibfnamefont{S.~A.} \bibnamefont{Kivelson}},
  \bibnamefont{and} \bibinfo{author}{\bibfnamefont{X.~P.~A.}
  \bibnamefont{Gao}}, \bibinfo{journal}{Reviews of Modern Physics}
  \textbf{\bibinfo{volume}{82}}, \bibinfo{pages}{1743} (\bibinfo{year}{2010}),
  ISSN \bibinfo{issn}{1539-0756},
  \urlprefix\url{http://dx.doi.org/10.1103/RevModPhys.82.1743}.

\bibitem[{\citenamefont{Ioffe and Larkin}(1989)}]{larkin}
\bibinfo{author}{\bibfnamefont{L.~B.} \bibnamefont{Ioffe}} \bibnamefont{and}
  \bibinfo{author}{\bibfnamefont{A.~I.} \bibnamefont{Larkin}},
  \bibinfo{journal}{Phys. Rev. B} \textbf{\bibinfo{volume}{39}},
  \bibinfo{pages}{8988} (\bibinfo{year}{1989}),
  \urlprefix\url{https://link.aps.org/doi/10.1103/PhysRevB.39.8988}.

\bibitem[{\citenamefont{Fisher et~al.}(1990)\citenamefont{Fisher, Grinstein,
  and Girvin}}]{UC1}
\bibinfo{author}{\bibfnamefont{M.~P.~A.} \bibnamefont{Fisher}},
  \bibinfo{author}{\bibfnamefont{G.}~\bibnamefont{Grinstein}},
  \bibnamefont{and} \bibinfo{author}{\bibfnamefont{S.~M.}
  \bibnamefont{Girvin}}, \bibinfo{journal}{Phys. Rev. Lett.}
  \textbf{\bibinfo{volume}{64}}, \bibinfo{pages}{587} (\bibinfo{year}{1990}),
  \urlprefix\url{https://link.aps.org/doi/10.1103/PhysRevLett.64.587}.

\bibitem[{\citenamefont{Fazio and Zappal\`a}(1996)}]{UCt2}
\bibinfo{author}{\bibfnamefont{R.}~\bibnamefont{Fazio}} \bibnamefont{and}
  \bibinfo{author}{\bibfnamefont{D.}~\bibnamefont{Zappal\`a}},
  \bibinfo{journal}{Phys. Rev. B} \textbf{\bibinfo{volume}{53}},
  \bibinfo{pages}{R8883} (\bibinfo{year}{1996}),
  \urlprefix\url{https://link.aps.org/doi/10.1103/PhysRevB.53.R8883}.

\bibitem[{\citenamefont{\ifmmode~\check{S}\else \v{S}\fi{}makov and
  S\o{}rensen}(2005)}]{UCt3}
\bibinfo{author}{\bibfnamefont{J.}~\bibnamefont{\ifmmode~\check{S}\else
  \v{S}\fi{}makov}} \bibnamefont{and}
  \bibinfo{author}{\bibfnamefont{E.}~\bibnamefont{S\o{}rensen}},
  \bibinfo{journal}{Phys. Rev. Lett.} \textbf{\bibinfo{volume}{95}},
  \bibinfo{pages}{180603} (\bibinfo{year}{2005}),
  \urlprefix\url{https://link.aps.org/doi/10.1103/PhysRevLett.95.180603}.

\bibitem[{\citenamefont{Witczak-Krempa
  et~al.}(2014)\citenamefont{Witczak-Krempa, S?rensen, and Sachdev}}]{cond3}
\bibinfo{author}{\bibfnamefont{W.}~\bibnamefont{Witczak-Krempa}},
  \bibinfo{author}{\bibfnamefont{E.~S.} \bibnamefont{S?rensen}},
  \bibnamefont{and} \bibinfo{author}{\bibfnamefont{S.}~\bibnamefont{Sachdev}},
  \bibinfo{journal}{Nature Physics} \textbf{\bibinfo{volume}{10}},
  \bibinfo{pages}{361} (\bibinfo{year}{2014}), ISSN \bibinfo{issn}{1745-2481},
  \urlprefix\url{http://dx.doi.org/10.1038/nphys2913}.

\bibitem[{\citenamefont{Chen et~al.}(2014{\natexlab{a}})\citenamefont{Chen,
  Liu, Deng, Pollet, and Prokof'ev}}]{UCt4}
\bibinfo{author}{\bibfnamefont{K.}~\bibnamefont{Chen}},
  \bibinfo{author}{\bibfnamefont{L.}~\bibnamefont{Liu}},
  \bibinfo{author}{\bibfnamefont{Y.}~\bibnamefont{Deng}},
  \bibinfo{author}{\bibfnamefont{L.}~\bibnamefont{Pollet}}, \bibnamefont{and}
  \bibinfo{author}{\bibfnamefont{N.}~\bibnamefont{Prokof'ev}},
  \bibinfo{journal}{Phys. Rev. Lett.} \textbf{\bibinfo{volume}{112}},
  \bibinfo{pages}{030402} (\bibinfo{year}{2014}{\natexlab{a}}),
  \urlprefix\url{https://link.aps.org/doi/10.1103/PhysRevLett.112.030402}.

\bibitem[{\citenamefont{Chester et~al.}(2020)\citenamefont{Chester, Landry,
  Liu, Poland, Simmons-Duffin, Su, and Vichi}}]{UCt5}
\bibinfo{author}{\bibfnamefont{S.~M.} \bibnamefont{Chester}},
  \bibinfo{author}{\bibfnamefont{W.}~\bibnamefont{Landry}},
  \bibinfo{author}{\bibfnamefont{J.}~\bibnamefont{Liu}},
  \bibinfo{author}{\bibfnamefont{D.}~\bibnamefont{Poland}},
  \bibinfo{author}{\bibfnamefont{D.}~\bibnamefont{Simmons-Duffin}},
  \bibinfo{author}{\bibfnamefont{N.}~\bibnamefont{Su}}, \bibnamefont{and}
  \bibinfo{author}{\bibfnamefont{A.}~\bibnamefont{Vichi}},
  \bibinfo{journal}{Journal of High Energy Physics}
  \textbf{\bibinfo{volume}{2020}}, \bibinfo{pages}{142} (\bibinfo{year}{2020}),
  \urlprefix\url{http://dx.doi.org/10.1007/JHEP06(2020)142}.

\bibitem[{\citenamefont{Haviland et~al.}(1989)\citenamefont{Haviland, Liu, and
  Goldman}}]{UCe1}
\bibinfo{author}{\bibfnamefont{D.~B.} \bibnamefont{Haviland}},
  \bibinfo{author}{\bibfnamefont{Y.}~\bibnamefont{Liu}}, \bibnamefont{and}
  \bibinfo{author}{\bibfnamefont{A.~M.} \bibnamefont{Goldman}},
  \bibinfo{journal}{Phys. Rev. Lett.} \textbf{\bibinfo{volume}{62}},
  \bibinfo{pages}{2180} (\bibinfo{year}{1989}),
  \urlprefix\url{https://link.aps.org/doi/10.1103/PhysRevLett.62.2180}.

\bibitem[{\citenamefont{Liu et~al.}(1991)\citenamefont{Liu, McGreer, Nease,
  Haviland, Martinez, Halley, and Goldman}}]{UCe2}
\bibinfo{author}{\bibfnamefont{Y.}~\bibnamefont{Liu}},
  \bibinfo{author}{\bibfnamefont{K.~A.} \bibnamefont{McGreer}},
  \bibinfo{author}{\bibfnamefont{B.}~\bibnamefont{Nease}},
  \bibinfo{author}{\bibfnamefont{D.~B.} \bibnamefont{Haviland}},
  \bibinfo{author}{\bibfnamefont{G.}~\bibnamefont{Martinez}},
  \bibinfo{author}{\bibfnamefont{J.~W.} \bibnamefont{Halley}},
  \bibnamefont{and} \bibinfo{author}{\bibfnamefont{A.~M.}
  \bibnamefont{Goldman}}, \bibinfo{journal}{Phys. Rev. Lett.}
  \textbf{\bibinfo{volume}{67}}, \bibinfo{pages}{2068} (\bibinfo{year}{1991}),
  \urlprefix\url{https://link.aps.org/doi/10.1103/PhysRevLett.67.2068}.

\bibitem[{\citenamefont{Lee and Ketterson}(1990)}]{UCe3}
\bibinfo{author}{\bibfnamefont{S.~J.} \bibnamefont{Lee}} \bibnamefont{and}
  \bibinfo{author}{\bibfnamefont{J.~B.} \bibnamefont{Ketterson}},
  \bibinfo{journal}{Phys. Rev. Lett.} \textbf{\bibinfo{volume}{64}},
  \bibinfo{pages}{3078} (\bibinfo{year}{1990}),
  \urlprefix\url{https://link.aps.org/doi/10.1103/PhysRevLett.64.3078}.

\bibitem[{\citenamefont{Witczak-Krempa
  et~al.}(2012)\citenamefont{Witczak-Krempa, Ghaemi, Senthil, and
  Kim}}]{resistivity2}
\bibinfo{author}{\bibfnamefont{W.}~\bibnamefont{Witczak-Krempa}},
  \bibinfo{author}{\bibfnamefont{P.}~\bibnamefont{Ghaemi}},
  \bibinfo{author}{\bibfnamefont{T.}~\bibnamefont{Senthil}}, \bibnamefont{and}
  \bibinfo{author}{\bibfnamefont{Y.~B.} \bibnamefont{Kim}},
  \bibinfo{journal}{Phys. Rev. B} \textbf{\bibinfo{volume}{86}},
  \bibinfo{pages}{245102} (\bibinfo{year}{2012}),
  \urlprefix\url{https://link.aps.org/doi/10.1103/PhysRevB.86.245102}.

\bibitem[{\citenamefont{Emery and Kivelson}(1995)}]{badmetal}
\bibinfo{author}{\bibfnamefont{V.~J.} \bibnamefont{Emery}} \bibnamefont{and}
  \bibinfo{author}{\bibfnamefont{S.~A.} \bibnamefont{Kivelson}},
  \bibinfo{journal}{Phys. Rev. Lett.} \textbf{\bibinfo{volume}{74}},
  \bibinfo{pages}{3253} (\bibinfo{year}{1995}),
  \urlprefix\url{https://link.aps.org/doi/10.1103/PhysRevLett.74.3253}.

\bibitem[{\citenamefont{Hussey et~al.}(2004)\citenamefont{Hussey, Takenaka, and
  Takagi}}]{badmetal2}
\bibinfo{author}{\bibfnamefont{N.~E.} \bibnamefont{Hussey}},
  \bibinfo{author}{\bibfnamefont{K.}~\bibnamefont{Takenaka}}, \bibnamefont{and}
  \bibinfo{author}{\bibfnamefont{H.}~\bibnamefont{Takagi}},
  \bibinfo{journal}{Philosophical Magazine} \textbf{\bibinfo{volume}{84}},
  \bibinfo{pages}{2847} (\bibinfo{year}{2004}),
  \eprint{https://doi.org/10.1080/14786430410001716944},
  \urlprefix\url{https://doi.org/10.1080/14786430410001716944}.

\bibitem[{\citenamefont{Peskin}(1978)}]{peskindual}
\bibinfo{author}{\bibfnamefont{M.~E.} \bibnamefont{Peskin}},
  \bibinfo{journal}{Annals of Physics} \textbf{\bibinfo{volume}{113}},
  \bibinfo{pages}{122 } (\bibinfo{year}{1978}), ISSN \bibinfo{issn}{0003-4916},
  \urlprefix\url{http://www.sciencedirect.com/science/article/pii/000349167890%
252X}.

\bibitem[{\citenamefont{Dasgupta and Halperin}(1981)}]{halperindual}
\bibinfo{author}{\bibfnamefont{C.}~\bibnamefont{Dasgupta}} \bibnamefont{and}
  \bibinfo{author}{\bibfnamefont{B.~I.} \bibnamefont{Halperin}},
  \bibinfo{journal}{Phys. Rev. Lett.} \textbf{\bibinfo{volume}{47}},
  \bibinfo{pages}{1556} (\bibinfo{year}{1981}),
  \urlprefix\url{https://link.aps.org/doi/10.1103/PhysRevLett.47.1556}.

\bibitem[{\citenamefont{Fisher and Lee}(1989)}]{leedual}
\bibinfo{author}{\bibfnamefont{M.~P.~A.} \bibnamefont{Fisher}}
  \bibnamefont{and} \bibinfo{author}{\bibfnamefont{D.~H.} \bibnamefont{Lee}},
  \bibinfo{journal}{Phys. Rev. B} \textbf{\bibinfo{volume}{39}},
  \bibinfo{pages}{2756} (\bibinfo{year}{1989}),
  \urlprefix\url{https://link.aps.org/doi/10.1103/PhysRevB.39.2756}.

\bibitem[{\citenamefont{Balents et~al.}(2005)\citenamefont{Balents, Bartosch,
  Burkov, Sachdev, and Sengupta}}]{vortexbalents}
\bibinfo{author}{\bibfnamefont{L.}~\bibnamefont{Balents}},
  \bibinfo{author}{\bibfnamefont{L.}~\bibnamefont{Bartosch}},
  \bibinfo{author}{\bibfnamefont{A.}~\bibnamefont{Burkov}},
  \bibinfo{author}{\bibfnamefont{S.}~\bibnamefont{Sachdev}}, \bibnamefont{and}
  \bibinfo{author}{\bibfnamefont{K.}~\bibnamefont{Sengupta}},
  \bibinfo{journal}{Progress of Theoretical Physics Supplement}
  \textbf{\bibinfo{volume}{160}}, \bibinfo{pages}{314} (\bibinfo{year}{2005}),
  ISSN \bibinfo{issn}{0375-9687},
  \urlprefix\url{http://dx.doi.org/10.1143/PTPS.160.314}.

\bibitem[{\citenamefont{Burkov and Balents}(2005)}]{burkovbalents}
\bibinfo{author}{\bibfnamefont{A.~A.} \bibnamefont{Burkov}} \bibnamefont{and}
  \bibinfo{author}{\bibfnamefont{L.}~\bibnamefont{Balents}},
  \bibinfo{journal}{Phys. Rev. B} \textbf{\bibinfo{volume}{72}},
  \bibinfo{pages}{134502} (\bibinfo{year}{2005}),
  \urlprefix\url{https://link.aps.org/doi/10.1103/PhysRevB.72.134502}.

\bibitem[{\citenamefont{Moessner and Sondhi}(2001)}]{sondhiising}
\bibinfo{author}{\bibfnamefont{R.}~\bibnamefont{Moessner}} \bibnamefont{and}
  \bibinfo{author}{\bibfnamefont{S.~L.} \bibnamefont{Sondhi}},
  \bibinfo{journal}{Phys. Rev. B} \textbf{\bibinfo{volume}{63}},
  \bibinfo{pages}{224401} (\bibinfo{year}{2001}),
  \urlprefix\url{https://link.aps.org/doi/10.1103/PhysRevB.63.224401}.

\bibitem[{\citenamefont{Xu and Sachdev}(2009)}]{xusachdevtriangle}
\bibinfo{author}{\bibfnamefont{C.}~\bibnamefont{Xu}} \bibnamefont{and}
  \bibinfo{author}{\bibfnamefont{S.}~\bibnamefont{Sachdev}},
  \bibinfo{journal}{Phys. Rev. B} \textbf{\bibinfo{volume}{79}},
  \bibinfo{pages}{064405} (\bibinfo{year}{2009}),
  \urlprefix\url{https://link.aps.org/doi/10.1103/PhysRevB.79.064405}.

\bibitem[{\citenamefont{Slagle and Xu}(2014)}]{kevintriangle}
\bibinfo{author}{\bibfnamefont{K.}~\bibnamefont{Slagle}} \bibnamefont{and}
  \bibinfo{author}{\bibfnamefont{C.}~\bibnamefont{Xu}}, \bibinfo{journal}{Phys.
  Rev. B} \textbf{\bibinfo{volume}{89}}, \bibinfo{pages}{104418}
  (\bibinfo{year}{2014}),
  \urlprefix\url{https://link.aps.org/doi/10.1103/PhysRevB.89.104418}.

\bibitem[{\citenamefont{Xu and Balents}(2011)}]{xubalents2011}
\bibinfo{author}{\bibfnamefont{C.}~\bibnamefont{Xu}} \bibnamefont{and}
  \bibinfo{author}{\bibfnamefont{L.}~\bibnamefont{Balents}},
  \bibinfo{journal}{Phys. Rev. B} \textbf{\bibinfo{volume}{84}},
  \bibinfo{pages}{014402} (\bibinfo{year}{2011}),
  \urlprefix\url{https://link.aps.org/doi/10.1103/PhysRevB.84.014402}.

\bibitem[{\citenamefont{{Wu} et~al.}(2008)\citenamefont{{Wu}, {Mondragon-Shem},
  and {Zhou}}}]{ringwu}
\bibinfo{author}{\bibfnamefont{C.}~\bibnamefont{{Wu}}},
  \bibinfo{author}{\bibfnamefont{I.}~\bibnamefont{{Mondragon-Shem}}},
  \bibnamefont{and} \bibinfo{author}{\bibfnamefont{X.-F.}
  \bibnamefont{{Zhou}}}, \bibinfo{journal}{arXiv e-prints}
  \bibinfo{eid}{arXiv:0809.3532} (\bibinfo{year}{2008}), \eprint{0809.3532}.

\bibitem[{\citenamefont{Wang et~al.}(2010)\citenamefont{Wang, Gao, Jian, and
  Zhai}}]{ringzhai}
\bibinfo{author}{\bibfnamefont{C.}~\bibnamefont{Wang}},
  \bibinfo{author}{\bibfnamefont{C.}~\bibnamefont{Gao}},
  \bibinfo{author}{\bibfnamefont{C.-M.} \bibnamefont{Jian}}, \bibnamefont{and}
  \bibinfo{author}{\bibfnamefont{H.}~\bibnamefont{Zhai}},
  \bibinfo{journal}{Phys. Rev. Lett.} \textbf{\bibinfo{volume}{105}},
  \bibinfo{pages}{160403} (\bibinfo{year}{2010}),
  \urlprefix\url{https://link.aps.org/doi/10.1103/PhysRevLett.105.160403}.

\bibitem[{\citenamefont{{Zhang} and {Chen}}(2021)}]{ringchen}
\bibinfo{author}{\bibfnamefont{X.-T.} \bibnamefont{{Zhang}}} \bibnamefont{and}
  \bibinfo{author}{\bibfnamefont{G.}~\bibnamefont{{Chen}}},
  \bibinfo{journal}{arXiv e-prints} \bibinfo{eid}{arXiv:2102.09272}
  (\bibinfo{year}{2021}), \eprint{2102.09272}.

\bibitem[{\citenamefont{Lake et~al.}(2021)\citenamefont{Lake, Senthil, and
  Vishwanath}}]{ringlake}
\bibinfo{author}{\bibfnamefont{E.}~\bibnamefont{Lake}},
  \bibinfo{author}{\bibfnamefont{T.}~\bibnamefont{Senthil}}, \bibnamefont{and}
  \bibinfo{author}{\bibfnamefont{A.}~\bibnamefont{Vishwanath}},
  \bibinfo{journal}{Phys. Rev. B} \textbf{\bibinfo{volume}{104}},
  \bibinfo{pages}{014517} (\bibinfo{year}{2021}),
  \urlprefix\url{https://link.aps.org/doi/10.1103/PhysRevB.104.014517}.

\bibitem[{\citenamefont{Musser et~al.}(2021)\citenamefont{Musser, Senthil, and
  Chowdhury}}]{debanjan2021}
\bibinfo{author}{\bibfnamefont{S.}~\bibnamefont{Musser}},
  \bibinfo{author}{\bibfnamefont{T.}~\bibnamefont{Senthil}}, \bibnamefont{and}
  \bibinfo{author}{\bibfnamefont{D.}~\bibnamefont{Chowdhury}},
  \emph{\bibinfo{title}{Theory of a continuous bandwidth-tuned wigner-mott
  transition}} (\bibinfo{year}{2021}), \eprint{2111.09894}.

\bibitem[{\citenamefont{{Grover} and {Vishwanath}}(2012)}]{groveredge}
\bibinfo{author}{\bibfnamefont{T.}~\bibnamefont{{Grover}}} \bibnamefont{and}
  \bibinfo{author}{\bibfnamefont{A.}~\bibnamefont{{Vishwanath}}},
  \bibinfo{journal}{arXiv e-prints} \bibinfo{eid}{arXiv:1206.1332}
  (\bibinfo{year}{2012}), \eprint{1206.1332}.

\bibitem[{\citenamefont{Xu et~al.}(2020{\natexlab{c}})\citenamefont{Xu, Wu,
  Jian, and Xu}}]{edgexu1}
\bibinfo{author}{\bibfnamefont{Y.}~\bibnamefont{Xu}},
  \bibinfo{author}{\bibfnamefont{X.-C.} \bibnamefont{Wu}},
  \bibinfo{author}{\bibfnamefont{C.-M.} \bibnamefont{Jian}}, \bibnamefont{and}
  \bibinfo{author}{\bibfnamefont{C.}~\bibnamefont{Xu}}, \bibinfo{journal}{Phys.
  Rev. B} \textbf{\bibinfo{volume}{101}}, \bibinfo{pages}{184419}
  (\bibinfo{year}{2020}{\natexlab{c}}),
  \urlprefix\url{https://link.aps.org/doi/10.1103/PhysRevB.101.184419}.

\bibitem[{\citenamefont{Jian et~al.}(2021)\citenamefont{Jian, Xu, Wu, and
  Xu}}]{edgexu2}
\bibinfo{author}{\bibfnamefont{C.-M.} \bibnamefont{Jian}},
  \bibinfo{author}{\bibfnamefont{Y.}~\bibnamefont{Xu}},
  \bibinfo{author}{\bibfnamefont{X.-C.} \bibnamefont{Wu}}, \bibnamefont{and}
  \bibinfo{author}{\bibfnamefont{C.}~\bibnamefont{Xu}},
  \bibinfo{journal}{SciPost Phys.} \textbf{\bibinfo{volume}{10}},
  \bibinfo{pages}{33} (\bibinfo{year}{2021}),
  \urlprefix\url{https://scipost.org/10.21468/SciPostPhys.10.2.033}.

\bibitem[{\citenamefont{Polchinski}(1994)}]{polchinskinfl}
\bibinfo{author}{\bibfnamefont{J.}~\bibnamefont{Polchinski}},
  \bibinfo{journal}{Nuclear Physics B} \textbf{\bibinfo{volume}{422}},
  \bibinfo{pages}{617 } (\bibinfo{year}{1994}), ISSN \bibinfo{issn}{0550-3213},
  \urlprefix\url{http://www.sciencedirect.com/science/article/pii/055032139490%
4499}.

\bibitem[{\citenamefont{Nayak and Wilczek}(1994{\natexlab{a}})}]{nayaknfl1}
\bibinfo{author}{\bibfnamefont{C.}~\bibnamefont{Nayak}} \bibnamefont{and}
  \bibinfo{author}{\bibfnamefont{F.}~\bibnamefont{Wilczek}},
  \bibinfo{journal}{Nuclear Physics B} \textbf{\bibinfo{volume}{417}},
  \bibinfo{pages}{359 } (\bibinfo{year}{1994}{\natexlab{a}}), ISSN
  \bibinfo{issn}{0550-3213},
  \urlprefix\url{http://www.sciencedirect.com/science/article/pii/055032139490%
4774}.

\bibitem[{\citenamefont{Nayak and Wilczek}(1994{\natexlab{b}})}]{nayaknfl2}
\bibinfo{author}{\bibfnamefont{C.}~\bibnamefont{Nayak}} \bibnamefont{and}
  \bibinfo{author}{\bibfnamefont{F.}~\bibnamefont{Wilczek}},
  \bibinfo{journal}{Nuclear Physics B} \textbf{\bibinfo{volume}{430}},
  \bibinfo{pages}{534 } (\bibinfo{year}{1994}{\natexlab{b}}), ISSN
  \bibinfo{issn}{0550-3213},
  \urlprefix\url{http://www.sciencedirect.com/science/article/pii/055032139490%
1589}.

\bibitem[{\citenamefont{Lee}(2009)}]{nfl2}
\bibinfo{author}{\bibfnamefont{S.-S.} \bibnamefont{Lee}},
  \bibinfo{journal}{Phys. Rev. B} \textbf{\bibinfo{volume}{80}},
  \bibinfo{pages}{165102} (\bibinfo{year}{2009}),
  \urlprefix\url{https://link.aps.org/doi/10.1103/PhysRevB.80.165102}.

\bibitem[{\citenamefont{Mross et~al.}(2010)\citenamefont{Mross, McGreevy, Liu,
  and Senthil}}]{nfl3}
\bibinfo{author}{\bibfnamefont{D.~F.} \bibnamefont{Mross}},
  \bibinfo{author}{\bibfnamefont{J.}~\bibnamefont{McGreevy}},
  \bibinfo{author}{\bibfnamefont{H.}~\bibnamefont{Liu}}, \bibnamefont{and}
  \bibinfo{author}{\bibfnamefont{T.}~\bibnamefont{Senthil}},
  \bibinfo{journal}{Phys. Rev. B} \textbf{\bibinfo{volume}{82}},
  \bibinfo{pages}{045121} (\bibinfo{year}{2010}),
  \urlprefix\url{https://link.aps.org/doi/10.1103/PhysRevB.82.045121}.

\bibitem[{\citenamefont{Metlitski and Sachdev}(2010{\natexlab{a}})}]{nfl4}
\bibinfo{author}{\bibfnamefont{M.~A.} \bibnamefont{Metlitski}}
  \bibnamefont{and} \bibinfo{author}{\bibfnamefont{S.}~\bibnamefont{Sachdev}},
  \bibinfo{journal}{Phys. Rev. B} \textbf{\bibinfo{volume}{82}},
  \bibinfo{pages}{075127} (\bibinfo{year}{2010}{\natexlab{a}}),
  \urlprefix\url{https://link.aps.org/doi/10.1103/PhysRevB.82.075127}.

\bibitem[{\citenamefont{Metlitski and Sachdev}(2010{\natexlab{b}})}]{nfl5}
\bibinfo{author}{\bibfnamefont{M.~A.} \bibnamefont{Metlitski}}
  \bibnamefont{and} \bibinfo{author}{\bibfnamefont{S.}~\bibnamefont{Sachdev}},
  \bibinfo{journal}{Phys. Rev. B} \textbf{\bibinfo{volume}{82}},
  \bibinfo{pages}{075128} (\bibinfo{year}{2010}{\natexlab{b}}),
  \urlprefix\url{https://link.aps.org/doi/10.1103/PhysRevB.82.075128}.

\bibitem[{\citenamefont{Metlitski et~al.}(2015)\citenamefont{Metlitski, Mross,
  Sachdev, and Senthil}}]{nflpairing1}
\bibinfo{author}{\bibfnamefont{M.~A.} \bibnamefont{Metlitski}},
  \bibinfo{author}{\bibfnamefont{D.~F.} \bibnamefont{Mross}},
  \bibinfo{author}{\bibfnamefont{S.}~\bibnamefont{Sachdev}}, \bibnamefont{and}
  \bibinfo{author}{\bibfnamefont{T.}~\bibnamefont{Senthil}},
  \bibinfo{journal}{Phys. Rev. B} \textbf{\bibinfo{volume}{91}},
  \bibinfo{pages}{115111} (\bibinfo{year}{2015}),
  \urlprefix\url{https://link.aps.org/doi/10.1103/PhysRevB.91.115111}.

\bibitem[{\citenamefont{Wang and Chubukov}(2015)}]{nflpairing2}
\bibinfo{author}{\bibfnamefont{Y.}~\bibnamefont{Wang}} \bibnamefont{and}
  \bibinfo{author}{\bibfnamefont{A.~V.} \bibnamefont{Chubukov}},
  \bibinfo{journal}{Phys. Rev. B} \textbf{\bibinfo{volume}{92}},
  \bibinfo{pages}{125108} (\bibinfo{year}{2015}),
  \urlprefix\url{https://link.aps.org/doi/10.1103/PhysRevB.92.125108}.

\bibitem[{\citenamefont{Mandal}(2016)}]{nflpairing3}
\bibinfo{author}{\bibfnamefont{I.}~\bibnamefont{Mandal}},
  \bibinfo{journal}{Phys. Rev. B} \textbf{\bibinfo{volume}{94}},
  \bibinfo{pages}{115138} (\bibinfo{year}{2016}),
  \urlprefix\url{https://link.aps.org/doi/10.1103/PhysRevB.94.115138}.

\bibitem[{\citenamefont{Lederer et~al.}(2015)\citenamefont{Lederer, Schattner,
  Berg, and Kivelson}}]{nflpairing4}
\bibinfo{author}{\bibfnamefont{S.}~\bibnamefont{Lederer}},
  \bibinfo{author}{\bibfnamefont{Y.}~\bibnamefont{Schattner}},
  \bibinfo{author}{\bibfnamefont{E.}~\bibnamefont{Berg}}, \bibnamefont{and}
  \bibinfo{author}{\bibfnamefont{S.~A.} \bibnamefont{Kivelson}},
  \bibinfo{journal}{Phys. Rev. Lett.} \textbf{\bibinfo{volume}{114}},
  \bibinfo{pages}{097001} (\bibinfo{year}{2015}),
  \urlprefix\url{https://link.aps.org/doi/10.1103/PhysRevLett.114.097001}.

\bibitem[{\citenamefont{Wang et~al.}(2016)\citenamefont{Wang, Abanov,
  Altshuler, Yuzbashyan, and Chubukov}}]{nflpairing5}
\bibinfo{author}{\bibfnamefont{Y.}~\bibnamefont{Wang}},
  \bibinfo{author}{\bibfnamefont{A.}~\bibnamefont{Abanov}},
  \bibinfo{author}{\bibfnamefont{B.~L.} \bibnamefont{Altshuler}},
  \bibinfo{author}{\bibfnamefont{E.~A.} \bibnamefont{Yuzbashyan}},
  \bibnamefont{and} \bibinfo{author}{\bibfnamefont{A.~V.}
  \bibnamefont{Chubukov}}, \bibinfo{journal}{Phys. Rev. Lett.}
  \textbf{\bibinfo{volume}{117}}, \bibinfo{pages}{157001}
  (\bibinfo{year}{2016}),
  \urlprefix\url{https://link.aps.org/doi/10.1103/PhysRevLett.117.157001}.

\bibitem[{\citenamefont{Lederer et~al.}(2017)\citenamefont{Lederer, Schattner,
  Berg, and Kivelson}}]{nflpairing6}
\bibinfo{author}{\bibfnamefont{S.}~\bibnamefont{Lederer}},
  \bibinfo{author}{\bibfnamefont{Y.}~\bibnamefont{Schattner}},
  \bibinfo{author}{\bibfnamefont{E.}~\bibnamefont{Berg}}, \bibnamefont{and}
  \bibinfo{author}{\bibfnamefont{S.~A.} \bibnamefont{Kivelson}},
  \bibinfo{journal}{Proceedings of the National Academy of Sciences}
  \textbf{\bibinfo{volume}{114}}, \bibinfo{pages}{4905} (\bibinfo{year}{2017}),
  \eprint{http://www.pnas.org/content/114/19/4905.full.pdf},
  \urlprefix\url{http://www.pnas.org/content/114/19/4905.abstract}.

\bibitem[{\citenamefont{Zou and Chowdhury}(2020)}]{zounfl}
\bibinfo{author}{\bibfnamefont{L.}~\bibnamefont{Zou}} \bibnamefont{and}
  \bibinfo{author}{\bibfnamefont{D.}~\bibnamefont{Chowdhury}},
  \bibinfo{journal}{Phys. Rev. Research} \textbf{\bibinfo{volume}{2}},
  \bibinfo{pages}{023344} (\bibinfo{year}{2020}),
  \urlprefix\url{https://link.aps.org/doi/10.1103/PhysRevResearch.2.023344}.

\bibitem[{\citenamefont{Mandal}(2020)}]{mandal}
\bibinfo{author}{\bibfnamefont{I.}~\bibnamefont{Mandal}},
  \bibinfo{journal}{Phys. Rev. Research} \textbf{\bibinfo{volume}{2}},
  \bibinfo{pages}{043277} (\bibinfo{year}{2020}),
  \urlprefix\url{https://link.aps.org/doi/10.1103/PhysRevResearch.2.043277}.

\bibitem[{\citenamefont{Calabrese et~al.}(2003)\citenamefont{Calabrese,
  Pelissetto, and Vicari}}]{vicari}
\bibinfo{author}{\bibfnamefont{P.}~\bibnamefont{Calabrese}},
  \bibinfo{author}{\bibfnamefont{A.}~\bibnamefont{Pelissetto}},
  \bibnamefont{and} \bibinfo{author}{\bibfnamefont{E.}~\bibnamefont{Vicari}},
  \bibinfo{journal}{arXiv:cond-mat/0306273}  (\bibinfo{year}{2003}).

\bibitem[{\citenamefont{Isakov et~al.}(2007)\citenamefont{Isakov, Paramekanti,
  and Kim}}]{isakov}
\bibinfo{author}{\bibfnamefont{S.~V.} \bibnamefont{Isakov}},
  \bibinfo{author}{\bibfnamefont{A.}~\bibnamefont{Paramekanti}},
  \bibnamefont{and} \bibinfo{author}{\bibfnamefont{Y.~B.} \bibnamefont{Kim}},
  \bibinfo{journal}{Phys. Rev. B} \textbf{\bibinfo{volume}{76}},
  \bibinfo{pages}{224431} (\bibinfo{year}{2007}),
  \urlprefix\url{https://link.aps.org/doi/10.1103/PhysRevB.76.224431}.

\bibitem[{\citenamefont{Senthil
  et~al.}(2004{\natexlab{a}})\citenamefont{Senthil, Vishwanath, Balents,
  Sachdev, and Fisher}}]{deconfine1}
\bibinfo{author}{\bibfnamefont{T.}~\bibnamefont{Senthil}},
  \bibinfo{author}{\bibfnamefont{A.}~\bibnamefont{Vishwanath}},
  \bibinfo{author}{\bibfnamefont{L.}~\bibnamefont{Balents}},
  \bibinfo{author}{\bibfnamefont{S.}~\bibnamefont{Sachdev}}, \bibnamefont{and}
  \bibinfo{author}{\bibfnamefont{M.~P.~A.} \bibnamefont{Fisher}},
  \bibinfo{journal}{Science} \textbf{\bibinfo{volume}{303}},
  \bibinfo{pages}{1490} (\bibinfo{year}{2004}{\natexlab{a}}), ISSN
  \bibinfo{issn}{0036-8075},
  \urlprefix\url{https://science.sciencemag.org/content/303/5663/1490}.

\bibitem[{\citenamefont{Senthil
  et~al.}(2004{\natexlab{b}})\citenamefont{Senthil, Balents, Sachdev,
  Vishwanath, and Fisher}}]{deconfine2}
\bibinfo{author}{\bibfnamefont{T.}~\bibnamefont{Senthil}},
  \bibinfo{author}{\bibfnamefont{L.}~\bibnamefont{Balents}},
  \bibinfo{author}{\bibfnamefont{S.}~\bibnamefont{Sachdev}},
  \bibinfo{author}{\bibfnamefont{A.}~\bibnamefont{Vishwanath}},
  \bibnamefont{and} \bibinfo{author}{\bibfnamefont{M.~P.~A.}
  \bibnamefont{Fisher}}, \bibinfo{journal}{Physical Review B}
  \textbf{\bibinfo{volume}{70}} (\bibinfo{year}{2004}{\natexlab{b}}), ISSN
  \bibinfo{issn}{1550-235X},
  \urlprefix\url{http://dx.doi.org/10.1103/PhysRevB.70.144407}.

\bibitem[{\citenamefont{Wang et~al.}(2021)\citenamefont{Wang, Cheng,
  Witczak-Krempa, and Meng}}]{chengfrac}
\bibinfo{author}{\bibfnamefont{Y.-C.} \bibnamefont{Wang}},
  \bibinfo{author}{\bibfnamefont{M.}~\bibnamefont{Cheng}},
  \bibinfo{author}{\bibfnamefont{W.}~\bibnamefont{Witczak-Krempa}},
  \bibnamefont{and} \bibinfo{author}{\bibfnamefont{Z.~Y.} \bibnamefont{Meng}},
  \bibinfo{journal}{Nature Communications} \textbf{\bibinfo{volume}{12}}
  (\bibinfo{year}{2021}), ISSN \bibinfo{issn}{2041-1723},
  \urlprefix\url{http://dx.doi.org/10.1038/s41467-021-25707-z}.

\bibitem[{\citenamefont{Kalmeyer and Laughlin}(1987)}]{csl1}
\bibinfo{author}{\bibfnamefont{V.}~\bibnamefont{Kalmeyer}} \bibnamefont{and}
  \bibinfo{author}{\bibfnamefont{R.~B.} \bibnamefont{Laughlin}},
  \bibinfo{journal}{Phys. Rev. Lett.} \textbf{\bibinfo{volume}{59}},
  \bibinfo{pages}{2095} (\bibinfo{year}{1987}),
  \urlprefix\url{https://link.aps.org/doi/10.1103/PhysRevLett.59.2095}.

\bibitem[{\citenamefont{Kalmeyer and Laughlin}(1989)}]{csl2}
\bibinfo{author}{\bibfnamefont{V.}~\bibnamefont{Kalmeyer}} \bibnamefont{and}
  \bibinfo{author}{\bibfnamefont{R.~B.} \bibnamefont{Laughlin}},
  \bibinfo{journal}{Phys. Rev. B} \textbf{\bibinfo{volume}{39}},
  \bibinfo{pages}{11879} (\bibinfo{year}{1989}),
  \urlprefix\url{https://link.aps.org/doi/10.1103/PhysRevB.39.11879}.

\bibitem[{\citenamefont{Cheng et~al.}(2016)\citenamefont{Cheng, Zaletel,
  Barkeshli, Vishwanath, and Bonderson}}]{LSMtopo}
\bibinfo{author}{\bibfnamefont{M.}~\bibnamefont{Cheng}},
  \bibinfo{author}{\bibfnamefont{M.}~\bibnamefont{Zaletel}},
  \bibinfo{author}{\bibfnamefont{M.}~\bibnamefont{Barkeshli}},
  \bibinfo{author}{\bibfnamefont{A.}~\bibnamefont{Vishwanath}},
  \bibnamefont{and}
  \bibinfo{author}{\bibfnamefont{P.}~\bibnamefont{Bonderson}},
  \bibinfo{journal}{Phys. Rev. X} \textbf{\bibinfo{volume}{6}},
  \bibinfo{pages}{041068} (\bibinfo{year}{2016}),
  \urlprefix\url{https://link.aps.org/doi/10.1103/PhysRevX.6.041068}.

\bibitem[{\citenamefont{Levin and Stern}(2009)}]{FQSH1}
\bibinfo{author}{\bibfnamefont{M.}~\bibnamefont{Levin}} \bibnamefont{and}
  \bibinfo{author}{\bibfnamefont{A.}~\bibnamefont{Stern}},
  \bibinfo{journal}{Phys. Rev. Lett.} \textbf{\bibinfo{volume}{103}},
  \bibinfo{pages}{196803} (\bibinfo{year}{2009}),
  \urlprefix\url{https://link.aps.org/doi/10.1103/PhysRevLett.103.196803}.

\bibitem[{\citenamefont{Barkeshli and McGreevy}(2014)}]{mcgreevy}
\bibinfo{author}{\bibfnamefont{M.}~\bibnamefont{Barkeshli}} \bibnamefont{and}
  \bibinfo{author}{\bibfnamefont{J.}~\bibnamefont{McGreevy}},
  \bibinfo{journal}{Phys. Rev. B} \textbf{\bibinfo{volume}{89}},
  \bibinfo{pages}{235116} (\bibinfo{year}{2014}),
  \urlprefix\url{https://link.aps.org/doi/10.1103/PhysRevB.89.235116}.

\bibitem[{\citenamefont{Isakov et~al.}(2012)\citenamefont{Isakov, Melko, and
  Hastings}}]{hastingsz2}
\bibinfo{author}{\bibfnamefont{S.~V.} \bibnamefont{Isakov}},
  \bibinfo{author}{\bibfnamefont{R.~G.} \bibnamefont{Melko}}, \bibnamefont{and}
  \bibinfo{author}{\bibfnamefont{M.~B.} \bibnamefont{Hastings}},
  \bibinfo{journal}{Science} \textbf{\bibinfo{volume}{335}},
  \bibinfo{pages}{193} (\bibinfo{year}{2012}), ISSN \bibinfo{issn}{1095-9203},
  \urlprefix\url{http://dx.doi.org/10.1126/science.1212207}.

\bibitem[{\citenamefont{Pufu and Sachdev}(2013)}]{monopole2}
\bibinfo{author}{\bibfnamefont{S.~S.} \bibnamefont{Pufu}} \bibnamefont{and}
  \bibinfo{author}{\bibfnamefont{S.}~\bibnamefont{Sachdev}},
  \bibinfo{journal}{Journal of High Energy Physics}
  \textbf{\bibinfo{volume}{2013}}, \bibinfo{pages}{127} (\bibinfo{year}{2013}),
  ISSN \bibinfo{issn}{1029-8479},
  \urlprefix\url{http://dx.doi.org/10.1007/JHEP09(2013)127}.

\bibitem[{\citenamefont{Dyer et~al.}(2015)\citenamefont{Dyer, Mezei, Pufu, and
  Sachdev}}]{monopole3}
\bibinfo{author}{\bibfnamefont{E.}~\bibnamefont{Dyer}},
  \bibinfo{author}{\bibfnamefont{M.}~\bibnamefont{Mezei}},
  \bibinfo{author}{\bibfnamefont{S.~S.} \bibnamefont{Pufu}}, \bibnamefont{and}
  \bibinfo{author}{\bibfnamefont{S.}~\bibnamefont{Sachdev}},
  \bibinfo{journal}{Journal of High Energy Physics}
  \textbf{\bibinfo{volume}{2015}}, \bibinfo{pages}{37} (\bibinfo{year}{2015}).

\bibitem[{\citenamefont{K\"{o}nig et~al.}(2007)\citenamefont{K\"{o}nig,
  Wiedmann, Br\"{u}ne, Roth, Buhmann, Molenkamp, Qi, and Zhang}}]{qshex}
\bibinfo{author}{\bibfnamefont{M.}~\bibnamefont{K\"{o}nig}},
  \bibinfo{author}{\bibfnamefont{S.}~\bibnamefont{Wiedmann}},
  \bibinfo{author}{\bibfnamefont{C.}~\bibnamefont{Br\"{u}ne}},
  \bibinfo{author}{\bibfnamefont{A.}~\bibnamefont{Roth}},
  \bibinfo{author}{\bibfnamefont{H.}~\bibnamefont{Buhmann}},
  \bibinfo{author}{\bibfnamefont{L.~W.} \bibnamefont{Molenkamp}},
  \bibinfo{author}{\bibfnamefont{X.-L.} \bibnamefont{Qi}}, \bibnamefont{and}
  \bibinfo{author}{\bibfnamefont{S.-C.} \bibnamefont{Zhang}},
  \bibinfo{journal}{Science} \textbf{\bibinfo{volume}{318}},
  \bibinfo{pages}{766} (\bibinfo{year}{2007}), ISSN \bibinfo{issn}{1095-9203},
  \urlprefix\url{http://dx.doi.org/10.1126/science.1148047}.

\bibitem[{\citenamefont{Lee and Nagaosa}(1992)}]{leenagaosa}
\bibinfo{author}{\bibfnamefont{P.~A.} \bibnamefont{Lee}} \bibnamefont{and}
  \bibinfo{author}{\bibfnamefont{N.}~\bibnamefont{Nagaosa}},
  \bibinfo{journal}{Phys. Rev. B} \textbf{\bibinfo{volume}{46}},
  \bibinfo{pages}{5621} (\bibinfo{year}{1992}),
  \urlprefix\url{https://link.aps.org/doi/10.1103/PhysRevB.46.5621}.

\bibitem[{\citenamefont{Hartnoll et~al.}(2018)\citenamefont{Hartnoll, Lucas,
  and Sachdev}}]{HoloQ}
\bibinfo{author}{\bibfnamefont{S.~A.} \bibnamefont{Hartnoll}},
  \bibinfo{author}{\bibfnamefont{A.}~\bibnamefont{Lucas}}, \bibnamefont{and}
  \bibinfo{author}{\bibfnamefont{S.}~\bibnamefont{Sachdev}},
  \emph{\bibinfo{title}{Holographic Quantum Matter}} (\bibinfo{publisher}{The
  MIT Press}, \bibinfo{year}{2018}).

\bibitem[{\citenamefont{Nave and Lee}(2007)}]{lee2007}
\bibinfo{author}{\bibfnamefont{C.~P.} \bibnamefont{Nave}} \bibnamefont{and}
  \bibinfo{author}{\bibfnamefont{P.~A.} \bibnamefont{Lee}},
  \bibinfo{journal}{Phys. Rev. B} \textbf{\bibinfo{volume}{76}},
  \bibinfo{pages}{235124} (\bibinfo{year}{2007}),
  \urlprefix\url{https://link.aps.org/doi/10.1103/PhysRevB.76.235124}.

\bibitem[{\citenamefont{Chen et~al.}(2014{\natexlab{b}})\citenamefont{Chen,
  Kee, and Kim}}]{gang2014}
\bibinfo{author}{\bibfnamefont{G.}~\bibnamefont{Chen}},
  \bibinfo{author}{\bibfnamefont{H.-Y.} \bibnamefont{Kee}}, \bibnamefont{and}
  \bibinfo{author}{\bibfnamefont{Y.~B.} \bibnamefont{Kim}},
  \bibinfo{journal}{Phys. Rev. Lett.} \textbf{\bibinfo{volume}{113}},
  \bibinfo{pages}{197202} (\bibinfo{year}{2014}{\natexlab{b}}),
  \urlprefix\url{https://link.aps.org/doi/10.1103/PhysRevLett.113.197202}.

\bibitem[{\citenamefont{{Kovtun}}(2012)}]{Kovtun2012}
\bibinfo{author}{\bibfnamefont{P.}~\bibnamefont{{Kovtun}}},
  \bibinfo{journal}{Journal of Physics A Mathematical General}
  \textbf{\bibinfo{volume}{45}}, \bibinfo{eid}{473001} (\bibinfo{year}{2012}),
  \eprint{1205.5040}.

\bibitem[{\citenamefont{Delacretaz}(2020)}]{luca}
\bibinfo{author}{\bibfnamefont{L.~V.} \bibnamefont{Delacretaz}},
  \bibinfo{journal}{SciPost Phys.} \textbf{\bibinfo{volume}{9}},
  \bibinfo{pages}{34} (\bibinfo{year}{2020}),
  \urlprefix\url{https://scipost.org/10.21468/SciPostPhys.9.3.034}.

\bibitem[{\citenamefont{Schattner et~al.}(2016)\citenamefont{Schattner,
  Lederer, Kivelson, and Berg}}]{numerical1}
\bibinfo{author}{\bibfnamefont{Y.}~\bibnamefont{Schattner}},
  \bibinfo{author}{\bibfnamefont{S.}~\bibnamefont{Lederer}},
  \bibinfo{author}{\bibfnamefont{S.~A.} \bibnamefont{Kivelson}},
  \bibnamefont{and} \bibinfo{author}{\bibfnamefont{E.}~\bibnamefont{Berg}},
  \bibinfo{journal}{Phys. Rev. X} \textbf{\bibinfo{volume}{6}},
  \bibinfo{pages}{031028} (\bibinfo{year}{2016}),
  \urlprefix\url{https://link.aps.org/doi/10.1103/PhysRevX.6.031028}.

\bibitem[{\citenamefont{Xu et~al.}(2017)\citenamefont{Xu, Sun, Schattner, Berg,
  and Meng}}]{numerical2}
\bibinfo{author}{\bibfnamefont{X.~Y.} \bibnamefont{Xu}},
  \bibinfo{author}{\bibfnamefont{K.}~\bibnamefont{Sun}},
  \bibinfo{author}{\bibfnamefont{Y.}~\bibnamefont{Schattner}},
  \bibinfo{author}{\bibfnamefont{E.}~\bibnamefont{Berg}}, \bibnamefont{and}
  \bibinfo{author}{\bibfnamefont{Z.~Y.} \bibnamefont{Meng}},
  \bibinfo{journal}{Phys. Rev. X} \textbf{\bibinfo{volume}{7}},
  \bibinfo{pages}{031058} (\bibinfo{year}{2017}),
  \urlprefix\url{https://link.aps.org/doi/10.1103/PhysRevX.7.031058}.

\bibitem[{\citenamefont{Jiang and Devereaux}(2019)}]{numerical3}
\bibinfo{author}{\bibfnamefont{H.-C.} \bibnamefont{Jiang}} \bibnamefont{and}
  \bibinfo{author}{\bibfnamefont{T.~P.} \bibnamefont{Devereaux}},
  \bibinfo{journal}{Science} \textbf{\bibinfo{volume}{365}},
  \bibinfo{pages}{1424} (\bibinfo{year}{2019}),
  \urlprefix\url{https://doi.org/10.1126%2Fscience.aal5304}.

\bibitem[{\citenamefont{Szasz et~al.}(2020{\natexlab{b}})\citenamefont{Szasz,
  Motruk, Zaletel, and Moore}}]{numerical4}
\bibinfo{author}{\bibfnamefont{A.}~\bibnamefont{Szasz}},
  \bibinfo{author}{\bibfnamefont{J.}~\bibnamefont{Motruk}},
  \bibinfo{author}{\bibfnamefont{M.~P.} \bibnamefont{Zaletel}},
  \bibnamefont{and} \bibinfo{author}{\bibfnamefont{J.~E.} \bibnamefont{Moore}},
  \bibinfo{journal}{Phys. Rev. X} \textbf{\bibinfo{volume}{10}},
  \bibinfo{pages}{021042} (\bibinfo{year}{2020}{\natexlab{b}}),
  \urlprefix\url{https://link.aps.org/doi/10.1103/PhysRevX.10.021042}.

\bibitem[{\citenamefont{Kaul and Sachdev}(2008)}]{sachdevkaul}
\bibinfo{author}{\bibfnamefont{R.~K.} \bibnamefont{Kaul}} \bibnamefont{and}
  \bibinfo{author}{\bibfnamefont{S.}~\bibnamefont{Sachdev}},
  \bibinfo{journal}{Phys. Rev. B} \textbf{\bibinfo{volume}{77}},
  \bibinfo{pages}{155105} (\bibinfo{year}{2008}),
  \urlprefix\url{https://link.aps.org/doi/10.1103/PhysRevB.77.155105}.

\bibitem[{\citenamefont{Benvenuti and Khachatryan}(2019)}]{EPQED}
\bibinfo{author}{\bibfnamefont{S.}~\bibnamefont{Benvenuti}} \bibnamefont{and}
  \bibinfo{author}{\bibfnamefont{H.}~\bibnamefont{Khachatryan}},
  \bibinfo{journal}{Journal of High Energy Physics}
  \textbf{\bibinfo{volume}{2019}} (\bibinfo{year}{2019}), ISSN
  \bibinfo{issn}{1029-8479},
  \urlprefix\url{http://dx.doi.org/10.1007/JHEP05(2019)214}.

\end{thebibliography}

\appendix

\onecolumngrid

\section{Field theories for $N = 6$ and $N = 12$ of scenario (1)}

In the next section we will derive the projective symmetry group
transformation for the low energy vortex modes of scenario (1).
For $N = 6$, with symmetries $R_{2\pi/3}$, translation, $P_x
\mathcal{T}$, and $P_y$, the PSG-invariant interactions between
the vortex fields $\psi_a$ beyond Eq.~\ref{cpn} take the following
form:
\begin{equation}
\begin{aligned}
\mathcal{L}^{(1)\prime}[\psi_a] = &  u_1\sum_{a=0}^{2}(|\psi_{2a}|^2+|\psi_{2a+1}|^2)^2 + u_2 \left(\sum_{a=0}^{5}|\psi_a|^2\right)^2 \\
&+ v_1\left(\sum_{a=0}^{5}\psi_a^2\right)\left(\sum_{a=0}^{5}(\psi_a^*)^2\right) + v_2\sum_{a=0}^{2}(\psi_{2a}^2+\psi_{2a+1}^2)((\psi_{2a}^*)^2+(\psi_{2a+1}^*)^2) \\
&+w_1\sum_{a=0}^2(|\psi_{2a}|^2-|\psi_{2a+1}|^2)(\psi_{2a+2}\psi^*_{2a+3}+\psi^*_{2a+2}\psi_{2a+3})+w_2\left\{\sum_{a=0}^2
(\psi_{2a}^2-\psi_{2a+1}^2)\psi^*_{2a+2}\psi^*_{2a+3}+c.c.\right\}+\dots
\label{int6}
\end{aligned}
\end{equation}
Here the dots stand for terms higher than the quartic order. The
parameters $\{u_1, u_2, v_1, v_2, w_1, w_2\}$ in \eqref{int6} are
all real, and the index $a$ for $\psi_a$ is regarded as cyclic
modulo $6$.

In addition to the quartic terms, the gauge invariant density wave
order parameter can couple to the Fermi surface of the fermionic
partons, and quartic terms of $\psi_a$ with singularity in the
frequency space can be generated as was pointed out by
Ref.~\onlinecite{debanjan2021}, such as $|\omega| |S_{\omega,q}|^2
$, where $S_{\omega,q}$ is a bilinear of $\psi_a$. This coupling
only arises for scenario (1). For scenario (2) discussed in the
main text, the 3D XY$^\ast$ fixed point should be stable against
symmetry allowed perturbations; the field theory Eq.~\ref{qedcs}
is also stable against coupling to the fermionic parton Fermi
surface.

Although we do not aim to give a full discussion of the fate of
the infrared limit of scenario (1), in the current work we
establish the formalism for this problem that one can use in the
future. As we explained in the previous paragraph, after
integrating out the fermion that is connected by the finite
momentum of the density wave order parameter, a term is generated
$\sim |\omega| |S_{\omega,q}|^2$, where $S = \psi^\dagger T \psi$
and $T$ is an $N \times N$ matrix. One can introduce a new field
$\Phi$ through the Hubbard-Stratonovich transformation, and
$\psi_a$ will interact with the $\Phi$ field~\cite{sachdevkaul}.
We start with the first line of Eq.~\ref{int6}. The field theory
Eq.~\ref{cpn} with $u_1$ and $u_2$ in Eq.~\ref{int6} can be
reformulated by introducing multiple Lagrange multipliers
$\lambda_i$: \beqn \mathcal{L}^{(1)} &=& \sum_{a = 0}^{N-1}
|(\partial - \ii A) \psi_a|^2 + \ii \sum_{i = 1}^{N_1} \lambda_i
\left( \sum_{\tau = 1}^{N_2} |\psi_{\tau, i}|^2 \right) + \ii \Phi
\psi^\dagger T \psi; \cr\cr && \langle \lambda_i(\vec{q})
\lambda_{i'}(- \vec{q}) \rangle = \frac{8}{N_2}|q|\delta_{i,i'},
\cr\cr && \langle A_{\mu}(\vec{q}) A_{\nu}(-\vec{q}) \rangle =
\frac{16}{N}\left( \frac{\delta_{\mu\nu} - q_\mu
q_\nu/q^2}{|q|}\right), \cr\cr && \langle \Phi(\vec{q}) \Phi(-
\vec{q}) \rangle = g|\omega|. \label{n6HS} \eeqn Here $N = N_1
N_2$, and for the real system with $N= 6$, $N_1 = 3$ and $N_2 =
2$. Introducing $\lambda_i$ for each index $i$ physically means
that we are investigating the theory near the point with a
SU$(N_2)$ symmetry for each index $i$, rather than the original
CP$^{N-1}$ theory with a large SU$(N)$ flavor symmetry. This is
analogous to the ``easy-plane bosonic QED$_3$" considered in
Ref.~\onlinecite{EPQED}. The actions of $\lambda_i$ and the
transverse component of gauge field $A$ are generated by
integrating out the fields $\psi_a$. One possible way to proceed
with the calculation is that, we can fix $N_1$, and take $1/N_2$
as a small parameter. When $g$ is the same order of $1/N_2$, the
interaction between $\psi_a$ and the $\Phi$ field will lead to the
contribution comparable with that arising from coupling to
$\lambda_i$ and $A$. The calculation would be analogous to the one
formulated in Ref.~\onlinecite{edgexu2}, where the nonlocal
interaction on top of a bosonic QED flows to a new fixed point.
One can evaluate the scaling behaviors (such as
relevance/irrelevance in the IR) of the $v$ and $w$ terms in the
second and third lines in Eq.~\ref{n6HS} at this new fixed point.
By exploring the parameter space of $g$, $1/N_2$, and different
choice of matrix $T$, it is possible to identify a finite region
where Eq.~\ref{n6HS} corresponds to a stable fixed point where the
$v$ and $w$ terms in Eq.~\ref{int6} are irrelevant.

The same strategy can be applied to the situation with $N = 12$.
With long moir\'{e} lattice constants, the 6-fold rotation
$R_{\pi/3}$ also becomes a good approximate symmetry. Together
with $R_{\pi/3}$, the quartic terms in the field theory for $N =
12$ (please refer to the phase diagram in Fig.~\ref{BZ}) are:
\beqn \mathcal{L}^{(1)\prime}[\psi_{\sigma, \tau, i}] &=&
u_1\sum_{\sigma, i}\left(\sum_\tau|\psi_{\sigma, \tau,
i}|^2\right)^2 + u_2 \left(\sum_{\sigma\tau i}|\psi_{\sigma\tau
i}|^2\right)^2 \cr\cr &+& v_1\sum_{\sigma, i \neq i'} \left(
\sum_{\tau}|\psi_{\sigma, \tau, i}|^2 \right) \left(
\sum_{\tau'}|\psi_{\sigma, \tau', i'}|^2 \right) + v_2 \sum_{i}
\left( \sum_{\tau}|\psi_{+,\tau, i}|^2 \right) \left(\sum_{\tau'}
|\psi_{-, \tau', i}|^2 \right) \cr \cr &+& w_1\left\vert\sum_{i,
\tau}\psi_{+, \tau, i}\psi_{-, \tau,i}\right\vert^2 + \ii
w_2\left(\sum_{i, \tau, \tau'}\psi^*_{+, \tau ,i+1}\psi^*_{-,
\tau, i+1}\psi_{+, \tau', i}\psi_{-, \tau', i}-h.c.\right)
\label{n12} \eeqn Here the 12 modes are labelled by $\psi_{\sigma,
\tau, i}$ in which $\tau=\pm$ labels two degenerate modes at the
same momentum, $\sigma=\pm$ labels two sets of momenta that are
each connected by $R_{2\pi/3}$, and $i=0,1,2 \mod 3$ labels these
three momenta within each set.

We can again start with the first line of Eq.~\ref{n12}, and
introduce Lagrange multiplier $\lambda_{\sigma,i}$ which couples
to the $\psi_a$ fields as $\sum_{\tau = 1}^{N_2}
\lambda_{\sigma,i}|\psi_{\sigma, \tau, i}|^2$. Notice that we have
generalized $\tau$ to $1 \cdots N_2$. Then the
Hubbard-Stratonovich transformation can introduce new fields that
couple to $\psi_a$ to account for the singular terms generated
through interacting with the Fermi surface. A combined
perturbation theory of $1/N_2$ and $g$ can again determine the
relevance/irrelevance of the second and third lines of
Eq.~\ref{n12}. In particular, the two terms in the second line of
Eq.~\ref{n12} are indeed irrelevant with large-$N_2$, as the
scaling dimension of $\sum_\tau |\psi_{\sigma, \tau, i}|^2$ is $2$
with large-$N_2$.

\section{The PSG transformation for $N = 6$ in scenario (1)}

Under the boson-vortex duality, the dual vortex theory on the
hexagonal lattice takes the form
\begin{equation}
H = \sum_{\langle ij \rangle} - t_{ij} \phi^*_i
\phi_{j}+H'_\phi+V_\phi+\dots, \ \ \ t_{ij} = t e^{-\ii A_{ij}}
\end{equation}
Here $H'_\phi$ describes hopping terms between further neighbors.
The potential $V_\phi$ includes a quadratic term $\sum_i
r|\phi_i|^2$ which tunes through the phase transition.

\begin{figure}
\includegraphics[width=0.45\textwidth]{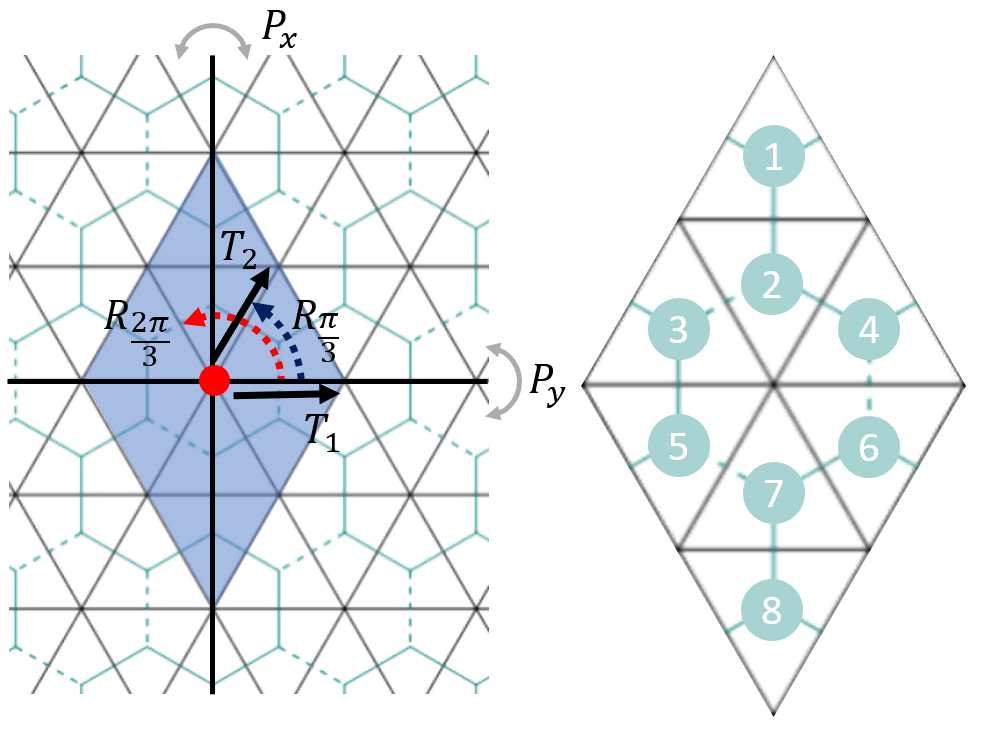}
\caption{Crystal symmetry of the triangular lattice, the nearest
neighbor hopping amplitudes of the vortices, and the unit cell
after taking into account of the sign of $t_{ij}$.}
\label{vbs_lattice}
\end{figure}

When $t_{ij}$ is nonzero only for nearest neighbor links on the
dual honeycomb lattice, and it takes positive sign on the solid
links and negative sign on the dashed links in Fig.~\ref{lattice}
due to the $\pi$ flux of $A_\mu$ through each hexagon, there are
four minima of the vortex band structure in the Brillouin zone
(Fig.~\ref{BZ}). We label the four minimum modes from 0 to 3, each
have momentum $(k_x,k_y)$
\begin{equation}
\mathbf{Q}_{0,1}=\mathbf{K}=\left(\frac{2\pi}{3\sqrt{3}},0\right),
\ \ \
\mathbf{Q}_{2,3}=\mathbf{K}'=\left(\frac{\pi}{3\sqrt{3}},\frac{\pi}{3}\right).
\end{equation}
With further neighbor vortex hopping (please refer to the phase
diagram in Fig.~\ref{BZ}), the minima of the vortex band structure
can shift to the $M$ points, similar to
Ref.~\onlinecite{kevintriangle}. When the degenerate minima are
shifted to the $M$ points (Fig.~\ref{BZ}), the six corresponding
momenta are
\begin{equation}
\mathbf{Q}_{0,1}=\left(\frac{\pi}{2\sqrt{3}},-\frac{\pi}{6}\right),\
\mathbf{Q}_{2,3}=\left(\frac{\pi}{2\sqrt{3}},\frac{\pi}{6}\right),\
\mathbf{Q}_{4,5}=\left(0,\frac{\pi}{3}\right).
\end{equation}
Similar to the four minima case, the vortex field can be expanded
using these six modes as
\begin{equation}
\phi_{n,\mathbf{r}} \sim \sum_{a=0}^5 \psi_a v_{a,n}e^{\ii
\mathbf{Q}_a\cdot\mathbf{r}}. \label{eq:expand}
\end{equation}
The coefficients $v_{a,n}$ are solved from the band structure.

The symmetries of the theory for one single valley must include
translation $T_1, T_2$, three-fold rotation $R_{2\pi/3}$, $P_x
\mathcal{T}$. These transformations do not mix the two valleys. In
the following we derive the PSG matrices of these symmetries. We
first need the form of the transformations when acting on the 8
sites in each unit cell:
\begin{align}
            & T_{1,2}(\phi_{n,\mathbf{k}})=\sum_m(t_{1,2})_{nm}\phi_{m,\mathbf{k}},\  t_1=\begin{pmatrix}
            0&0&0&0&0&0&1&0\\
            0&0&0&0&0&0&0&1\\
            0&0&0&-1&0&0&0&0\\
            0&0&1&0&0&0&0&0\\
            0&0&0&0&0&1&0&0\\
            0&0&0&0&-1&0&0&0\\
            -1&0&0&0&0&0&0&0\\
            0&-1&0&0&0&0&0&0\\
        \end{pmatrix},\ t_2=\begin{pmatrix}
        0&0&1&0&0&0&0&0\\
        0&0&0&0&1&0&0&0\\
        -1&0&0&0&0&0&0&0\\
        0&0&0&0&0&0&-1&0\\
        0&-1&0&0&0&0&0&0\\
        0&0&0&0&0&0&0&1\\
        0&0&0&1&0&0&0&0\\
        0&0&0&0&0&-1&0&0\\
    \end{pmatrix}
\end{align}
\begin{align}
    &R_{2\pi/3}(\phi_{n,\mathbf{k}})=(r_{\pi/3})_{nm}\phi_{m,R_{2\pi/3}\mathbf{k}},\
r_{2\pi/3}=\begin{pmatrix}
        1&0&0&0&0&0&0&0\\
        0&0&0&0&1&0&0&0\\
        0&0&0&0&0&0&1&0\\
        0&0&1&0&0&0&0&0\\
        0&0&0&0&0&1&0&0\\
        0&1&0&0&0&0&0&0\\
        0&0&0&1&0&0&0&0\\
        0&0&0&0&0&0&0&1\\
    \end{pmatrix},
\end{align}
\begin{align}
   &P_x\mathcal{T}(\phi_{n,\mathbf{k}})=(p_xt)_{nm}\phi_{m,-P_x\mathbf{k}},\
(p_xt)_{ab}=\begin{pmatrix}
    1&0&0&0&0&0&0&0\\
    0&1&0&0&0&0&0&0\\
    0&0&0&-1&0&0&0&0\\
    0&0&-1&0&0&0&0&0\\
    0&0&0&0&0&1&0&0\\
    0&0&0&0&1&0&0&0\\
    0&0&0&0&0&0&-1&0\\
    0&0&0&0&0&0&0&-1
\end{pmatrix},
\end{align}

Besides these symmetries, here we argue that, if the system does
have an effective Hubbard model description with two local Wannier
orbitals per unit cell (one for each valley), $P_y$ is also a good
symmetry of the Hubbard model, as long as the valley mixing is
negligible, which is a justified assumption with long wavelength
moir\'{e} potential modulation. Let us first assume there is no
valley mixing, then for each valley the band structure of the
moir\'{e} mini band is described by a tight binding model with one
orbital per site on the moir\'{e} triangular lattice. The hopping
amplitude $t(\theta)$ along angle $\theta$ must satisfy the
following relations based on the explicit $P_x \mathcal{T}$ and
translation symmetry: \beqn t(\theta) = t^\ast(\pi - \theta), \ \
\ t^\ast(\theta) = t(\pi + \theta), \eeqn we can easily show that
$t(\theta) = t(-\theta)$, namely the system should have a $P_y$
symmetry.

However, when there is valley mixing, $t$ becomes a $2 \times 2$
matrix with off-diagonal terms that mix two valleys. A $2\times 2$
hopping matrix $t$ should satisfy four symmetries, $P_x$,
$\mathcal{T}$, translation, and $R_{2\pi/3}$ rotation. A natural
choice of $P_x$ and $\mathcal{T}$ on $t$ is \beqn P_x: t(\theta)
\rightarrow \sigma^x t(\pi - \theta) \sigma^x; \ \ \ \mathcal{T}:
t(\theta) \rightarrow (\ii \sigma^y) t^\ast (-\ii\sigma^y); \eeqn
and the translation symmetry plus hermicity demands
$t^\dagger(\theta) = t(\pi + \theta)$. $P_y$ does not change the
valley indices; if $P_y$ takes $t(\theta)$ to $t(-\theta)$, there
exists a valley mixing term $t(\theta) \sim \ii \sigma^x
\sin(3\theta)$ that preserves all the symmetries mentioned above,
but breaks $P_y$; while if $P_y$ takes $t(\theta)$ to $\sigma^z
t(-\theta) \sigma^z$ this term becomes $t(\theta) \sim \ii
\sigma^y \cos(3\theta)$.

$P_y$ acts on the $\phi$ bosons as
\begin{align}
&P_y(\phi_{n,\mathbf{k}})=(p_y)_{nm}\phi^*_{m,-P_y\mathbf{k}},\
(p_y)_{ab}=\begin{pmatrix}
0&0&0&0&0&0&0&1\\
0&0&0&0&0&0&1&0\\
0&0&0&0&1&0&0&0\\
0&0&0&0&0&1&0&0\\
0&0&1&0&0&0&0&0\\
0&0&0&1&0&0&0&0\\
0&1&0&0&0&0&0&0\\
1&0&0&0&0&0&0&0
\end{pmatrix}.
\end{align}

Furthermore, in the case with long moir\'{e} lattice constant, we
additionally have the six-fold rotation $R_{\pi/3}$
\begin{align}
        R_{\pi/3}(\phi_{n,\mathbf{k}})=(r_{\pi/3})_{nm}\phi_{m,R_{\pi/3}\mathbf{k}},\
        (r_{\pi/3})_{ab}=\begin{pmatrix}
        0&0&0&0&0&0&0&-1\\
        0&0&-1&0&0&0&0&0\\
        0&0&0&0&1&0&0&0\\
        0&1&0&0&0&0&0&0\\
        0&0&0&0&0&0&-1&0\\
        0&0&0&-1&0&0&0&0\\
        0&0&0&0&0&1&0&0\\
        1&0&0&0&0&0&0&0\\
    \end{pmatrix}.
    \end{align}
In the position space, the transformation rules can be summarized
as \begin{equation} G(\phi_{n,\mathbf{r}})=\sum_{m=1}^8
g_{n,m}\phi_{m,\mathbf{r}'_{m}}
\end{equation}
in which $\mathbf{r}'_{m}$ is the center of the unit cell of field
$\phi_m$ which is obtained by certain site in the original unit
cell (centered at $\mathbf{r}$) after transformation under
symmetry operation $G$. For example, under $T_1$, $\mathbf{r}'_{7}
= \mathbf{r}'_{8} =\mathbf{r}+2\mathbf{a}_2$, because sites 1 and
2 at unit cell $\mathbf{r}$ are transformed into sites 7 and 8 in
the nearby enlarged unit cell which is centered at
$\mathbf{r}+2\mathbf{a}_2$. In general, we can write the
transformation as $\mathbf{r}_m'=G\mathbf{r}+\vec{\Delta}_{G,m}$,
in which $\vec{\Delta}_{G,m}$ is a constant that does not depend
on $\mathbf{r}$, and $G\mathbf{r}$ is the coordinate of the center
of the unit cell after spacial symmetry $G$.

Now we plug in the low energy expansions of $\phi_{n_\mathbf{k}}$
around the minima into the equation, which yields
\begin{equation}
        \sum_{a=0}^{N-1} G(\psi_a)v_{a,n}e^{\ii\mathbf{Q}_a\cdot\mathbf{r}} = \sum_{a=0}^{N-1}\sum_
        {m=1}^8\psi_a g_{nm}v_{m,a}e^{\ii\mathbf{Q}_a\cdot \mathbf{r}_m'}.
\end{equation}
The relation can be viewed as a vector identity with $n$ being the
vector index on both sides. Because all the vectors $v_{a,n}
(a=0,\dots,N-1)$ are orthogonal to each other, we can multiply the
conjugated vector $v^*_{b,n}$ on both sides and sum over $n$:
\begin{equation}
\label{psgeqn} G(\psi_b)e^{\ii\mathbf{Q}_b\cdot\mathbf{r}} =
\sum_{a=0}^{N-1} \sum_ {m,n=1}^8\psi_a
v^*_{b,n}g_{n,m}v_{a,m}e^{\ii\mathbf{Q}_a\cdot \mathbf{r}_m'}.
\end{equation}
For this equation to hold for all $\mathbf{r}$, the RHS needs to
have the same momentum. This requires
$\mathbf{Q}_b=G^{-1}\mathbf{Q}_a$, which can only be satisfied by
two possible choices of $a$ (recall that in the convention of
eight-site unit cell, each momentum $\mathbf{Q}_a$ always has two
fold degeneracy for all $N$), denoted by $a_1$ and $a_2$. Thus we
eventually have
\begin{equation}
G(\psi_b)=\sum_{m,n=1}^{8}v^\dagger_{b,n} g_{nm}
v_{a_1,m}e^{\ii\mathbf{Q}_{a_1}\cdot\vec{\Delta}_{G,m}}\times\psi_{a_1}+\sum_{m,n=1}^{8}v^\dagger_{b,n}
g_{nm}
v_{a_2,m}e^{\ii\mathbf{Q}_{a_2}\cdot\vec{\Delta}_{G,m}}\times\psi_{a_2}
\end{equation}
The final results can be organized into $N\times N$ matrices. For
$N=6$, the transformations read
\begin{align}
         &T_{1,2}(\psi_a)=(\mathfrak{t}_{1,2})_{ab}\psi_b,\ (\mathfrak{t}_1)_{ab}=\begin{pmatrix}
            -1&0&0&0&0&0\\
            0&1&0&0&0&0\\
            0&0&0&1&0&0\\
            0&0&1&0&0&0\\
            0&0&0&0&0&1\\
            0&0&0&0&-1&0\\
         \end{pmatrix},\ (\mathfrak{t}_2)_{ab}=\begin{pmatrix}
            0&1&0&0&0&0\\
            -1&0&0&0&0&0\\
            0&0&1&0&0&0\\
            0&0&0&-1&0&0\\
            0&0&0&0&0&-1\\
            0&0&0&0&-1&0\\
         \end{pmatrix},
\end{align}
\begin{align}
    &R_{2\pi/3}(\psi_a)=(\mathfrak{R}_{2\pi/3})_{ab}\psi_b,\ (\mathfrak{R}_{2\pi/3})_{ab}=\begin{pmatrix}
        0&0&0&0&1&0\\
        0&0&0&0&0&1\\
        -1&0&0&0&0&0\\
        0&-1&0&0&0&0\\
        0&0&1&0&0&0\\
        0&0&0&1&0&0\\
    \end{pmatrix},
\end{align}
\begin{align}
    &P_x\mathcal{T}(\psi_a)=(\mathfrak{P}_{x}\mathfrak{T})_{ab}\psi_b,\ (\mathfrak{P}_{x}\mathfrak{T})_{ab}=\frac{1}{\sqrt{2}}\begin{pmatrix}
    0&0&1&-1&0&0\\
    0&0&-1&-1&0&0\\
    1&-1&0&0&0&0\\
    -1&-1&0&0&0&0\\
    0&0&0&0&1&-1\\
    0&0&0&0&-1&-1\\
    \end{pmatrix}.
\end{align}
\begin{align}
     &P_y(\psi_a)=(\mathfrak{P}_y)_{ab}\psi^*_b,\ (\mathfrak{P}_y)_{ab}=\frac{1}{\sqrt{2}}\begin{pmatrix}
    0&0&1&1&0&0\\
    0&0&1&-1&0&0\\
    1&1&0&0&0&0\\
    1&-1&0&0&0&0\\
    0&0&0&0&1&1\\
    0&0&0&0&1&-1\\
    \end{pmatrix}.
    \end{align}
\begin{align}
    R_{\pi/3}(\psi_a)=(\mathfrak{R}_{\pi/3})_{ab}\psi_b,\ (\mathfrak{R}_{\pi/3})_{ab}=
    \begin{pmatrix}
        0&0&0&-1&0&0\\
        0&0&1&0&0&0\\
        0&0&0&0&0&1\\
        0&0&0&0&-1&0\\
        0&1&0&0&0&0\\
        -1&0&0&0&0&0\\
    \end{pmatrix}.
\end{align}

\begin{figure}
\includegraphics[width=4.6cm]{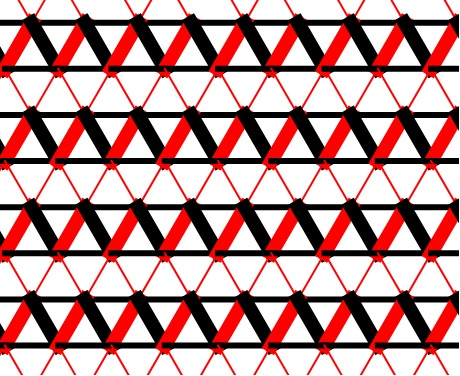}
\includegraphics[width=4.75cm]{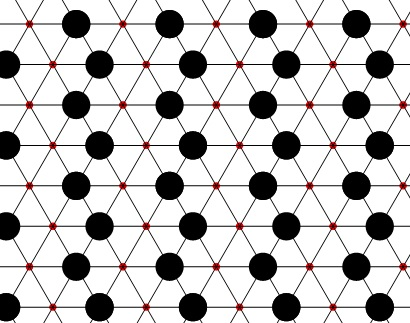}
\caption{Some possible density wave patterns of the original boson
that correspond to different condensate of $\psi_a $ with $a = 0,
\cdots 5$. The left and right patterns correspond to $\vec{\Psi}
\sim (1, 0, 0, 0, 0, 0)$ and $\vec{\Psi} \sim (0, 1/\sqrt{2}, 1/2,
-1/2, 0, 0)$ respectively. } \label{pattern}
\end{figure}

Deep inside the vortex condensate phase with $r\ll 0$ in equation
Eq.~\ref{cpn}, the vector
$\vec{\Psi}=(\psi_0,\psi_1,\psi_2,\psi_3,\psi_4,\psi_5)$ can have
different condensates depending on the parameters in
Eq.~\ref{int6}.
%We assume $u>0$ such that the sum $\sum_{a=0}^5
%|\psi_a|^2$ can be viewed as a constraint and all the other
%quartic terms as perturbations.
Without loss of generality we set
$\sum_{a=0}^5 |\psi_a|^2=1$. The two figures in Eq.~\ref{pattern}
illustrate the density waves of the bosonic parton centered at the
bonds and the sites on the moir\'e triangular lattice that
correspond to two different possible condensates of $\vec{\Psi}$.
The density on the bond $l$ is inferred from $t_{ij}\langle
\phi_i^*\phi_j\rangle$, with $ij$ being the link on the dual
honeycomb lattice that is dual to $l$, and $t_{ij}$ takes the sign
according to the gauge convention of Fig.~\ref{lattice}. The
operator $t_{ij}\langle \phi_i^*\phi_j\rangle$ is the energy
density in terms of vortex fields, and the modulation of this
operator should correspond to the valence bond solid of the
original bosonic parton. We also consider an operator centered on
site $p$ of the original lattice (plaquette of the dual lattice):
$\sum_{\langle ij\rangle\in p} t_{ij} \langle \phi_i^*\phi_j
\rangle,$ with the summation over the links that surround the
plaquette $p$ on the dual honeycomb lattice, whose center hosts
the site $p$ of the original moir\'e triangular lattice. In both
cases, $\langle \phi_i^*\phi_j\rangle$ is evaluated using
Eq.~\ref{eq:expand} and the value of $\vec{\Psi}$ which minimizes
the quartic energy. The left pattern in Eq.~\ref{pattern} is a
rather common valence bond solid configuration for either spin-1/2
system or hard core boson on the triangular lattice. If one
started with the construction-I of the parton construction, the
discussion in this section corresponds to the original electron
system with an average 1/2 electron per unit cell (the filling
considered in Ref.~\onlinecite{tmdinsulator2}); while for
construction-II, the discussion here applies to one electron per
unit cell, and the analysis in this section corresponds to one of
the two spin/valley flavors of the system.

\section{Dual of the vortex theory}

Here we derive the Lagrangian written in terms of the fractionally
charged bosonic partons for scenario (1). We start with
Eq.~\ref{cpn} in our paper: \beqn \mathcal{L}^{(1)} = \sum_{j =
0}^{N-1} |(\partial_\mu - \ii A_\mu)\psi_j|^2 + r |\psi_j|^2 +
\frac{\ii}{2\pi} A \wedge d (a + e A_{\mathrm{ext}}) + \cdots
\label{cpn2} \eeqn To facilitate the calculation of the DC
resistivity which will be discussed in the next subsection, we
need to ``dual back" to the charge-carriers, which requires
deforming Eq.~\ref{cpn2} with an easy-plane anisotropy $\sum_j
|\psi_j|^4$. The bosonic fractional charge carriers $\varphi_j$
are the vortices of the vortex fields $\psi_j$. We first take the
standard duality for $\psi_j$, and Eq.~\ref{cpn2} becomes: \beqn
\mathcal{L}^{(1)} = \sum_{j = 0}^{N-1} |(\partial - \ii
\tilde{A}_j)\varphi_j|^2 + \tilde{r} |\varphi_j|^2 +
\frac{\ii}{2\pi} \tilde{A}_j \wedge d A + \frac{\ii}{2\pi} A
\wedge d (a + e A_{\mathrm{ext}}) + \cdots \eeqn The basic duality
relation is that the current of $\psi_j$, $i.e.$ $J_{\psi_j} \sim
d \tilde{A}_j$. Now integrating out $A$ would lead to the
following constraint for the rest of the gauge fields: \beqn
\sum_j \tilde{A}_j - a - e A_{\mathrm{ext}} = 0. \label{eq:tildeA}
\eeqn From this constraint we can take $\tilde{A}_j$ as \beqn
\tilde{A}_j = \tilde{a}_j + \frac{1}{N} a + \frac{e}{N}
A_{\mathrm{ext}}, \ \ \ \sum_j \tilde{a}_j = 0. \eeqn Hence the
dual of the dual theory becomes \beqn \mathcal{L}^{(1)} = \sum_{j
= 0}^{N-1} |(\partial - \ii \tilde{a}_j - \ii\frac{1}{N} a - \ii
\frac{e}{N} A_{\mathrm{ext}})\varphi_j|^2 + \tilde{r}
|\varphi_j|^2 + \cdots . \label{cpndual} \eeqn The gauge fields
$\tilde{a}_j$ are still subject to the constraint $\sum_j
\tilde{a}_j = 0$. $\varphi_j$ carries $e/N$ charge of external EM
gauge field; it also carries charge $1/N$ of gauge field $a$ which
is shared with the fermionic parton $f_\alpha$.

For scenario (2) the theory in terms of fractional parton
$\varphi$ is much simpler: there is only one flavor of $\varphi$
for each valley, and there is no extra continuous gauge fields
$\tilde{a}$ besides gauge field $a$: Following the calculation in
Ref.~\onlinecite{resistivity2}, one can generalize this one flavor
of $\varphi$ in each valley to an $\bf{N}$ component of bosons:
\beqn \mathcal{L}^{(2)} = \sum_{l = 1}^{\bf{N}} |(\partial -
\ii\frac{1}{N} a - \ii \frac{e}{N} A_{\mathrm{ext}})\varphi^l|^2 +
\ii \lambda |\varphi^l|^2 + \cdots \label{cpndual2}  \eeqn and the
bosons will scatter with both gauge field $a$ and field $\lambda$
which is introduced as a Lagrange multiplier. The fact that
$\varphi^l$ carries charge $1/N$ of gauge field $a$ does not
change the scattering rate through the large-$\bf{N}$ calculation,
as the gauge charge cancels out in the calculation of scattering
rate through the large-$\bf{N}$ approach. Compared with scenario
(2), in scenario (1) the parton $\varphi_j$ is also coupled to
extra gauge fields $\tilde{a}_j$, which will lead to extra
scattering to the charge carriers.

When computing the resistivity, especially the DC resistivity of
scenario (1), we also rely on a large$-{\bf N}$ generalization,
namely we need to introduce an extra $l = 1 \cdots {\bf N}$ index
for each component of fractional charge field: $\varphi_j^l$.

\section{DC resistivity jump in scenario (1)}

In this section we present a detailed computation of the DC
resistivity jump in the scenario (1) of MIT, i.e. the scenario
when the insulator has a density wave. We start with
Eq.~\ref{cpndual}. The resistivity jump at the MIT is given by the
universal resistivity of the bosonic sector of the system $\rho_b$
at the MIT. First of all, one can prove a generalized Ioffe-Larkin
rule, which combines the resistivity of each parton $\varphi_j$
into $\rho_b$: \beqn \rho_b = \frac{\hbar}{e^2} \left( \sum_{j =
0}^{N-1} \rho_{b,j} \right), \eeqn where $\rho_{b,j}$ is the
resistivity of each parton $\varphi_j$, when the charge of
$\varphi_j$ is taken to be 1. This generalized Ioffe-Larkin rule
can be proven by formally integrating out $\varphi_j$, gauge
fields $\tilde{a}_j$ and $a$ from Eq.~\ref{cpndual}, and
eventually arriving at a response function of $A_{\mathrm{ext}}$.
At each level of the path integral, we keep a quadratic form of
the action, i.e. the random phase approximation. This Ioffe-Larkin
rule is independent of the assignment of electric charges on each
parton.

To compute $\rho_b$, we formulate the quantum Boltzmann equation
(QBE) for the $\varphi_j$ fields of a given valley. The
computation follows that for $\rho_b$ at the MIT without charge
fractionalization~\cite{resistivity2}, where the gauge field
dynamics needs to be modified due to the charge fractionalization,
which we explain in detail below for comparison. Note that
$\rho_b$ can be finite without momentum relaxation due to the
emergent particle-hole symmetry. Furthermore, the two-in two-out
scatterings among the $\varphi_j$ fields are enough to relax the
current and generate finite DC resistivity. For simplicity, we
consider the scattering between the $\varphi_j$ and emergent gauge
fields in Eq.~\eqref{cpndual}, where the gauge fields are in
thermal equilibrium and their dynamics is acquired due to the
coupling with the matter fields $\varphi_j$ and $f$. Here, we
argue that treating the gauge fields as in thermal equilibrium is
a legitimate approximation. First, the gauge field $a$ couples to
the spinon field $f$, which is sensitive to impurities and relaxes
momentum fast. Second, diagrammatically, the two-in two-out
scatterings between the $\varphi_j$ fields that give finite DC
resistivity can be captured by the $\varphi_j$ scattering with the
emergent gauge fields.

To simplify the computation of the gauge field dynamics, it is
convenient to express Eq.~\ref{cpndual} in terms of the gauge
field $\tilde{A}_j$ (Eq.~\ref{eq:tildeA}), together with the
effective action for the spinon field, the dual theory reads
\begin{align}
\mathcal{L}^{(1)}= \sum_{j=0}^{N-1} |(\partial - \ii \tilde{A}_j
)\varphi_j |^2 + \tilde{r} |\varphi_j|^2 +
\bar{f}\left(\partial_\tau - \mu  - \ii \sum_{j=0}^{N-1}
\tilde{A}_{j,0} +  \ii e A_{{\rm ext},0}  +\frac{1}{2m} (\nabla -
\ii \sum_{j=0}^{N-1} \tilde{\bm{A}}_{j} + \ii e \bm{A}_{\rm
ext})^2 \right) f + \cdots. \label{eq:ddaction}
\end{align}
Integrating out $\varphi_j$ and $f$ fields, the gauge field
propagators read
\begin{align}
D^{(\tilde{A})}_{ij} = - \ii \langle {\rm{T}}_t \tilde{A}_i
\tilde{A}_j \rangle =
\begin{cases}
\frac{\Pi^{J}_b + (N-1) \Pi^{J}_f}{ (\Pi_b^{J})^2 + N \Pi^{J}_b \Pi^{J}_f} & \text{ if } i=j \\
\frac{-\Pi^{J}_f}{ (\Pi_b^{J})^2 + N \Pi^{J}_b \Pi_f} & \text{ if } i \neq j \\
\end{cases},
\end{align}
where $\Pi_{b}^J, \Pi_{f}^J$ is the current-current correlation
function for $\varphi_j$ and $f$ fields, respectively.

For a controlled systematic calculation of transport, we introduce
a large number of (complex) rotor and spinon flavors ${\bf N}$
with the constraint $\sum_{l=1}^{\bf N} |\varphi_j^{l}|^2 =1$ for
all $j=0,1, ..., N-1$, and only the $l=1$ component couples to
$A_{\rm ext}$. The ${\bf N}=1$ limit will be taken at the end. The
effective action for the extended model becomes
\begin{align}
\mathcal{L} =  &\sum_{j=0}^{N-1} \left(\sum_{l=1}^{{\bf N}}|(\partial - \ii \tilde{A}_{j})
\varphi_j^l |^2 + \ii \lambda_j (\sum_{l=1}^{{\bf N}} |\varphi_j^l|^2 -1) +
\frac{1}{2g^2} (\epsilon_{\mu \nu \lambda} \partial_\nu \tilde{A}_{j,\lambda})^2 \right) \nonumber\\
& + \sum_{l=1}^{\bf N} \bar{f}_l \left(\partial_\tau - \mu  - \ii
\sum_{j=0}^{N-1} \tilde{A}_{j,0} +  \ii e A_{{\rm ext},0}
\delta_{l,1}  +\frac{1}{2m} (\nabla -  \ii \sum_{j=0}^{N-1}
\tilde{\bm{A}}_{j} + \ii e \bm{A}_{\rm ext} \delta_{l,1})^2
\right) f_l + \cdots .
\end{align}
Using the Fourier expansion for the electrically charged rotor
$\varphi_j^{l=1}$ in terms of the holons (+) and doublons (-),
\begin{equation}
\varphi_j^{l=1} = \int_{\bm k} \alpha_{+,j} (t, \bm k) e^{\ii \bm
k \cdot \bm x} + \alpha_{-,j} (t, \bm k) e^{-\ii \bm k \cdot \bm
x},
\end{equation}
the conductivity $\sigma_{b,j}=\rho_{b,j}^{-1}$ can be obtained as
\begin{align}
\sigma_{b,j} = \langle J_{x,j} \rangle / E_x, \quad  \langle
J_{x,j} \rangle =\int_{\bm k} \sum_{s=\pm} s \frac{\bm
k}{\epsilon_{\bm k}} f_{s,j} (t, \bm k),
\end{align}
where we define the distribution for holon ($s=+$) and doublon
($s=-$) as $f_{s,j}=\langle \alpha_{s,j}^\dagger (t, \bm k)
\alpha_{s,j} (t, \bm k) \rangle $, and they satisfy the QBE as
\begin{align}
(\partial_t + s \bm E \cdot \partial_{\bm k} ) f_{s,j}(t, \bm k) =
\frac{1}{ 2 {\bf N}} (I_{\lambda_j}[f_{\pm,j}] +
I_{\tilde{A}_j}[f_{\pm,j}]). \label{eq:QBE}
\end{align}
Note that the gauge choice in Eq.~\ref{eq:ddaction} ensures that
$f_{s,j}$ are decoupled and equal for different $j$ within the
approximation that $\tilde{A}_j$ is in thermal equilibrium, so the
subindex $j$ will be dropped unless there is ambiguity. The RHS of
Eq.~\ref{eq:QBE} reads \begin{align} \mathrm{RHS} = & \frac{1}{2
{\bf N}} \int_0^{\infty} \frac{\dd \Omega }{\pi} \int \frac{\dd^2
\bm q}{(2\pi)^2}
\{ \tau_\lambda \Im D^{(\lambda)} (\Omega, \bm q) + \tau_{\tilde{A}}\Im D^{(\tilde{A})}_{ii}(\Omega, \bm q) \} \nonumber \\
& \times \{ \frac{2\pi \delta(\epsilon_{\bm k} -\epsilon_{\bm k +
\bm q}+\Omega)}{4 \epsilon_{\bm k} \epsilon_{\bm k + \bm q}}
[ f_s(t, \bm k)(1+f_s (t, \bm k + \bm q) )n_{\bm q} (\Omega) - (1 + f_s(t, \bm k) ) f_s (t, \bm k + \bm q) ( 1 + n_{\bm q} (\Omega) ) ]  \nonumber \\
&+ \qquad \frac{2\pi \delta(\epsilon_{\bm k} -\epsilon_{\bm k +
\bm q}-\Omega)}{4 \epsilon_{\bm k} \epsilon_{\bm k + \bm q}}
[ f_s(t, \bm k)(1+f_s (t, \bm k + \bm q) ) ( 1 + n_{\bm q} (\Omega) ) - (1 + f_s(t, \bm k) ) f_s (t, \bm k + \bm q)  n_{\bm q} (\Omega)]  \nonumber \\
&+ \qquad \frac{2\pi \delta(-\epsilon_{\bm k} -\epsilon_{\bm k +
\bm q}+\Omega)}{4 \epsilon_{\bm k} \epsilon_{\bm k + \bm q}} [
f_s(t, \bm k) f_s (t, \bm k + \bm q) ( 1 + n_{\bm q} (\Omega)) -
(1 + f_s(t, \bm k) ) ( 1 + f_s (t, \bm k + \bm q) ) n_{\bm q}
(\Omega) ] \}, \end{align} where $\tau_\lambda = -1$ and
$\tau_{\tilde{A}} = (2\bm k \times \hat{\bm q})^2$ come from the
bare vertex functions.

\begin{table}[th]
%\begin{sidewaystable}
\begin{center}
%\begin{small}
\begin{tabular}{|c||c|c|c|c|c|c|c|c|}
\hline
$ N$ & 1 & 2 & 3 & 4 & 5 & 6 & ... & $\infty$  \\
\hline
%\hhline{|===============|}
%
$\sigma_{b,j} (e^2/\hbar)$ & 0.021 &  0.029 &  0.034 & 0.036 &  0.038 & 0.039&  & 0.047 \\
$\rho_b (h/e^2)$ & 3.72 & 5.41 & 7.09 & 8.76 & 10.44 & 12.11 & & $\left(3.62 + 1.68(N-1)\right)$ \\
\hline
\end{tabular}
\caption{\label{tab:rhocase1} Rotor conductivity ($\sigma_{b,j}$)
and resistivity jump $\rho_b$ at the MIT with fractionally charged
bosonic parton $e_\ast = e/N$. }
\end{center}
\end{table}

$\Im D^{(\lambda)}, \Im D^{(\tilde{A})}$ physically denote the
density of states of the emergent fields that scatter with
$\varphi$, which are broad in the $(\Omega, \bm q)$ space due to
the couplings with the $\varphi$ fields. Below, we ignore the bare
dynamics. $D^{(\lambda), (\tilde{A})}$ in the large-$\bf{N}$ limit
reads
\begin{align}
D^{(\lambda)} (\Omega, \bm q) &=  \frac{1}{\Pi_{b}} ,\nonumber \\
D^{(\tilde A)}_{ii} (\Omega, \bm q) &= \frac{\Pi^{J}_b + (N-1)
\Pi^{J}_f}{ (\Pi_b^{J})^2 + N \Pi^{J}_b \Pi^{J}_f} = \frac{N-1}{N}
\frac{1}{\Pi_b^J}+ \frac{1}{N} \frac{1}{\Pi_b^J + N \Pi_f^{J}} ,
\end{align}
where $D^{(\tilde A)}_{ii}$ reduces to the MIT without charge
fractionalization as discussed in Ref.~\onlinecite{resistivity2}
when $N=1$. For $N>1$, as only the linear combination of
$\tilde{A}_j$, i.e. $\sum_{j=0}^{N-1} \tilde{A}_j$ couples to the
spinon field $f$ and is Landau damped, there is a factor
$\frac{1}{N}$ for the Landau damped component of the gauge field
propagator $D^{(\tilde A)}_{ii}$, which may also be understood as
the $a$ component of gauge field in Eq.~\eqref{cpndual}. The rest
part is not Landau damped, and is determined solely by $\Pi^J_b$.
Note that as $\Im \Pi_f^J \gg \Im \Pi_b^J$ in the limit $\mu \gg
T$, the Landau damped component can be approximated as
$\frac{1}{N} \frac{1}{\Pi_b^J + N\Pi_f^J } \approx \frac{1}{N}
\frac{1}{\Pi_b^J (\Omega=0, \bm q)+ N \Pi_f^J (\Omega, \bm q)}$,
and be treated in the same way as Ref.~\onlinecite{resistivity2}
for the gauge field $a$. On the other hand, the first term in
$D^{(\tilde A)}_{ii}$ should be determined for generic $\Omega,
\bm q$. Using the standard expression for polarizations $\Pi$,
\begin{align}
\Pi_b (\Omega, \bm q) &= \frac{T}{2} \sum_m \int_{\bm k}
\tau_\lambda \frac{1}{(\nu_m + \Omega_n)^2 + \epsilon_{\bm k + \bm
q}^2 }\frac{1}{\nu_m^2 + \epsilon_{\bm k}^2}
|_{ \ii \Omega_n \rightarrow  \Omega + \ii \delta }\nonumber\\
\Pi_b^J (\Omega, \bm q) &=  \frac{T}{2} \sum_m \int_{\bm k}
\tau_{\tilde A}\frac{1}{(\nu_m + \Omega_n)^2 + \epsilon_{\bm k +
\bm q}^2 }\frac{1}{\nu_m^2 + \epsilon_{\bm k}^2}
|_{ \ii \Omega_n \rightarrow  \Omega + \ii \delta }\nonumber\\
\Pi_{f}^J (\Omega, \bm q) &= - \frac{T}{2} \sum_m \int_{\bm k}
\frac{(2\bm k \times \hat{\bm q})^2}{(2 m)^2} \frac{1}{\ii
(\omega_m + \Omega_n) - \xi_{\bm k + \bm q} } \frac{1}{\ii
\omega_m  - \xi_{\bm k } } |_{ \ii \Omega_n \rightarrow \Omega +
\ii \delta },
\end{align}
Eq.~\eqref{eq:QBE} can be solved self-consistently. In
Tab.~\ref{tab:rhocase1}, we show $\sigma_{b,j}$ and the final
resistivity $\rho_b = (N \sigma_{b,j}^{-1})/2$ at different $N$,
again the factor of $1/2$ arises from the two spin/valley flavors.
$\rho_b$ increases roughly linearly with $N$, and the fit of the
data points at different $N$ gives \beqn \rho_b = \left(R^{(0)} +
R^{(1)} (N - 1) \right) \frac{h}{e^2} = \left(3.62 +
1.68(N-1)\right) \frac{h}{e^2}. \eeqn

\end{document}